\documentclass[aps,preprint,showkeys,nofootinbib,prb,onecolumn]{revtex4-1}%
\usepackage{amssymb}
\usepackage{amsfonts}
\usepackage{amsmath}
\usepackage{float}
\usepackage{graphicx}
\usepackage{natbib}
\usepackage{hyperref}
\usepackage{caption}%
\providecommand{\U}[1]{\protect\rule{.1in}{.1in}}

\begin{document}
\preprint{Review Paper}
\title[Susceptibilty of Dirac Fermions and Dilute Alloys]{Magnetic  
       Susceptibility of Dirac Fermions, Bi-Sb Alloys, Interacting Bloch
       Fermions, Dilute Nonmagnetic Alloys, and Kondo Alloys }
\author{${}^{\ast\dagger\ddagger}$Felix A. Buot, ${}^{\dagger}$Roland E. S. 
        Otadoy, and ${}^{\dagger}$Karla B. Rivero}
\affiliation{${}^{\ast}$Computational Materials Science Center, George Mason 
             University, Fairfax, VA 22030, U.S.A,}
\affiliation{${}^{\dagger}$TCSE Center, Spintronics Group, Physics Department, 
             University of San Carlos, Talamban, Cebu 6000, Philippines}         
\affiliation{${}^{\ddagger}$C\&LB Research Institute, Carmen, Cebu 6005, 
             Philippines}

\begin{abstract}
Wide ranging interest in Dirac Hamiltomian is due to the emergence of novel
materials, namely, graphene, topological insulators and superconductors, the
newly-discovered Weyl semimetals, and still actively-sought after Majorana
fermions in real materials. We give a brief review of the relativistic Dirac
quantum mechanics and its impact in the developments of modern physics. The
quantum band dynamics of Dirac Hamiltonian is crucial in resolving the giant
diamagnetism of bismuth and Bi-Sb alloys. Quantitative agreement of the theory
with the experiments on Bi-Sb alloys has been achieved, and physically
meaningful contributions to the diamagnetism has been identified. We also
treat relativistic Dirac fermion as an interband dynamics in uniform magnetic
fields. For the interacting Bloch electrons, the role of translation symmetry
for calculating the magnetic susceptibility avoids any approximation to second
order in the field. The magnetic susceptibility of Hubbard model and those of
Fermi liquids are readily obtained as limiting cases. The expressions for
magnetic susceptibility of dilute nonmagnetic alloys give a firm theoretical
foundation of the empirical formulas used in fitting experimental results. For
completeness, the magnetic susceptibility of dilute magnetic or Kondo alloys
is also given for high and low temperature regimes.
\end{abstract}

\email{fbuot@gmu.edu}

\maketitle
\tableofcontents

\section{Introduction}

There is a growing interest from the nanoscience and nanotechnology community
of the Dirac Hamiltonian in condensed matter owing to the emergence of novel
materials that mimic the relativistic Dirac quantum mechanical behavior. The
purpose of this review is to present a unified treatment of Dirac Hamiltonian
in solids and relativistic Dirac quantum mechanics from the point of view of
energy-band quantum dynamics\cite{wannierBT} coupled with the lattice Weyl
transformation techniques.\cite{buot4c} This unified view seems to explicitly
emerge in the calculation of the magnetic susceptibility of bismuth and Bi-Sb
alloys.\cite{buotMcClure} Large diamagnetism in solids has been attributed to
interband quantum dynamics,\cite{adams} often giving large $g$-factor due to
pseudo-spin degrees of freedom and induced magnetic field. These are inherent
in interband quantum dynamics. In most cases we are referring to two bands
only which could be Kramer's degenerate bands.\cite{cohenblount,adams,
buotMcClure} On the other hand, the classic Landau-Peierls diamagnetism is
purely a single-band dynamical (orbital) effect. More recently, Fukuyama et
al\cite{review2} give a review on diamagnetism of Dirac electrons in solids
from a theoretical perspective of many-body Green's function technique.
However, no theoretical calculations were made and compared with the beautiful
experiments of Wherli\cite{wherli} on the diamagnetism of Bi-Sb alloys.

Here we employ a theoretical perspective of band dynamics that has a long
history even before the time of Peierls,\cite{adams} who introduced the
Peierls phase factor, and Wannier who introduced the Wannier 
function.\cite{blount} This band dynamical treatment is generalized to the
Dirac relativistic quantum mechanics and many-body condensed matter
physics.\cite{erikkols, suttorpdegroot, buot3, buot4} A detailed
calculation\cite{supplement} of the diamagnetism of Bi-Sb alloys using the
theory of Buot and McClure\cite{buotMcClure} yields outstanding quantitative
agreement with experimental results.\cite{wherli}We also give a review of the
calculations of the magnetic susceptibility of other systems.

\subsection{Historical Background}

Firstly, in this section we will give a background on relativistic Dirac
fermions, the beautiful Dirac equation and Dirac's declaration of anti-matter
and discovery of positrons. We focus on its impact in motivating the
development of modern physics, in particular condensed matter physics leading
to a plethora of quasiparticle excitations with exotic properties. Because of
this, condensed matter physics has become a low-energy proving ground on some
of theoretical concepts in quantum field theory, high-energy elementary
particle physics, and cosmology, where the Dirac equation has been extended
and consistently deformed in ways exposing novel excitations/quasiparticles in
physical systems.

In the space-time domain of condensed matter physics, it is interesting that
the relativistic Dirac-like equation was first recognized in the $\vec{k}%
\cdot\vec{p}$ \ band theory of bismuth and Bi-Sb alloys.\cite{wolf} This
scientific historical event is like a repeat of what has happened in ordinary
space-time with the quantum theory of relativistic electrons published\ in
1928 by Dirac\cite{dirac1928} in the form of what is now known as the Dirac
equation with \textit{four-component} fields. \ The following year,
Weyl\cite{weyl} showed that for massless fermions, a simpler equation would
suffice, involving \textit{two-component} fields as opposed to the
four-component fields of Dirac equation. These massless fermions is now known
as the Weyl spin $\frac{1}{2}$ fermions. About nine years later, in 1937,
Majorana\cite{majorana} was searching for a real version of the Dirac equation
which is still Lorentz invariant. Thus, by imposing reality constraint of the
Dirac equation, other solutions were obtained by Majorana still describing
spin $\frac{1}{2}$ fermions, whose outstanding unique property is that they
are their own anti-particles. By virtue of the fact that the complex field of
the Dirac fermions is replaced by real fields, one refers to Dirac fermions as
consisting of two Majorana fermions. Thus, Majorana fermions are often
referred to as half-femions.

Developments in physics in the early $20$th century is not only confined to
relativistic fermions but also to relativistic bosons which act as
force-fields between particles. These particle-particle interactions are
usually mediated by massless bosons such as photons, gluons, and gravitons.
However, relativistic massive bosons, the so-called Proca particles, mediate
the weak interactions between elementary particles. These are, for example,
the $W^{\pm}$ and $Z^{0}$ spin-$1$ heavy vector bosons. Twenty five years
later after Majorana, Skyrme\cite{skyrme} proposed a topological soliton in
quantum field theory which is now referred to as skyrmion. Then in 1978,
Callan et al\cite{callan} proposed another topological objects known as
merons. In magnetic systems, skyrmions, merons and bimerons are closely 
related.

Thus, whereas Weyl demonstrated the existence of massless relativistic
spin-$\frac{1}{2}$ fermions, Proca\cite{proca} demonstrated in 1936 the
existence of massive relativistic spin-$1$ bosons. In crystals, phonons are
generally classified as the Nambu-Goldstone modes, but in the interaction
between Cooper pairs in BCS superconductivity theory no massless phonons are
present, only massive plasmalike excitations.\cite{frampton}In gauge-field
theory of standard model, the Proca action is the gauge-fixed version of the
Stueckelberg action which is a special case of Higgs mechanism through which
the boson acquires mass.

Weyl fermions are irreducible representations of the proper Lorentz group,
they are considered as building blocks of any kind of fermion field. Weyl
fermions are either right chiral or left chiral but can not have both
components. A general fermion field can be described by two Weyl fields, one
left-chiral and one right-chiral. It is worth mentioning that helicity and
chirality coincide for massless fermions. By combining massless Weyl-fermion
fields of different chiralities, one has not really generated a mass but has
created a group-theoretical framework where mass can be allowed in the Dirac
Lagrangian since the mass term must contain two different chiralities. Thus, a
\textit{massive} fermion must have a left-chiral as well as a right-chiral 
component.

\subsubsection{Parallel events in condensed matter physics}

Surprisingly the above chain of scientific events in ordinary space-time have
been followed, although much later experimentally, by corresponding events in
the space-time domain of condensed matter systems. Although, as early as 1937,
Herring\cite{herring} have already theoretically predicted the possibility of
Weyl points in band theory of solid state physics. In more recent years
Nielsen and Ninomiya\cite{nielsenninomiya} have suggested that excess of
particles with a particular chirality were associated with Weyl fermions and
could be observable in solids.

As mentioned before, the explicit form of Dirac Hamiltonian first appeared in
bismuth and Bi-Sb alloys in a paper by Wolff in 1964 upon Blount's
suggestion.\cite{wolf} Actually Cohen\cite{cohen} gave the same form of the
Hamiltonian of bismuth four years earlier in 1960, but did not cast into the
form of Dirac Hamiltonian. Then, with the discovery of graphene 
\cite{graphenediscover} in 2004, massless Dirac fermions were identified in the 
energy-band structure at the $K$-points of the Brillouin zone. The following 
years are marked with understanding of materials with band structures that are 
``topologically'' protected, typically of materials with strong spin-orbit 
coupling. This understanding have taken roots much earlier from works on quantum 
Hall effect (QHE) and quantum spin Hall effect (QSHE). The term 'topological' 
refers to a concept whereby there is a `holographic' quasiparticle-state 
structure often localized at boundaries, domain walls or defects, which is 
topologically protected by the properties of the bulk. This maybe viewed as a 
sort of \textit{entanglement} of the excitation-state with the bulk structure 
and therefore a \textit{highly-nonlocal} property of excitation-state immune to 
local perturbation.\cite{footnote} 

The pertinent measure of this entanglement is the subject of exploding research 
activities on the so-called \textit{topological entanglement entropy}. 
\cite{kitaevpreskill} Indeed, a \textit{holographic interpretation} of 
topological entanglement entopy was given.\cite{ryutakayanagi} This is also 
known in the literature simply as generalized \textit{bulk-boundary 
correspondence}. These years are replete with findings of topological insulators 
and topological superconductors, based on the realization that band theory must 
take into account concepts such as Chern numbers and Berry phases, familiar in 
quantum field theory of elementary particles, quantum-Hall effect, Peierls phase 
of a plaquette, Aharonov-Bohm effect, and in Born-Oppenheimer approximations. 
Indeed, theorists are now engaged in the exciting fields of topological field 
theory (TFT) and topological band theory (TBT). This attests to the merging of 
elementary particle physics, cosmology, and condensed matter physics.

Topological superconductors are simply the analogue of topological insulators.
Whereas, topological insulators (TI) have a bulk band-gap with odd number of
\textit{relativistic Dirac fermions} and gapless modes on the surface,
topological superconductors (TSC) are certain type of full superconducting gap
TSC in the bulk, but due to the inherent particle-hole symmetry, have gapless
modes of chargeless \textit{Majorana `edge' states}, also associated with
\textit{Andreev bound state} (ABS)\cite{tanaka, otadoy} on its boundaries,
interfaces (e.g. interface between topological insulators and superconductors), 
and defects, supported by the bulk topological invariants. The emergence of 
topological insulators and superconductors also brought to light two types of 
quasiparticles properties, namely, (a) local or `trivial' quasiparticles, and 
(b) nonlocal, `nontrivial' or topological quasiparticles, often termed as 
\textit{topological charge}. The second type are robust states, these cannot be 
created or removed by any local operators. The topological charge is also called 
the topological quantum number and is sometimes called the winding number of the 
solution. The topological quantum numbers are topological invariants associated 
with topological defects or soliton-type solutions of some set of differential 
equations modeling a physical system.

For example, the first $3$D topological insulator was predicted for Bi$_{1 - 
x}$-Sb$_{x}$ system with $(0 \leq x \leq 0.04)$, where the theoretical band
structure calculation predicts the $3$D topological insulator phase in 
Bi$_{0.9}$Sb$_{0.1}$.\cite{hasan} These developments are followed around 
mid-year of $2015$ with the experimental discovery of Weyl fermions which were
identified in the so-called Weyl semimetals. Specifically, the historic 
angle-resolved photoemission spectroscopy (ARPES) experiments performed on Ta-As 
has revealed Weyl fermions in the bulk.\cite{ARPES1, ARPES} Likewise, similar 
experiments on photonic crystal have identified Weyl points (not Weyl fermions) 
inside the photonic crystal.\cite{photonicweyl}Weyl points differ from Dirac 
points since the former has two-component wavefunction whereas the later 
generally has four component wavefunction, e.g. in graphene the $K_{\pm}$ 
points in the Brillouin zone end ows the two chiralities for a Dirac Hamiltonian 
for graphene.

The excitement about the experimental discovery of Weyl semimetals has to do 
with its great potential for ultrafast devices. The absence of backscattering 
for Weyl fermions is related to the so-called Klein paradox in quantum 
electrodynamcis due to conservation of chirality. Because of this, Weyl fermions 
cannot be localized by random potential scattering in the form of Anderson 
localization\cite{Anderson} common to massive electrons. Moreover, with the 
eficient electron-hole pairs screening of impurities, mobilities of Weyl 
fermions are expected to be more than order of magnitude higher than the best Si 
transistors. In passing, we could say that the crossing of the bands of 
different symmetry properties in Bi$_{1-x}$-Sb$_{x}$ alloys \cite{supplement} 
for antimony concentration of $x=0.04$ might also serves as a Weyl semimetal, 
ignoring questions of topological stability.

\paragraph{The hunt for Majorana fermions}

On the other hand, the \textit{experimental} Majorana fermions in solid state 
systems remains a challenging pursuit, a bit of `holy grail' which one can 
perhaps draw a parallel with the search for Higgs particles in high-energy 
physics.\cite{higgs} In conventional condensed matter system, $c^{\dagger}$ and 
its Hermitian conjugate $c$ is a \textit{physically distinct operator} that 
annihilates electron or creates a hole. Since particles and antiparticles have 
opposite conserved charges, a Majorana fermion with its own antiparticle is a 
necessarily uncharged fermion. In his original paper, Majorana fermions can have 
arbitrary spin, so with spin zero it is still a fermion since the Majorana 
annihilation and creation operators still obey the anticommutation rule. For 
example, a mixture of particles and anti-particles of the form, 
\begin{align*}
 \gamma = \int dr \Big( f(r) \Psi_{\uparrow} +g(r) \Psi_{\downarrow} 
      + f^{\ast}(r) \Psi_{\downarrow}^{\dagger} +g^{\ast}(r) 
      \Psi_{\uparrow}^{\dagger} \Big) = \gamma^{\dagger}
\end{align*}
indicates a chargeless and spinless Majorana fermions, often referred to as a
\textit{featureless} Majorana fermions. Another form of $\gamma = 
uc_{\sigma}^{\dagger} + u^{\ast} c_{\sigma}$ with equal spin projection, say a 
triplet or effectively spinless since spin degree of freedom does not have to be
accounted for, is also a Majorana field operator, $\gamma = \gamma^{\dagger}$.
This particular $\gamma$-form arises by imposing the Majorana condition 
\cite{fradkin} on the Bogoliubov-de Gennes (BdG) equation of superconductivity.

A conventional quasiparticle in superconductor is a broken Cooper pair, an
excitation called a Bogoliubov quasiparticle and can have spin $\frac{1}{2}$.
It is simply a linear combination of creation and annihilation operators,
namely, $b_{\alpha} = u_{\alpha} a_{\alpha} + v_{\alpha} a_{-\alpha}^{\dagger}$ 
and $b_{-\alpha}^{\dagger} = u_{\alpha}^{\ast} a_{-\alpha}^{\dagger} + 
v_{\alpha}^{\ast} a_{\alpha}$, where $u_{\alpha}$ and $v_{\alpha}$ are the 
components of the wavefunction of the Bogoliubov-de Gennes (BdG) equation. The 
requirement of Bogoliubov canonical transformation is that $\Big[ b_{\alpha}, 
b_{-\alpha}^{\dagger} \Big] = \Big( a_{\alpha}a_{-\alpha}^{\dagger} + 
a_{-\alpha}^{\dagger} a_{\alpha}\Big)$, therefore we must have $\Big( \big| 
u_{\alpha} \big|^{2} + \big| v_{\alpha} \big|^{2} \Big) = 1$ and $\Big( 
u_{\alpha} v_{\alpha}^{\ast} + v_{\alpha}^{\ast} u_{\alpha}\Big) = 0$. The 
particle created by the operator $b_{\alpha}^{\dagger}$ is a fermion, the 
so-called Bogoliubov quasiparticle (or ``Bogoliubon''). It combines the 
properties of a negatively charged electron and a positively charged hole. 
Indeed, Majorana fermion must have a form of superposition of particle and 
anti-particle. However, the creation and annihilation operators for Bogoliubons 
are still distinct. Thus, whereas charge prevents Majorana from emerging in a 
metal, on the other hand \textit{distinct} creation and annihilation operators 
through superposition of electrons and holes with opposite spins is preventing 
Majorana quasiparticles in conventional $s$-wave superconductors. If Majorana 
fermion is to appear in the solid state it must therefore be in the form of 
\textit{still to be experimetally demonstrated }nontrivial emergent Majorana 
excitations in real materials. The attention is focused on topological 
superconductors.

After a theoretical demonstration of the existence of Majoranas at the ends of
a $p$-wave pairing Kitaev-chain, several theoretical demonstrations for the
existence of zero-mode Majorana bound states (MBS) follow. In Kitaev's
prediction, inducing some types of superconductivity, known as the proximity
effect, would cause the formation of Majoranas. These emergent particles are
stable (Majorana degenerate bound states) and do not annihilate each other
(unless the chain or wire is too short) because they are spatially separated.
Thus, $p$-superconductors provide a natural hunting ground for Majoranas.

The search for Majorana has also pave the way for the \textit{novel physics of 
zero modes} of the extended Dirac equation with inhomogeneous mass term that 
varies with position (corresponding to the momentum-dependent pairing potential 
in BdG equation), yielding a \textit{kink-soliton} solution in $1$-D, a 
\textit{vortex} solution in $2$-D, and a \textit{magnetic monopole} in $3$-D. 
\cite{jackiw} In condensed matter physics, the experimental search for Majorana 
is focused on exotic superconductors, namely, in triplet $p$-wave 
superconductivity in one dimension ($1$D), where MBS are located at both ends of 
the superconducting wire, and triplet $p+ip$-wave superconductivity in two 
dimensions ($2$D) where the MBS has been theoretically demonstrated to reside at 
the core of the vortex at an interface. In triplet $3$D superconductors, the MBS 
is at the core of the `hedgehog' configuration. These topological 
superconductors realize topological phases that support non-Abelian exotic 
excitations at their boundaries and at topological defects (e.g., hedgehog 
configuration). Most importantly, zero-energy modes localize at the ends of a 
$1$D topological $p$-wave superconductor, and bind to vortices in the $2$D 
topological $p + ip$-wave superconducting case. These zero-modes are precisely 
the condensed matter realization of Majorana fermions that are now being 
vigorously pursued. Moreover, engineered hererostructures using proximity effect 
with the $s$-wave superconductor, the so-called proximity-induced topological 
superconductor are correspondingly and vigorously also being pursued.

From the technological point of view these topologically-robust Majorana 
excitations are envisaged to implement quantum computing where braiding 
operation constitutes bits manipulation, analogous to the Yang-Baxter equations 
first introduced in statistical mechanics. The Majorana number density is 
limited to an integer (mode $2$), i.e., $0$ and $1$, thus ideally representing 
a quantum bit. An intrigung proposal is a superconductor-topological 
insulator-superconductor (STIS) junction that forms a nonchiral $1$D wire for 
Majorana fermions. These (STIS) junctions can be combined into circuits which 
allow for the creation, manipulation, and fusion of Majorana bound states for 
topological quantum computation.\cite{fukane} There are also proposals for 
interacting non-Abelian anyons as Majorana fermions in Kitaev's honeycomb 
lattice model.\cite{nonabelianmajor} Indeed, Majorana fermions obey non-Abelian 
statistics, since Majorana fermions can have arbitrary spin statistics.

Several groups have experimentally reported detecting Majorana fermions. 
\cite{das, deng, finck} More recently, a Princeton group\cite{princeton}, have 
reported detecting Majorana by following Kitaev's prediction that, under the 
correct conditions a Majorana fermion bound states would appear at each end of a 
superconducting wire or Kitaev chain.

In summary, it is worth emphasizing that condensed matter physics has become the 
low-energy playground for discovering various quasiparticles and exotic 
topological excitations, which were mostly first proposed in quantum field 
theory of elementary particles, namely, Dirac fermions, Weyl fermions, Proca 
particles, vortices, skyrmions, merons, bimerons and other 
topologically-protected quasiparticles obeying non-Abelian and anyon 
statistical properties.

\section{The Relativistic Dirac Hamiltonian}

In ordinary space-time and in its original form, the Dirac equation is given
by Dirac in the following forms\cite{patch}
\begin{equation}
 i\hbar \frac{\partial \psi (x,t)}{\partial t} = \Bigg[ \beta mc^{2} + c \Bigg(
      {\sum_{j}^{3}} \alpha_{j} p_{j} \Bigg) \Bigg] \psi (x,t)  \label{diraceq}
\end{equation}
where
\begin{equation*}
 \beta =
      \begin{pmatrix}
       I_{2} & 0\\
       0 & -I_{2}
      \end{pmatrix}
 = \gamma^{0}, \ \ \alpha_{\mu} = 
      \begin{pmatrix}
       0 & \sigma_{\mu}\\
       \sigma_{\mu} & 0
      \end{pmatrix}
 , \ \ \mu = 1, 2,3
\end{equation*}
Therefore the Dirac Hamiltonian is of the matrix form is,
\begin{equation}
 \mathcal{H}_{\textrm{Dirac}} = 
      \begin{pmatrix}
       mc^{2} & c \vec{p} \cdot \vec{\sigma}\\
       c \vec{p} \cdot \vec{\sigma} & -mc^{2}%
      \end{pmatrix}
 \label{diracform}
\end{equation}
Equation (\ref{diracform}) is the form that can occur in the $k\cdot p$
treatment of two-band theory of solids. The Dirac Hamiltonian has eigenvalues
given by
\begin{equation*}
 E = \pm \sqrt{m^{2}c^{4} + c^{2} |p|^{2}}
\end{equation*}
Equation (\ref{diraceq}) can be rewritten as
\begin{equation*}
 i \hbar \beta \frac{\partial \psi (x,t)}{\partial ct} = \Bigg[ mc + \Bigg( 
      \sum_{j}^{3} (\beta \alpha_{j}) p_{j} \Bigg) \Bigg] \psi (x,t)
\end{equation*}
and is usually given in its relativistic invariance form, as%
\begin{equation}
 i\hbar \gamma^{\mu} \partial_{\mu} \psi - mc \psi = 0 \label{eq1}%
\end{equation}
where in the Dirac $\gamma$-basis,%
\begin{equation}
 \gamma^{0} = \beta = 
      \begin{pmatrix}
       I_{2} & 0\\
       0 & -I_{2}
      \end{pmatrix}
 , \label{diracBasis}
\end{equation}
\begin{equation}
 \gamma^{\mu} = \gamma^{0} \alpha^{\mu} = 
      \begin{pmatrix}
       0 & \sigma_{\mu} \\
       -\sigma_{\mu} & 0
      \end{pmatrix}
 \label{gama_mu}
\end{equation}

These $\gamma$-matrices satisfy the relations of Clifford algebra,
\begin{equation*}
 \{ \gamma^{\mu},\gamma^{\upsilon} \} = 2 \eta^{\mu \upsilon}
\end{equation*}
where the curly bracket stands for anticommutator. The anticommutator of
$\{ \sigma_{x},\sigma_{y} \} = 2 \delta_{xy} I$. Thus
\begin{equation}
 \eta^{ij} = 
      \begin{pmatrix}
       1 & 0 & 0 & 0\\
       0 & -1 & 0 & 0\\
       0 & 0 & -1 & 0\\
       0 & 0 & 0 & -1
      \end{pmatrix}
 \label{DiracMetric}%
\end{equation}
defining Clifford algebra over a pseudo-orthogonal $4$-D space with metric 
signature $(1,3)$ given by the matrix $\eta^{ij}$. It is a constant in special 
relativity but a function of space-time in general relativity. Equation 
(\ref{eq1}) is an eigenvalue equation for the $4$-momentum operator, $i \hbar 
\gamma^{\mu} \partial_{\mu}$ for the free Dirac electrons with eigenvalue equal 
to $mc$.

The Dirac $\gamma$-basis has the chirality operator given by,
\begin{equation*}
 \gamma^{5} = 
      \begin{pmatrix}
       0 & I_{2}\\
       I_{2} & 0
      \end{pmatrix}
\end{equation*}
The number $5$ is a remnant of old notation in which $\gamma^{0}$ was called 
``$\gamma^{4}$''. Although $\gamma^{5}$ is not one of the gamma matrices of 
Clifford algebra over a pseudo-orthogonal $4$-D space, this matrix is useful in 
discussions of quantum-mechanical chirality. For example, using the 
$\gamma$-matrices in the Dirac basis, a Dirac field can be projected onto its 
\textit{left-handed} and \textit{right-handed} components by,
\begin{align*}
 \psi_{L} & =\frac{1}{2} \ \Big(1 - \gamma^{5} \Big) \psi\\
 \psi_{R} & =\frac{1}{2} \ \Big(1 + \gamma^{5} \Big) \psi
\end{align*}
Thus, we have
\begin{align*}
 \gamma^{5} \psi_{L} & = - \psi_{L}\\
 \gamma^{5} \psi_{R} & = + \psi_{R}
\end{align*}
with eigenvalues $\pm 1$. The $\gamma^{5}$ anticommutes with all the 
$\gamma^{\mu}$ matrices.

On the other hand, the set $\{\gamma^{0}, \gamma^{1}, \gamma^{2}, \gamma^{3}, 
i\gamma^{5} \}$ forms the basis of the Clifford algebra in $5$-spacetime 
dimensions for the metric signature $(1,4)$. The higher-dimensional 
$\gamma$-matrices generalize the $4$-dimensional $\gamma$-matrices of Dirac to 
arbitrary dimensions. The higher-dimensional $\gamma$-matrices are utilized in 
\textit{relativistically invariant} wave equations for fermions spinors in 
arbitrary space-time dimensions, notably in string theory and supergravity.

\subsection{Weyl representation of Dirac equation}

The Weyl representation of the $\gamma$-matrices is also known as the
\textit{chiral} basis, in which $\gamma^{k} (k = 1, 2, 3)$ remains the same but 
$\gamma^{0}$ is different, and so $\gamma^{5}$ is also different and diagonal. A 
possible choice of the Weyl basis is
\begin{equation*}
 \gamma^{0} = 
      \begin{pmatrix}
       0 & -I_{2}\\
       -I_{2} & 0
      \end{pmatrix}
 , \ \gamma^{k} = 
      \begin{pmatrix}
       0 & \sigma^{k}\\
       -\sigma^{k} & 0
      \end{pmatrix}
 , \ \gamma^{5} =
      \begin{pmatrix}
       I_{2} & 0\\
       0 & -I_{2}
      \end{pmatrix}
\end{equation*}
In the \textit{Weyl representation}, the Dirac equations reads%
\begin{align}
 E\psi_{1} = & c \vec{\sigma} \cdot \vec{p} \ \psi_{1} + mc^{2} \psi_{2} 
               \nonumber\\
 E\psi_{2} = & -c \vec{\sigma} \cdot \vec{p} \ \psi_{2} + mc^{2} \psi_{1} 
               \label{DiracWeyl}%
\end{align}
It is worthwhile to point out that Eq.(\ref{DiracWeyl}) is interesting and bears 
resemblance to the eigenvalue equation for graphene if the $\pm$ chirality 
degree of freedom of the zero-mode dispersions from the two inequivalent 
$K_{\pm}$ points in the Brillouin zone (BZ) is taken into account. The isospin 
degree of freedom arises from the degeneracy of these inequivalent $K_{\pm}$ 
points at the BZ corners. Thus, the $K_{\pm}$ points Dirac electrons in graphene 
fits the Weyl representation of relativitic Dirac equations. It is the isospin 
degree of freedom that gives each $K$ point in BZ a definite chirality. This has 
several exotic physical consequences as will be discussed in a separate paper by 
the authors.

Thus, in accounting for the $K_{\pm}$ points of the Brillouin zone of graphene 
its Hamiltonian exactly resembles the relativistic Dirac Hamiltonian in the Weyl 
representation with zero mass. Note that if $m = 0$, we only need to solve one 
of the $2 \times 2$ matrix equations, yielding massless Weyl fermions with 
definite chirality (note also that chirality and helicity are both good quantum 
labels for massless fermions). This is clarified in what follows.

In matrix form we have,
\begin{equation*}
 E 
      \begin{pmatrix}
       \psi_{1}\\
       \psi_{2}
      \end{pmatrix}
 = c \sigma \cdot p \gamma^{5} 
      \begin{pmatrix}
       \psi_{1}\\
       \psi_{2}
      \end{pmatrix}
 - mc^{2} \gamma^{0} 
      \begin{pmatrix}
       \psi_{1}\\
       \psi_{2}
      \end{pmatrix}
\end{equation*}
The eigenvalues are,
\begin{equation*}
 E = \pm \sqrt{c^{2}p^{2} + m^{2}c^{4}}
\end{equation*}
Here, $\psi_{1}$ and $\psi_{2}$ are the eigenstates of the chirality operator
$\gamma^{5}$. The Weyl basis has the advantage that its chiral projections
take a simple form,
\begin{align*}
 \psi_{R} & = \frac{1}{2} \Big(1 - \gamma^{5} \Big) 
      \begin{pmatrix}
       \psi_{1}\\
       \psi_{2}
      \end{pmatrix}
 = 
      \begin{pmatrix}
       \psi_{1}\\
       0
      \end{pmatrix}
 \\
 \psi_{L} & = \frac{1}{2} \Big(1 + \gamma^{5} \Big) 
      \begin{pmatrix}
       \psi_{1}\\
       \psi_{2}
      \end{pmatrix}
 =
      \begin{pmatrix}
       0\\
       \psi_{2}
      \end{pmatrix}
\end{align*}
Hence, in Weyl chirality $\gamma$-basis, we have
\begin{equation*}
 \gamma^{5} \psi_{R} = \psi_{R}, \ \ \ \ \gamma^{5} \psi_{L} = -\psi_{L}
\end{equation*}
Thus chirality and helicity are a good quantum numbers for Weyl massless fermions.

\subsection{Majorana representation of Dirac equation}

The Majorana representation of Dirac equation can occur in $p$-wave
superconductors. In the the Majorana $\gamma$-basis, all of the Dirac matrices
are imaginary and spinors $\Psi$ are real. We have
\begin{align}
 \hat{\gamma}^{0} & =
      \begin{pmatrix}
       0 & \sigma_{2}\\
       \sigma_{2} & 0
      \end{pmatrix}
 , \ \ \label{maj1}\\
 \hat{\alpha}^{1} & = \gamma^{0} \gamma^{1} = -
      \begin{pmatrix}
       0 & \sigma^{1}\\
       \sigma^{1} & 0
      \end{pmatrix}
 , \ \hat{\gamma}^{1} = 
      \begin{pmatrix}
       i \sigma_{3} & 0\\
       0 & i\sigma_{3}
      \end{pmatrix}
 , \label{maj2}\\
 \hat{\alpha}^{2} & = \gamma^{0} \gamma^{2} = 
      \begin{pmatrix}
       I_{2} & 0\\
       0 & -I_{2}
      \end{pmatrix}
 , \ \ \hat{\gamma}^{2} = 
      \begin{pmatrix}
       0 & -\sigma_{2}\\
       \sigma_{2} & 0
      \end{pmatrix}
 , \ \label{maj3}\\
 \hat{\alpha}^{3} & = \gamma^{0} \gamma^{3} = -
      \begin{pmatrix}
       0 & \sigma^{3}\\
       \sigma^{3} & 0
      \end{pmatrix}
 , \ \ \hat{\gamma}^{3} = 
      \begin{pmatrix}
       -i \sigma_{1} & 0\\
       0 & -i\sigma_{1}
      \end{pmatrix}
 . \label{maj4}
\end{align}
The gamma matrices are imaginary to obtain the particle-physics metric $(1,3)$, 
i.e., $(+,-,-,-)$ in which squared masses are positive.

The Majorana relativistic equation is thus given by
\begin{equation*}
 i \hbar \gamma^{\mu} \partial_{\mu} \psi - mc \psi = 0
\end{equation*}
Using the relation $\alpha_{\mu} = \gamma^{0} \gamma^{\mu}$, we obtain after 
multiplying by $\gamma^{0}$,
\begin{equation*}
 i \hbar \gamma^{0} \gamma^{\mu} \partial_{\mu} \psi - \gamma^{0} mc \psi = 0
\end{equation*}
which reduces to
\begin{equation*}
 i \hbar \frac{\partial}{\partial t} \Psi = \Big( c\hat{\alpha} \cdot p + 
      \hat{\gamma}^{0} mc^{2} \Big) \Psi
\end{equation*}
where $p=-i\hbar \nabla$ is imaginary, $\hat{\alpha}$ is real, and $\hat
{\gamma}^{0} = \beta_{\textrm{majorana}}$ is imaginary. Thus, the Majorana 
relativistic equation is real, giving real solution $\Psi$, which ensures charge 
neutrality of spin $\frac{1}{2}$ particle which is its own antiparticle. Note 
that in Dirac equation the \textit{Dirac mass} couples left- and right-handed
chirality, whereas in Majorana equation, the \textit{Majorana mass} couples
particle with antiparticle.

In terms of matrix equation, we have
\begin{align}
 i & \hbar \frac{\partial}{\partial t} 
      \begin{pmatrix}
       \Psi_{1}\\
       \Psi_{2}
      \end{pmatrix}
      \nonumber\\
 & = \Big(c \hat{\alpha}^{\mu} p_{\mu} + \hat{\gamma}^{0} mc^{2} \Big) \Psi 
      \nonumber\\
 & = 
      \begin{pmatrix}
       I_{2} cp_{y} & -\sigma^{1} cp_{x} - \sigma^{3} cp_{z} 
         + \sigma_{2} mc^{2}\\
       -\sigma^{1} cp_{x} - \sigma^{3} cp_{z} + \sigma_{2} mc^{2} & -I_{2}cp_{y}
      \end{pmatrix}
      \begin{pmatrix}
       \Psi_{1}\\
       \Psi_{2}
      \end{pmatrix}
      \label{majoranaMatrixEquation}
\end{align}
Therefore we have coupled set of equations,
\begin{align*}
 i \hbar \frac{\partial}{\partial t} \Psi_{1} & = I_{2} cp_{y} \Psi_{1} - 
      (\sigma^{1} cp_{x} + \sigma^{3} cp_{z} - \sigma_{2} mc^{2}) \Psi_{2}\\
 i \hbar \frac{\partial}{\partial t} \Psi_{2} & = -I_{2} cp_{y} \Psi_{2} - 
      (\sigma^{1} cp_{x} + \sigma^{3} cp_{z} - \sigma_{2} mc^{2}) \Psi_{1}
\end{align*}
In $(2+1)$-dimensional version, the matrix Hamiltonian of 
Eq.(\ref{majoranaMatrixEquation}) can be written as
\begin{equation*}
 \mathcal{H}_{M} = 
      \begin{pmatrix}
       I_{2} cp_{y} & -\sigma^{1} cp_{x} + \sigma_{2}mc^{2}\\
       -\sigma^{1} cp_{x} + \sigma_{2} mc^{2} & -I_{2} cp_{y}%
      \end{pmatrix}
\end{equation*}
By multiplying the wavefunction by a global phase equal to $\pi$, this can
also be given by an equivalent expression,
\begin{equation*}
 \mathcal{H}_{M} = 
      \begin{pmatrix}
       -I_{2} cp_{y} & \sigma^{1} cp_{x} - \sigma_{2} mc^{2}\\
       \sigma^{1} cp_{x} - \sigma_{2} mc^{2} & I_{2} cp_{y}%
      \end{pmatrix}
 .
\end{equation*}

In the case of Majorana fermions in superconductor, the Majorana mass term
$mc^{2}$ corresponds to absolute value of the pair potential $|\Delta|$. 
However, in general $\Delta = \Delta_{R} + i \Delta_{I}$. Thus, to have a real 
Majorana equation in $p$-wave superconductor, we can expect the following form 
for the self-adjoint Majorana Hamiltonian in superconductor,\cite{chamonetal}%
\begin{equation*}
 \mathcal{H}_{M} = 
      \begin{pmatrix}
       -I_{2} p_{y} & \sigma^{1} p_{x} + iI_{2} \Delta_{I} 
         - \sigma_{2} \Delta_{R}\\
       \sigma^{1} p_{x} - iI_{2} \Delta_{I} - \sigma_{2} \Delta_{R} 
         & I_{2} p_{y}
      \end{pmatrix}
\end{equation*}
where we have factored out the constant $c$ or equated to unity. This can be
substituted by a constant group velocity, $v$, for zero-gap or `massless' 
states.

\section{Superconductor: Bogoliubov-de Gennes equation}

There is a formal analogy between the Dirac relativistic equation, BCS theory
of superconductivity and BdG equation. We shall see later that their 
Hamiltonians have resemblance with the Hamiltonian of Bi and Bi-Sb alloys.

We can intuitively understand how the phonon-mediated electron-electron 
scattering in metals results in attractive interaction, i.e., by exchange of 
bosons leading to Cooper pairing. The instantenous emission and absorption of 
highly-energetic phonons by interacting pair of electrons near the Fermi surface 
with opposite initial momentum, $-k$ and $k$, and with final momentum states of 
the Cooper pair in the form
\begin{equation*}
 -k + q \leftrightsquigarrow k - q
\end{equation*}
where $q$ is the phonon wavevector will endow \textit{opposite impulses} to the 
pair. On the average, this becomes an attractive-binding force between them, 
resulting in a zero-total-momentum BCS bound state. In general, this attractive 
interaction dominates in highly-dense-electron metal system with efficient 
Coulomb-potential screening. This condition yields nonzero mean-field average 
(`pairing') for $\Big\langle \psi (x) \psi (x^{\prime}) \Big\rangle$ and its 
complex conjugate $\Big\langle \psi^{\dagger}(x) \psi^{\dagger} (x^{\prime}) 
\Big\rangle$.

\subsection{The BCS theory of superconductivity}

In BCS theory, not only the momentum will have opposite sign but pairs must have 
opposite spin as well to maximize interaction, because the exchange interaction 
between parallel spins will reduce the attractive phonon-mediated interaction. 
Thus, the inital momentum of phonon-mediated interaction between pair of 
electron is of the set $\Big\{ \hbar \vec{k}_{\uparrow}, -\hbar 
\vec{k}_{\downarrow} \Big\}$. This may also be interpreted as conservation of 
helicity for the pair. There are of course other boson-mediated pairing 
mechanisms which are more complex. For example, depending on the band structure 
a non-BCS pairing with nonzero total momentum of the pair in the form
\begin{equation*}
 -k + q \leftrightsquigarrow k + \kappa - q
\end{equation*}
where $q$ is the phonon wavevector, or
\begin{equation*}
 -k + \kappa + q \leftrightsquigarrow k + \kappa - q
\end{equation*}
via spin-singlet channel are referred to as the Fulde-Ferrel-Larkin-Ovchinnikov 
(FFLO) pairing.\cite{cblm, zgw} The FFLO pairing was also proposed for 
\textit{doped} Weyl semimetals which have a shifted Fermi surface brought by 
doping. The pairing theory was generalized to nonzero relative angular momentum 
type of pairing, such as the $p$-wave pairing, to be discussed later in 
connection with topological superconductors.

\subsubsection{Effective BCS Hamiltonian}

The simplest mean-field effective BCS many-body Hamiltonian can be rewritten in 
the equivalent \textit{first-quantized} version through the BdG formalism. In 
the BdG formalism, the \textit{eigenvaue problem} is essentially a 
first-quantized version of the second quantized effective BCS Hamiltonian 
formalism.

Consider a Hamiltonian of many-fermion system interacting through a 
spin-independent potential $\Phi (x)$,
\begin{align}
 H = & \int \psi_{\sigma}^{\dagger}(x) H_{0} \psi_{\sigma}(x)
       \label{gen_bcsH} \nonumber\\
     & +\frac{1}{2} \sum_{\sigma \sigma^{\prime}} \int \int 
       d^{3}x d^{3}x^{\prime} \Big[ \psi_{\sigma}^{\dagger}(x) 
       \psi_{\sigma^{\prime}}^{\dagger}(x^{\prime}) \Phi(x - x^{\prime})  
       \psi_{\sigma^{\prime}}(x^{\prime}) \psi_{\sigma}(x) \Big]
\end{align}
where $\sigma$ is the spin index, $\Phi(x - x^{\prime})$ is the 
translationally-invariant electron-phonon interaction potential, and
\begin{equation*}
 H_{0} = -\frac{\hbar^{2} \nabla^{2}}{2m} - \mu
\end{equation*}
since only the electrons near the Fermi surface can be redistributed or 
disturbed by the electron-electron interaction. Taking the Fourier transform to 
momentum space in finite volume $V$, we have
\begin{align*}
 \psi_{\sigma}(x) &  = \frac{1}{\sqrt{V}} \sum_{k} a_{k} e^{ik \cdot x}\\
 V(x) & = \frac{1}{\sqrt{V}} \sum_{q} \Phi_{q} e^{iq \cdot x}
\end{align*}
then we obtain
\begin{equation*}
 H = \sum_{k \sigma} \xi_{k} a_{k \sigma}^{\dagger} a_{k \sigma} + \frac{1}{2V}
     \sum_{\sigma \sigma^{\prime}} \sum_{k, k^{\prime}, q} \Phi_{q} 
     a_{k + q, \sigma}^{\dagger} a_{k^{\prime} - q, \sigma^{\prime}}^{\dagger} 
     a_{k^{\prime} \sigma^{\prime}} a_{k \sigma}
\end{equation*}
where
\begin{equation*}
 \xi_{k} = E(k) - \mu
\end{equation*}
Restricting to pairing of fermions with zero total momentum and opposite spin, 
such that if $k_{\uparrow}$ is occupied so is $-k_{\downarrow}$, we get the 
BCS Hamiltonian
\begin{equation*}
 H_{\textrm{BCS}} = \sum_{k \sigma} \xi_{k} a_{k \sigma}^{\dagger} a_{k \sigma} 
      + \frac{1}{V} \sum_{k, k^{\prime}} \Phi_{k - k^{\prime}} 
      a_{k^{\prime},\uparrow}^{\dagger} a_{-k^{\prime}, \downarrow}^{\dagger} 
      a_{-k,\downarrow} a_{k,\uparrow}
\end{equation*}
The bound pairs are not bose particles. We can define a creation and 
annihilation operators for pairs as follows
\begin{equation*}
 c_{k} = a_{-k, \downarrow} a_{k,\uparrow}, \ \ 
 c_{k}^{\dagger} = a_{k, \uparrow}^{\dagger} a_{-k,\downarrow}^{\dagger}, \ \ 
 c_{k}^{\dagger} c_{k} = n_{k,\uparrow} n_{-k,\downarrow}
\end{equation*}
we have
\begin{align*}
 \Big[ c_{k}, c_{k^{\prime}}^{\dagger} \Big]_{-} & = (1 - n_{k \uparrow} - n_{-k 
      \downarrow}) \delta_{k k^{\prime}}\\
 \big[ c_{k}, c_{k^{\prime}} \Big]_{-} & = 0
\end{align*}
where $n_{k \sigma} = a_{k, \sigma}^{\dagger} a_{k, \sigma}$, but the 
anticommutator given by
\begin{equation*}
 \Big[ c_{k}, c_{k^{\prime}} \Big]_{+} = 2c_{k} c_{k^{\prime}} \ (1 - \delta_{k 
      k^{\prime}})
\end{equation*}
is different from those of Bose particles. This is due to the terms, $(n_{k 
\uparrow} + n_{k \downarrow})$ in $(1 - n_{k \uparrow} - n_{k \downarrow})$ and 
$\delta_{kk^{\prime}}$ in $(1 - \delta_{kk^{\prime}})$ which comes from the 
Pauli exclusion principle. The Hamiltonian in terms of the $c$'s can be 
rewritten as
\begin{align}
 H_{\textrm{reduced}} = & \sum_{k \sigma} \xi_{k} c_{k \sigma}^{\dagger} 
      c_{k \sigma} + \frac{1}{V} \sum_{k, k^{\prime}} \Phi_{k - k^{\prime}} 
      c_{k^{\prime}}^{\dagger} c_{k}\nonumber\\
 = & \sum_{k > k_{F}, \sigma} \epsilon_{k} c_{k \sigma}^{\dagger} c_{k \sigma} 
      + \sum_{k < k_{F}, \sigma} |\epsilon_{k}| c_{k \sigma} 
      c_{k\sigma}^{\dagger} \nonumber\\
 & + \frac{1}{V} \sum_{k, k^{\prime}} \Phi_{k - k^{\prime}} 
      c_{k^{\prime}}^{\dagger} c_{k} - \sum_{k < k_{F}, \sigma} \epsilon_{k} 
      (1 - n_{k \uparrow} - n_{-k \downarrow})  \label{reducedH}
\end{align}

The effective BCS Hamiltonian in the mean-field approximation for $c_{k}$ is
obtained by writing
\begin{equation*}
 \Delta_{k} = -\frac{1}{V} \sum_{k^{\prime}} \Phi_{k - k^{\prime}} \langle 
      c_{k} \rangle
\end{equation*}
At finite temperature, the expression for the thermal average $\langle
c_{k^{\prime}} \rangle = \frac{\Delta_{k^{\prime}}}{2E(k^{\prime})} \tanh 
\Big( \frac{\beta E_{k^{\prime}}}{2} \Big)$ yields the self-consistency 
condition for $\Delta_{k}$, namely,
\begin{equation}
 \Delta_{k} = -\frac{1}{V} \sum_{k^{\prime}} \Phi_{k - k^{\prime}} 
      \frac{\Delta_{k^{\prime}}}{2E(k^{\prime})} \tanh \Bigg( 
      \frac{\beta E_{k^{\prime}}}{2} \Bigg)  \label{gapeq}
\end{equation}
Therefore Eq. (\ref{reducedH}) becomes in the mean field approximation,
\begin{align}
 H_{MF} & = \sum_{k \sigma} \xi_{k} a_{k \sigma}^{\dagger} a_{k \sigma} +
            \sum_{k} \Delta_{k}^{\ast} a_{-k,\downarrow} a_{k,\uparrow} + 
            H.c. \nonumber\\
        & = \frac{1}{2} \sum_{k \sigma} \bigg( \xi_{k} a_{k \sigma}^{\dagger} 
            a_{k \sigma} - \xi_{k} a_{-k \sigma} a_{-k \sigma}^{\dagger}\bigg) 
            + \sum_{k} \Delta_{k}^{\ast} a_{-k,\downarrow} a_{k,\uparrow} + H.c. 
            + \frac{1}{2} \sum_{k} \xi_{k} \label{Hmf1}
\end{align}

The spectrum of the last Hamiltonian can readily be found using the Nambu 
spinor,
\begin{equation}
 A_{k} =
      \begin{pmatrix}
       a_{k,\uparrow}\\
       a_{-k,\downarrow}^{\dagger}
      \end{pmatrix}
      \label{nambu}
\end{equation}
In terms of the Nambu spinor, the BCS Hamiltonian reads, by discarding
irrelevant constant terms, as
\begin{align*}
 H_{nambu} & = \sum_{k} A_{k}^{\dagger}
      \begin{pmatrix}
       \xi_{k} & \Delta_{k}\\
       \Delta_{k}^{\ast} & -\xi_{k}%
      \end{pmatrix}
 A_{k}\\
 & = \frac{1}{2} \sum_{k\sigma} \bigg( \xi_{k} a_{k \sigma}^{\dagger} 
     a_{k\sigma} - \xi_{k} a_{-k\sigma} a_{-k \sigma}^{\dagger} \bigg) +
     \sum_{k} \Delta_{k}^{\ast} a_{-k,\downarrow} a_{k,\uparrow} + H.c.
\end{align*}
The $k$-dependent spectrum, $\epsilon_{k}$, can readily be calculated using
the BdG first quantized equation, namely,
\begin{equation*}
      \begin{pmatrix}
       \xi_{k} & \Delta_{k}\\
       \Delta_{k}^{\ast} & -\xi_{k}%
      \end{pmatrix}
      \begin{pmatrix}
       u \\
       v
      \end{pmatrix}
 = \epsilon_{k} 
      \begin{pmatrix}
       u \\
       v
      \end{pmatrix}
\end{equation*}
which yields
\begin{equation*}
 \epsilon_{k} = \pm \sqrt{\xi_{k}^{2} + |\Delta_{k}|^{2}}
\end{equation*}

\subsection{Bogoliubov Quasiparticles}

We now expand $A_{k}$ in terms of the eigenfunctions of the Hamiltonian. This
is known as the \textit{Bogoliubov transformation}. We have,
\begin{equation*}
 \hat{\Gamma} = UA_{k}
\end{equation*}
where $U$ is the matrix of the eigenfunctions
\begin{equation*}
 U = 
      \begin{pmatrix}
       u & -v\\
       v & u
      \end{pmatrix}
\end{equation*}
where $|u|^{2} + |v|^{2} = 1$ by normality condition, and $u^{\ast} v - v^{\ast} 
u = 0$ by the othogonality condition. We therefore have
\begin{align}
      \begin{pmatrix}
       \gamma_{k,\uparrow} \\
       \gamma_{-k,\downarrow}
      \end{pmatrix}
 & = 
      \begin{pmatrix}
       u a_{k,\uparrow} - v a_{-k,\downarrow}^{\dagger} \\
       u a_{-k,\downarrow}^{\dagger} + v a_{k,\uparrow}
      \end{pmatrix}
      \nonumber\\
      \begin{pmatrix}
       \gamma_{k,\uparrow}^{\dagger}\\
       \gamma_{-k,\downarrow}^{\dagger}
      \end{pmatrix}
 & =
      \begin{pmatrix}
       u^{\ast} a_{k,\uparrow}^{\dagger} - v^{\ast} a_{-k,\downarrow} \\
       u^{\ast} a_{-k,\downarrow}+v^{\ast} a_{k,\uparrow}^{\dagger}
      \end{pmatrix}
      \label{bogotransform}
\end{align}
with inverse transformation as
\begin{equation*}
      \begin{pmatrix}
       a_{k,\uparrow} \\
       a_{-k,\downarrow}^{\dagger}
      \end{pmatrix}
 = 
      \begin{pmatrix}
       u \gamma_{k,\uparrow} + v \gamma_{-k,\downarrow}\\
       u \gamma_{-k,\downarrow} - v \gamma_{k,\uparrow}
      \end{pmatrix}
\end{equation*}
Note that the Bogoliubov $\gamma$'s are explicitly combinations of
\textit{particle} and \textit{antiparticle} operators, this is inherent in
superconductivity physics. The commutation relation is
\begin{eqnarray*}
 \Big\{ \gamma_{k,\uparrow}, \gamma_{k,\uparrow}^{\dagger} \Big\} = 
      |u|^{2} + |v|^{2} = 1 \\
 \big\{ \gamma_{k,\uparrow}, \gamma_{-k,\downarrow}^{\dagger} \Big\} = 0
\end{eqnarray*}
In terms of $\gamma$'s the BCS Hamiltonian can now be written as%
\begin{equation}
 H = \sum_{k} \Big\{ \gamma_{k,\uparrow}^{\dagger} \xi_{k} \gamma_{k,\uparrow} 
     - \gamma_{-k,\downarrow}^{\dagger} \xi_{k} \gamma_{-k,\downarrow} + 
     \gamma_{k,\uparrow}^{\dagger} \Delta_{k} \gamma_{-k,\downarrow} +       
     \gamma_{-k,\downarrow}^{\dagger} \Delta_{k}^{\ast} \gamma_{k,\uparrow}
     \Big\} \label{BquasipartH}
\end{equation}
which is the Hamiltonian for the Bogoliubov quasiparticles.

\subsubsection{The Heisenberg equation of motion: \textit{first-quantized BdG
equation}}
\begin{align}
 i \hbar \frac{\partial}{\partial t} \gamma_{k,\uparrow} & = 
      [\gamma_{k,\uparrow}, H] \nonumber\\
 i \hbar\frac{\partial}{\partial t} \gamma_{-k,\downarrow} & = 
      [\gamma_{-k,\downarrow}, H]  \label{heisenberg_eq}
\end{align}
We readily obtain the effective Schr\"{o}dinger equation,
\begin{align}
 i \hbar \frac{\partial}{\partial t} \gamma_{k,\uparrow} & = 
      [\gamma_{k,\uparrow}, H] \nonumber\\
 & = \xi_{k} \gamma_{k,\uparrow} + \Delta_{k} \gamma_{-k,\downarrow}
      \label{heisenberg_eq1} \\
 \nonumber \\  
 i \hbar \frac{\partial}{\partial t} \gamma_{-k,\downarrow} & = 
      [\gamma_{-k,\downarrow}, H] \nonumber\\
 & = (-\xi_{k}) \gamma_{-k,\downarrow} + \Delta_{k}^{\ast} \gamma_{k,\uparrow} 
      \label{heisenberg_eq2}
\end{align}
Therefore in matrix form, we have the \textit{first-quantized Schr\"{o}dinger
equation} known as the BdG equation,
\begin{equation}
 i \hbar \frac{\partial}{\partial t} 
      \begin{pmatrix}
       \gamma_{k,\uparrow}\\
       \gamma_{-k,\downarrow}
      \end{pmatrix}
 =
      \begin{pmatrix}
       \xi_{k} & \Delta_{k}\\
       \Delta_{k}^{\ast} & -\xi_{k}
      \end{pmatrix}
      \begin{pmatrix}
       \gamma_{k,\uparrow}\\
       \gamma_{-k,\downarrow}
      \end{pmatrix}
      \label{heisenberg_matrixeq}%
\end{equation}
with eigenvalues
\begin{equation*}
 \varepsilon_{k} = \pm \sqrt{\xi_{k}^{2} + |\Delta_{k}|^{2}}
\end{equation*}

\subsubsection{Eigenfunctions}

We can determine the eigenfunctions by the BdG matrix equation, Eq.
(\ref{heisenberg_matrixeq})
\begin{equation*}
 \varepsilon_{k}
      \begin{pmatrix}
       \gamma_{k,\uparrow}\\
       \gamma_{-k,\downarrow}
      \end{pmatrix}
 =
      \begin{pmatrix}
       \xi_{k} & \Delta_{k}\\
       \Delta_{k}^{\ast} & -\xi_{k}
      \end{pmatrix}
      \begin{pmatrix}
       \gamma_{k,\uparrow}\\
       \gamma_{-k,\downarrow}
      \end{pmatrix}
\end{equation*}
which yields,
\begin{eqnarray}
 \frac{\gamma_{-k,\downarrow}}{\gamma_{k,\uparrow}} =
      - \frac{(\xi_{k} - \varepsilon_{k})}{\Delta_{k}} = 
      - \frac{\bigg( \xi_{k} - \pm \sqrt{\xi_{k}^{2} + |\Delta_{k}|^{2}} 
      \bigg)}{\Delta_{k}} \label{result1}\\
 \frac{\gamma_{k,\uparrow}}{\gamma_{-k,\downarrow}} = 
      \frac{(\xi_{k} + \varepsilon_{k})}{\Delta_{k}^{\ast}} = 
      \frac{\bigg( \xi_{k} + \pm \sqrt{\xi_{k}^{2} + |\Delta_{k}|^{2}} 
      \bigg)}{\Delta_{k}^{\ast}} \label{result2}
\end{eqnarray}
The above determines the component of the eigenfunction in terms of its ratio only.

\subsubsection{Diagonalization by an orthogonal transformation}

Consider the `geometric' transformation $U$ given by
\begin{equation*}
 U = 
      \begin{pmatrix}
       \cos\theta & -\sin\theta\\
       \sin\theta & \cos\theta
      \end{pmatrix}
\end{equation*}
We identify the following expressions:
\begin{equation*}
 (\xi_{k} - \epsilon_{k}) \cos \theta + \Delta \sin \theta = 0
\end{equation*}
which yields
\begin{eqnarray*}
 \cos^{2} \theta = \frac{\Delta^{2}}{\Delta^{2} + (\xi_{k} - 
                   \epsilon_{k})^{2}}\\
 \sin \theta \cos \theta = \Bigg( \frac{-(\xi_{k} - 
                   \epsilon_{k})\Delta}{\Delta^{2} + (\xi_{k} - 
                   \epsilon_{k})^{2}} \Bigg)
\end{eqnarray*}
and the alternate expressions
\begin{equation*}
 \Delta^{\ast} \cos \theta - (\xi_{k} + \epsilon_{k}) \sin \theta = 0
\end{equation*}
which yields,%
\begin{eqnarray*}
 \cos^{2} \theta = \Bigg( \frac{(\xi_{k} + 
                   \epsilon_{k})^{2}}{(\xi_{k} + \epsilon_{k})^{2} + 
                   \Delta^{\ast 2}} \Bigg) \\
 \sin \theta \cos \theta = \Bigg( \frac{(\xi_{k} + \epsilon_{k}) 
                   \Delta^{\ast}}{(\xi_{k} + \epsilon_{k})^{2} + \Delta^{\ast 
                   2}} \Bigg)
\end{eqnarray*}
Note that
\begin{equation*}
 1 + \frac{\Delta^{2 \ast}}{(\xi_{k} + \epsilon_{k})^{2}} = 1 + 
      \frac{(\xi_{k} - \epsilon_{k})^{2}}{\Delta^{2}}
\end{equation*}
so we have two equvalent expressions for $\cos^{2} \theta$ and $\sin \theta
\cos \theta$, which will be handy in the diagonalization that follows.

Having obtained the expression for the \ cosine and sine functions of $\theta$, 
we now proceed to diagonalize the mean-field BdG Hamiltonian as
\begin{align*}
 & 
      \begin{pmatrix}
       \cos \theta & \sin \theta \\
       -\sin \theta & \cos \theta
      \end{pmatrix}
      \begin{pmatrix}
       \xi_{k} & \Delta_{k}\\
       \Delta_{k}^{\ast} & -\xi_{k}
      \end{pmatrix}
      \begin{pmatrix}
       \cos \theta & -\sin \theta\\
       \sin \theta & \cos \theta
      \end{pmatrix}
 \\
 & =
      \begin{pmatrix}
       \begin{Bmatrix}
        [\xi_{k} \cos^{2} \theta + \Delta \cos \theta \sin \theta] \\
        + [\Delta_{k}^{\ast} \sin \theta \cos \theta - \xi_{k} \sin^{2} \theta] 
       \end{Bmatrix}
       &
       \begin{Bmatrix}
        [-\xi_{k} \cos \theta \sin \theta + \Delta_{k} \cos^{2} \theta] \\
        + [-\Delta_{k}^{\ast} \sin^{2} \theta - \xi_{k} \sin \theta \cos 
           \theta]   
       \end{Bmatrix}
       \\
       \begin{Bmatrix}
        - [\xi_{k} \sin \theta \cos \theta + \Delta_{k} \sin^{2} \theta] \\
        + [\Delta_{k}^{\ast} \cos^{2} \theta - \xi_{k} \cos \theta \sin \theta] 
       \end{Bmatrix}
       &
       \begin{Bmatrix}
        - [-\xi_{k} \sin^{2} \theta + \Delta_{k} \sin \theta \cos \theta] \\
        + [-\Delta_{k}^{\ast} \cos \theta \sin \theta - \xi_{k} \cos^{2} 
           \theta] 
       \end{Bmatrix}
      \end{pmatrix}
\end{align*}
We have for the diagonal elements
\begin{equation}
 \Big[ \xi_{k} \cos^{2} \theta + \Delta \cos \theta \sin \theta \Big] + 
      \Big[ \Delta_{k}^{\ast} \sin \theta \cos \theta - \xi_{k} \sin^{2} 
\theta \Big] = \Bigg\{ \frac{(\xi_{k} + \epsilon_{k})^{2} + 
     \Delta_{k}^{2\ast}}{(\xi_{k} + \epsilon_{k})^{2} + \Delta^{\ast2}} \Bigg\}
     \epsilon_{k} = \epsilon_{k} \label{diag1}
\end{equation}
and
\begin{align}
 \Big[ \xi_{k} \sin^{2} \theta & - \Delta_{k} \sin \theta \cos \theta \Big]
      - \Big[ \Delta_{k}^{\ast} \cos \theta \sin \theta + \xi_{k} \cos^{2} 
      \theta \Big] \nonumber\\
 & = \Bigg( \frac{1}{(\xi_{k} + \epsilon_{k})^{2} + \Delta^{\ast2}} \Bigg)  
      \Big\{ -[\Delta_{k}^{2\ast} + (\xi_{k} + \epsilon_{k})^{2}] \epsilon_{k} 
      \Big\} \epsilon_{k} \nonumber \\
 & = -\epsilon_{k} \label{diag2}
\end{align}
One can also readily show that the off-diagonal elements are identically zero.

\subsubsection{Chirality: Doubling the degrees of freedom}

We can introduce chirality and helicity degrees of freedom\cite{footnote2} for
each energy band by extending the Nambu field operator, Eq. (\ref{nambu}), to
four components, namely,
\begin{equation*}
 \Psi_{a} \equiv 
      \begin{pmatrix}
       a_{k,\uparrow}\\
       a_{k,\downarrow}\\
       a_{-k,\uparrow}^{\dagger}\\
       a_{-k,\downarrow}^{\dagger}%
      \end{pmatrix}
\end{equation*}
Thus, aside from the original particle-hole degrees of freedom we have now
introduce the spin degrees of freedom. Consider simplifying the Hamiltonian as
follows,
\begin{align*}
 H_{a} = & \Psi_{a}^{\dagger}
       \begin{pmatrix}
        1 & 0 & 0 & 0\\
        0 & 1 & 0 & 0\\
        0 & 0 & -1 & 0\\
        0 & 0 & 0 & -1
       \end{pmatrix}
       \Psi_{a} \\
 H_{a} = & a_{k,\uparrow}^{\dagger} a_{k,\uparrow} 
           + a_{k,\downarrow}^{\dagger} a_{k,\downarrow} 
           - a_{-k,\uparrow} a_{-k,\uparrow}^{ \dagger}
           - a_{-k,\downarrow} a_{-k,\downarrow}^{\dagger}
\end{align*}
The BdG Hamiltonian for $|\Delta_{k}| = 0$ becomes
\begin{align*}
 H_{|\Delta_{k}| = 0} & = \frac{1}{2} 
      \begin{pmatrix}
       \xi_{k} & 0 & 0 & 0\\
       0 & \xi_{k} & 0 & 0\\
       0 & 0 & -\xi_{-k} & 0\\
       0 & 0 & 0 & -\xi_{-k}%
      \end{pmatrix}
      \\
 & = \frac{1}{2} \xi_{k} \sigma_{z} \otimes I_{2}
\end{align*}

\subsubsection{The pairing potential}

We write the pairing Hamiltonian as
\begin{align*}
 H_{\Delta} = & \Delta c_{k \uparrow}^{\dagger} c_{-k \downarrow}^{\dagger} 
              + \Delta^{\ast} c_{-k\downarrow} c_{k\uparrow}\\
 & = \frac{1}{2} \Bigg[ \Delta \Big(c_{k \uparrow}^{\dagger} 
     c_{-k\downarrow}^{\dagger} - c_{-k\downarrow}^{\dagger} 
     c_{k\uparrow}^{\dagger} \Big) + \Delta^{\ast} \Big( 
     c_{-k\downarrow} c_{k\uparrow} - c_{k\uparrow} c_{-k\downarrow} \Big)  
     \Bigg]
\end{align*}
In matrix notation, we have,
\begin{align*}
 H_{\Delta} & = \sum_{k} \Psi_{a}^{\dagger} 
      \begin{pmatrix}
       0 & 0 & 0 & \Delta\\
       0 & 0 & -\Delta & 0\\
       0 & -\Delta^{\ast} & 0 & 0\\
       \Delta^{\ast} & 0 & 0 & 0
      \end{pmatrix}
      \Psi_{a}\\
 & = \sum_{k} \Big\{ \Delta \Big( a_{k,\uparrow}^{\dagger} 
     a_{-k,\downarrow}^{\dagger} - a_{-k,\downarrow}^{\dagger} 
     a_{k,\uparrow}^{\dagger} \Big) + \Delta^{\ast} \Big( 
     a_{-k,\downarrow} a_{k,\uparrow} - a_{k,\uparrow} a_{-k,\downarrow} \Big) 
     \Big\}
\end{align*}
We have for complex $\Delta,$
\begin{align*}
 & 
      \begin{pmatrix}
       0 & 0 & 0 & \Delta\\
       0 & 0 & -\Delta & 0\\
       0 & -\Delta^{\ast} & 0 & 0\\
       \Delta^{\ast} & 0 & 0 & 0
      \end{pmatrix}
      \\
 & = 
      \begin{pmatrix}
       0 & 0 & 0 & \Delta_{R}\\
       0 & 0 & -\Delta_{R} & 0\\
       0 & -\Delta_{R} & 0 & 0\\
       \Delta_{R} & 0 & 0 & 0
      \end{pmatrix}
 + 
      \begin{pmatrix}
       0 & 0 & 0 & i\Delta_{I}\\
       0 & 0 & -i\Delta_{I} & 0\\
       0 & i\Delta_{I} & 0 & 0\\
       -i \Delta_{I} & 0 & 0 & 0
      \end{pmatrix}
 \\
 & = -\Delta_{R} \sigma_{y} \otimes \sigma_{y} - \Delta_{I} \sigma_{x} \otimes
     \sigma_{y}
\end{align*}
where
\begin{align*}
 \sigma_{y} \otimes \sigma_{y} & =
      \begin{pmatrix}
       0 & 0 & 0 & -1\\
       0 & 0 & 1 & 0\\
       0 & 1 & 0 & 0\\
       -1 & 0 & 0 & 0
      \end{pmatrix}
 \\
 \sigma_{x} \otimes \sigma_{y} & =
      \begin{pmatrix}
       0 & 0 & 0 & -i\\
       0 & 0 & i & 0\\
       0 & -i & 0 & 0\\
       i & 0 & 0 & 0
      \end{pmatrix}
\end{align*}

Therefore
\begin{equation*}
 H_{BdG} = \xi_{k} \sigma_{z}^{\prime} \otimes I_{2} - \Delta_{R} 
           \sigma_{y}^{\prime} \otimes \sigma_{y} - \Delta_{I} 
           \sigma_{x}^{\prime} \otimes \sigma_{y}
\end{equation*}
where the prime pertains to the particle-hole degrees of freedom.

The Bogoliubov transformation, Eq. (\ref{bogotransform}) can also be extended
to account for the chirality and spin degrees of freedom. The extended BdG
equation is
\begin{equation*}
      \begin{pmatrix}
       (\xi_{k} - \epsilon_{k}) & 0 & 0 & \Delta\\
       0 & (\xi_{k} - \epsilon_{k}) & -\Delta & 0\\
       0 & -\Delta^{\ast} & -(\xi_{k} + \epsilon_{k}) & 0\\
       \Delta^{\ast} & 0 & 0 & -(\xi_{k} + \epsilon_{k})
      \end{pmatrix}
      \begin{pmatrix}
       \gamma_{k,\uparrow}^{\dagger}\\
       \gamma_{k,\downarrow}^{\dagger}\\
       \gamma_{-k,\uparrow}^{\dagger}\\
       \gamma_{-k,\downarrow}^{\dagger}%
      \end{pmatrix}
 = 0
\end{equation*}
\begin{align*}
 (\xi_{k} - \epsilon_{k}) \gamma_{k,\uparrow}^{\dagger} + \Delta 
      \gamma_{-k,\downarrow}^{\dagger} & = 0 \Longrightarrow 
      - \frac{(\xi_{k} - \epsilon_{k})}{\Delta} = \frac{\gamma_{-k, 
      \downarrow}^{\dagger}}{\gamma_{k,\uparrow}^{\dagger}} \\
 (\xi_{k} - \epsilon_{k}) \gamma_{k,\downarrow}^{\dagger} - \Delta 
      \gamma_{-k,\uparrow}^{\dagger} & = 0 \Longrightarrow 
      \frac{(\xi_{k} - \epsilon_{k})}{\Delta} = \frac{\gamma_{-k, 
      \uparrow}^{\dagger}}{\gamma_{k,\downarrow}^{\dagger}} \\
 -\Delta^{\ast} \gamma_{k,\downarrow}^{\dagger} - (\xi_{k} + \epsilon_{k}) 
      \gamma_{-k,\uparrow}^{\dagger} & = 0 \Longrightarrow
      - \frac{(\xi_{k} + \epsilon_{k})}{\Delta^{\ast}} = 
      \frac{\gamma_{k,\downarrow}^{\dagger}}{\gamma_{-k,\uparrow}^{\dagger}}\\
 \Delta^{\ast} \gamma_{k,\uparrow}^{\dagger} - (\xi_{k} + \epsilon_{k})  
      \gamma_{-k,\downarrow}^{\dagger} & = 0 \Longrightarrow 
      \frac{(\xi_{k} + \epsilon_{k})}{\Delta^{\ast}} = 
      \frac{\gamma_{k,\uparrow}^{\dagger}}{\gamma_{-k,\downarrow}^{\dagger}}%
\end{align*}

\subsection{$(p_{x} + i p_{y})$-Wave Pairing for Topological
Superconductors}

In the original BCS treatment, pairing of particles was in a relative $s$-wave
state. However, the pairing theory was generalized to nonzero relative angular 
momentum type of pairing. Indeed, $p$-wave pairing was observed in He$^{3}$. The 
belief is that $d$-wave pairing occurs in heavy fermion and high-$T_{c}$ 
superconductors.

It is for nonzero relative angular momentum pairing that the resulting BdG 
equations yield the form of the Majorana representation of Dirac 
equations.\cite{readGreen} The effective BCS Hamiltonian in the mean-field 
approximation of the phonon-mediated interaction between electrons can thus be 
written in the form
\begin{equation}
 H_{F} = \int \psi^{\dagger}(x) H_{0} \psi(x) + \int \int \Big[ 
         \Delta^{\ast} (x, x^{\prime}) \psi(x) \psi(x^{\prime}) 
         + \Delta(x,x^{\prime}) \psi^{\dagger}(x) \psi^{\dagger}(x^{\prime}) 
         \Big] \label{BCSH}%
\end{equation}
The mean-field interaction via spin-singlet pairing, $\Delta(x, x^{\prime})$, is
\begin{equation*}
 \Delta(x,x^{\prime}) = -g \langle \psi(x) \psi(x^{\prime}) \rangle 
\end{equation*}
where $g$ is a coupling constant. For scattering problems, it is often 
convenient to cast the Hamiltonian in momentum space. We have for the mean-field 
interaction,
\begin{align}
 \Delta_{k} & = -g \langle a_{-k\downarrow} a_{k\uparrow} \rangle \qquad 
                \textrm{for } s \textrm{-wave superconductivity} \nonumber\\
            & = \Delta(k_{x} - ik_{y})\qquad \textrm{ for} (p_{x} + ip_{y})  
                \textrm{-wave superconductivity as } \vec{k} \Longrightarrow 0
 \label{delta_k}%
\end{align}
In the RHS of Eq. (\ref{delta_k}), $\Delta$ is a constant. Thus, BCS 
superconductors can be classified by the symmetry of the anomalous mean-field 
average in the boson-mediated electron-electron interaction $\Delta(x, 
x^{\prime})$. For a $2$-D $(p_{x} + ip_{y})$-wave topological triplet 
superconductors, we have,
\begin{equation*}
 H_{eff} = \int d^{2} k \ \Bigg[ (\varepsilon_{k} - \mu) c_{k}^{\dagger} c_{k}
           + \frac{1}{2} (\Delta_{k}^{\ast} c_{-k} c_{k} + \Delta_{k} 
           c_{k}^{\ast} c_{-k}^{\ast}) \Bigg]
\end{equation*}
where $\varepsilon_{k} \simeq \frac{k^{2}}{2m^{\ast}}$ is the quasiparticle 
kinetic energy and $\mu$ is the effective chemical potential. $\Delta_{k}$ is 
the gap function which is proportional to order parameter of the superconducting 
state. Again, we have the constraint provided by the Pauli exclusion principle, 
namely, only the electrons near the Fermi surface can be redistributed or 
disturbed by the electron-electron interaction. This in contrast, for example, 
for the case of excitons in semiconductors which involve the hydrogen-like 
pairing of holes on top of valence band and electrons at the bottom of the 
conduction band, although interband electron-electron pairing is quite possible 
in graphene\cite{lozoviksokolik} with zero energy-gap so that the Fermi surface 
coincide with the bottom of the conduction band and the top of the valence band. 
Indeed exotic superconductivity for graphene has been predicted upon doping with
carriers.\cite{levitov}

The effective Heisenberg equation of motion of the Hamiltonian, Eq. 
(\ref{BCSH}), is thus given by
\begin{align*}
 i \hbar \frac{\partial}{\partial t} \psi(z) & = [\psi(z), H_{F}] \\
 i \hbar \frac{\partial}{\partial t} \psi^{\dagger}(z) & = 
      [\psi^{\dagger}(z), H_{F}]
\end{align*}
We obtain,
\begin{align}
 i \hbar \frac{\partial}{\partial t} \psi(z) & = (\varepsilon_{k} - \mu) \psi(z) 
      + \int \Delta^{\ast}(z, x^{\prime}) \psi^{\dagger}(x^{\prime}) 
      + \int \psi^{\dagger}(x) \Delta(x, z) \label{schro1} \\
 i \hbar \frac{\partial}{\partial t} \psi^{\dagger} (z) & = 
      \psi^{\dagger}(z) (\varepsilon_{k} - \mu) - \Bigg\{ \int 
      \Delta^{\ast}(z, x^{\prime}) \psi(x^{\prime}) + \int \psi(x) 
      \Delta^{\ast}(x, z) \Bigg\} \label{schro2}%
\end{align}
The above set of coupled equations, Eqs. (\ref{schro1})-(\ref{schro2}) is still 
operator equations. By averting to first quantization, the above equations 
represent the BdG equations. This can be transformed to momentum space, e.g., 
using
\begin{align*}
 \psi(z) & = \frac{1}{\sqrt{V}} \sum_{k} e^{ik \cdot z} a_{k}, \\
 \psi^{\dagger}(z) & = \frac{1}{\sqrt{V}} \sum_{k} e^{-ik \cdot z} 
      a_{k}^{\dagger}.
\end{align*}
For the more complex $p$-wave pairing, solving for the quasiparticle spectrum
may require the use of Bethe \textit{ansatz}.\cite{readGreen}

\section{The Dirac Hamiltonian in Bismuth}

In the space-time domain of condensed matter populated by Bloch electrons
whose band dynamics is characterize by Wannier functions and Bloch functions,
a Dirac-like Hamiltonian appeared in a paper published in 1964 by 
Wolff.\cite{wolf} In the presence of spin-orbit coupling, time reversal, and
space inversion symmetry, the energy bands are doubly degenerate, known as
Kramer's conjugates.

We are interested in the $L$-point of the Brillouin zone where the direct gap
is small. Wolff give the Hamiltonian of bismuth, including spin-orbit
coupling, in a Dirac form as,
\begin{equation}
 \mathcal{H} = \beta \bigg( \frac{E_{g}}{2} \bigg) + \frac{(\Delta k)^{2}}{2m} I 
      + 
      \begin{pmatrix}
       0 & \mathcal{H}_{1}\\
       \mathcal{H}_{1} & 0
      \end{pmatrix}
      \label{eq2}
\end{equation}
where
\begin{equation*}
 \mathcal{H}_{1} = i \Delta k \sum_{\lambda = 1}^{3} K_{\lambda} 
      \sigma_{\lambda}
\end{equation*}
where $K_{\lambda}$'s are determined by the matrix elements, including 
spin-orbit effects, of the velocity operator given by
\begin{equation*}
 \vec{\pi} = \frac{\vec{p}}{m} + \frac{\mu_{o}}{4mc} \bigg( \vec{s} \times 
             \vec{\nabla} V \bigg),
\end{equation*}
where $\mu_{0}$ is the Bohr magneton. This gives the spin-orbit interaction 
correctly to order $\frac{e^{2} \hbar^{2}}{64m^{4}c^{4}}$ in the Hamiltonian, 
and is gauge invariant. Expanding $\mathcal{H}_{1}$ Eq. (\ref{eq2}), we have 
for $\mathcal{H}_{1}$ given by Wolff,
\begin{align*}
 \mathcal{H}_{1} & = \Delta k 
      \begin{pmatrix}
       \operatorname{Re}(t) + i \operatorname{Im}(t) &
         \operatorname{Re}(u) + i \operatorname{Im}(u) \\
       -[\operatorname{Re}(u) - i \operatorname{Im}(u)] & 
         \operatorname{Re}(t) - i \operatorname{Im}(t)
      \end{pmatrix}
 \\
 & = \Delta k 
      \begin{pmatrix}
       t & u\\
       -u^{\ast} & t^{\ast}
      \end{pmatrix}
 .
\end{align*}
The total Hamiltonian is of the form
\begin{equation*}
 \mathcal{H} =
      \begin{pmatrix}
       \Big( \frac{E_{g}}{2} \Big) + \frac{(\Delta k)^{2}}{2m} & 0 & t & u \\
       0 & \Big( \frac{E_{g}}{2} \Big) + \frac{(\Delta k)^{2}}{2m} & -u^{\ast} 
         & t^{\ast}\\
       t & u & \Big( -\frac{E_{g}}{2} \Big) + \frac{(\Delta k)^{2}}{2m} & 0 \\
       -u^{\ast} & t^{\ast} & 0 & \Big( -\frac{E_{g}}{2} \Big) 
         + \frac{(\Delta k)^{2}}{2m}
      \end{pmatrix}
\end{equation*}
Wolff eliminated the $\operatorname{Re}(t)$ by some unitary transformation 
applied to $\mathcal{H}_{1}$. Upon substituting the matrix elements in terms of 
the basis $u_{n}(L, \Delta k = 0)$, we end up with the expression, where we 
explicitly indicates the symmetry types of the corresponding band-edge 
wavefunctions at the $L$-point of the Brillouin zone as,
\begin{equation*}
 \mathcal{H} = 
      \begin{pmatrix}
       u_{n}(L, \Delta k = 0)  & L_{5} & L_{6} & L_{7} & L_{8} \\
       L_{5} & \Big( \frac{E_{g}}{2} \Big) + \frac{(\Delta k)^{2}}{2m} & 0 
         & \Delta k \cdot \langle L_{5} |\vec{\pi}| L_{7} \rangle 
         & \Delta k \cdot \langle L_{5} |\vec{\pi}| L_{8} \rangle \\
       L_{6} & 0 & \Big( \frac{E_{g}}{2} \Big) + \frac{(\Delta k)^{2}}{2m} 
         & \Delta k \cdot \langle L_{6} |\vec{\pi}| L_{7} \rangle  
         & \Delta k \cdot \langle L_{6} |\vec{\pi}| L_{8} \rangle \\
       L_{7} & \Delta k \cdot \langle L_{7} |\vec{\pi}| L_{5} \rangle  
         & \Delta k \cdot \langle L_{7} |\vec{\pi}| L_{6} \rangle  
         & \Big( -\frac{E_{g}}{2} \Big) + \frac{(\Delta k)^{2}}{2m} & 0 \\
       L_{8} & \Delta k \cdot \langle L_{8} |\vec{\pi}| L_{5} \rangle 
         & \Delta k \cdot \langle L_{8} |\vec{\pi}| L_{6} \rangle  
         & 0 & \Big( -\frac{E_{g}}{2} \Big) + \frac{(\Delta k)^{2}}{2m}
      \end{pmatrix}
\end{equation*}
The first two columns (rows) are for degenerate bands $\{L_{5}, L_{6}\}$ and for 
the last two columns (rows), for the degenerate $\{L_{7}, L_{8}\}$ at the 
$L$-point of the Brillouin zone. Observe that if $\Big( \pm \frac{E_{g}}{2} 
\Big) + \frac{(\Delta k)^{2}}{2m} \Longrightarrow 0$, we obtain the Weyl 
Hamiltonian. This condition holds at low energies with vanishing band-gap. This 
demonstrates the relative high-probability of finding Weyl fermions in solid 
state systems compared to finding Majorana fermions, which calls for more exotic
quasiparticles that is yet to be found in real materials.

Earlier, Cohen in 1960 wrote the bismuth Hamiltonian\cite{cohen} as%
\begin{equation*}
 \mathcal{H} = 
      \begin{pmatrix}
       K_{0} - E_{g} & 0 & t & u\\
       0 & K_{0} - E_{g} & -u^{\ast} & t^{\ast}\\
       t^{\ast} & -u & K_{1} & 0\\
       u^{\ast} & t & 0 & K_{1}%
      \end{pmatrix}
\end{equation*}
where the zero of energy is at the minimum of the conduction band.

For convenience in what follows, we recast the full $\vec{k} \cdot \vec{p}$
Hamiltonian as
\begin{equation*}
 \mathcal{H} = 
      \begin{pmatrix}
       u_{n}(L) & L_{5} & L_{6} & L_{7} & L_{8}\\
       L_{5} & K_{1} + \Delta & 0 & t & u^{\ast}\\
       L_{6} & 0 & K_{1} + \Delta & -u & t^{\ast}\\
       L_{7} & t^{\ast} & -u^{\ast} & K_{0} - \Delta & 0\\
       L_{8} & u & t & 0 & K_{0} - \Delta
      \end{pmatrix}
\end{equation*}
where the symmetry types of the corresponding band-edge wavefunctions of the 
first two columns (rows) are for degenerate bands $\{L_{5}, L_{6}\}$ and for the 
last two columns (rows), for the degenerate $\{L_{7}, L_{8}\}$ at the $L$-point 
of the Brillouin zone. Energies are measured from the center of the band gap, 
$\Delta = \frac{1}{2} E_{g}$, where $E_{g}$ is the direct band gap at $L$-point, 
$K_{1} = \frac{1}{2} k^{2} + R_{1}$, $K_{0} = \frac{1}{2} k^{2} + R_{0}$ where 
$R_{1}$ and $R_{0}$ are contributions quadratic in $k$ coming from bands other 
than the valence and conduction vands at $L$-point. The terms $t$ and $u$ are 
$\vec{k} \cdot \vec{\pi}$ matrix elements where $\vec{\pi} = \frac{\vec{p}}{m} + 
\frac{1}{2(mc)^{2}} \Big( \vec{s} \times \vec{\nabla} V \Big)$ includes the 
effect of spin-orbit coupling. The phases of $t$ and $u$ can be chosen 
independently without changing the form of the $\vec{k} \cdot \vec{p}$ 
Hamiltonian matrix. This fact allows for transformation of the above Hamiltonian 
to the Dirac form given by Wolff. We have
\begin{align*}
 u & = \langle L_{8} |\pi_{y}| L_{5} \rangle k_{y} + \langle L_{8} |\pi_{z}| 
       L_{5} \rangle k_{z}\\
 & \equiv q_{2} k_{y} + q_{3} k_{z} \\
 t & = \langle L_{8} |\pi_{x}| L_{6} \rangle k_{x}
\end{align*}
The coordinate axes are the binary (along $\Sigma$ symmetry line), bisetrix
(on the $\sigma$ plane) and trigonal (along $\Lambda$ symmetry line) crystal
direction ($b, b, t, \Longrightarrow x, y, z$ coordinate system). The 
eigenvalues of the $\vec{k} \cdot \vec{p}$ Hamiltonian matrix in the Lax 
two-band model, which neglect both $K_{1}$ and $K_{0}$ are
\begin{equation*}
 E = \pm \sqrt{\Delta^{2} + |t|^{2} + |u|^{2}}
\end{equation*}
where the $+$ and $-$ energy levels are doubly degenerate or Kramer conjugates. 
The reason for the neglect of $K_{1}$ and $K_{0}$ is that the most significant 
contribution to $\chi_{\perp}$ is $\chi_{L}^{22}$ when the magnetic field is 
along the bisetrix direction. Due to the large curvature of the energy bands in 
the binary and trigonal directions, the contribution of $K_{1}$ and $K_{0}$ are 
neglibly small. The Fermi surface is ellipsoidal and is tilted about the binary 
axis, there being a cross term in $k_{y}k_{z}$ from $|u|^{2}$. In the principal 
axes of the Fermi surface ellipsoid, the relation $\operatorname{Re} 
(q_{2}^{\prime} q_{3}^{\prime}) = 0$ holds. We therefore choose, $q_{1}^{\prime} 
= Q_{1}, q_{2}^{\prime} = -iQ_{2}$, and $q_{3}^{\prime} = Q_{3}$, where $Q_{1}, 
Q_{2}$, and $Q_{3}$ are all real valued. We have in the principal axes,
\begin{equation}
 \mathcal{H}^{\prime} =
      \begin{pmatrix}
       u_{n}(L) & L_{5}^{\prime} & L_{6}^{\prime} & L_{7}^{\prime} & 
         L_{8}^{\prime}\\
       L_{5}^{\prime} & \Delta & 0 & Q_{1} k_{x} & Q_{3} k_{z} + iQ_{2} k_{y}\\
       L_{6}^{\prime} & 0 & \Delta & -Q_{3} k_{z} + iQ_{2} k_{y} & Q_{1} k_{x}\\
       L_{7}^{\prime} & Q_{1} k_{x} & - Q_{3} k_{z} - iQ_{2} k_{y} & -\Delta & 
         0\\
       L_{8}^{\prime} & Q_{3} k_{z} - iQ_{2} k_{y} & Q_{1} k_{x} & 0 & -\Delta
      \end{pmatrix}
      \label{eq3}
\end{equation}
We observe that rearranging the basis $L_{7}^{\prime}$ and $L_{8}^{\prime}$ we
obtain
\begin{equation}
 \mathcal{H}^{\prime} = 
      \begin{pmatrix}
       u_{n}(L) & L_{5}^{\prime} & L_{6}^{\prime} & L_{8}^{\prime} & 
         L_{7}^{\prime}\\
       L_{5}^{\prime} & \Delta & 0 & Q_{3} k_{z} + iQ_{2} k_{y} & Q_{1} k_{x}\\
       L_{6}^{\prime} & 0 & \Delta & Q_{1} k_{x} & -Q_{3} k_{z} + iQ_{2} k_{y}\\
       L_{8}^{\prime} & Q_{3} k_{z} - iQ_{2} k_{y} & Q_{1} k_{x} & -\Delta & 0\\
       L_{7}^{\prime} & Q_{1} k_{x} & -Q_{3} k_{z} - iQ_{2} k_{y} & 0 & -\Delta
      \end{pmatrix}
\end{equation}
which has the form of the Bogoliubov-de Gennes (BdG) Hamiltonian of a $3$-D 
topological superconductor which supports surface states as Majorana fermions. 
\cite{qizhang} The physics we are concern here is of course entirely different 
since we do not deal with boson-mediated electron-electron Cooper pairing.

\subsection{Reduction of $4\times4$ Matrix to Diagonal Blocks of $2\times2$
Matrices}

Rewrite Eq. (\ref{eq3}) as
\begin{align*}
 \mathcal{H}^{\prime} & =
      \begin{pmatrix}
       u_{n}(L) & L_{5}^{\prime} & L_{6}^{\prime} & L_{7}^{\prime} & 
         L_{8}^{\prime}\\
       L_{5}^{\prime} & \Delta & 0 & \eta & \rho\\
       L_{6}^{\prime} & 0 & \Delta & -\rho^{\ast} & \eta\\
       L_{7}^{\prime} & \eta & -\rho & -\Delta & 0\\
       L_{8}^{\prime} & \rho^{\ast} & \eta & 0 & -\Delta
      \end{pmatrix}
      \\
 & = 
      \begin{pmatrix}
       D & A\\
       A^{-1} & -D
      \end{pmatrix}
\end{align*}
The transformation $\mathcal{U}$ is given by
\begin{equation*}
 \mathcal{U} = 
      \begin{pmatrix}
       aI & -bI\\
       bI & aI
      \end{pmatrix}
\end{equation*}
where unitarity condition renders
\begin{equation*}
 a^{2} + b^{2} = I
\end{equation*}
The other condition that determines $a$ and $b$ is the requirement that the 
transformed $A^{\prime}$ have zeros in the diagonal. Then one obtains
\begin{equation*}
 \mathcal{U}^{-1} \mathcal{H}^{\prime} \mathcal{U} = 
      \begin{pmatrix}
       u_{n}(L) & L_{5}^{^{\prime \prime}} & L_{6}^{^{\prime \prime}} 
         & L_{7}^{^{\prime \prime}} & L_{8}^{^{\prime \prime}}\\
       L_{5}^{^{\prime \prime}} & \varepsilon & 0 & 0 & \rho\\
       L_{6}^{^{\prime \prime}} & 0 & \varepsilon & -\rho^{\ast} & 0\\
       L_{7}^{^{\prime \prime}} & 0 & -\rho & -\varepsilon & 0\\
       L_{8}^{^{\prime \prime}} & \rho^{\ast} & 0 & 0 & -\varepsilon
      \end{pmatrix}
\end{equation*}
We can rearrange the labels as
\begin{equation}
 \mathcal{\tilde{H}} = 
      \begin{pmatrix}
       u_{n}(L) & L_{5}^{^{\prime\prime}} & L_{8}^{^{\prime\prime}} 
         & L_{6}^{^{\prime\prime}} & L_{7}^{^{\prime\prime}}\\
       L_{5}^{^{\prime\prime}} & \varepsilon & \rho & 0 & 0\\
       L_{8}^{^{\prime\prime}} & \rho^{\ast} & -\varepsilon & 0 & 0\\
       L_{6}^{^{\prime\prime}} & 0 & 0 & \varepsilon & \rho^{\ast}\\
       L_{7}^{^{\prime\prime}} & 0 & 0 & \rho & -\varepsilon
      \end{pmatrix}
      \label{eq10}
\end{equation}
where we have changed the sign of $L_{7}^{^{\prime\prime}}$, thus we obtain
\begin{equation}
 \Big( \mathcal{\tilde{H}}^{\prime} \Big) = 
      \begin{pmatrix}
       H_{1} & 0\\
       0 & H_{1}^{\dagger}
      \end{pmatrix}
      \label{eq10A}
\end{equation}
Therefore, we consider only the $2 \times 2$ Hamiltonian matrix $H_{1}$ in 
deriving the expression of the magnetic susceptibility. The $k \cdot p$ periodic 
eigenfunctions near the $L$-point of the $2 \times 2$ matrix $H_{1}$ are
\begin{align*}
 L_{c}(k) & = a L_{c} + b^{\ast} L_{v} \\
 L_{v}(k) & = a L_{v} - b L_{c}
\end{align*}
where
\begin{align*}
 a & = \frac{\sqrt{E(E + \varepsilon)}}{\sqrt{2} E}\\
 b & = \frac{Q_{1} k_{x} + iQ_{3} k_{z}}{\sqrt{2E(E + \varepsilon)}}
\end{align*}
The expression for the eigenvalues are
\begin{equation*}
 E_{\pm} = \pm \sqrt{\varepsilon^{2} + |\rho|^{2}}
\end{equation*}

\subsection{Magnetic Susceptibility of Dirac-Bloch Fermions}

For calculating the magnetic susceptibility of Bloch electrons in bismuth with 
strong spin-orbit coupling, it is important that this should be accounted for in 
all stages of the calculations. This is described fully by the formalism given 
by Roth.\cite{roth} The susceptibility expression given by Roth can be written 
as a group of terms proportional to the first and second powers of Dirac-spin 
Bohr magneton plus remaining expression similar to that of Wannier and 
Upadhyaya\cite{wannierupadya} with $\vec{p} + \vec{k}$ replaced by $\vec{\pi} = 
\Big( \vec{p} + \vec{k} + \frac{\beta}{c} \vec{s} \times \vec{\nabla} V \Big)$ 
differing only in taking of traces due to the spin states in the wave function, 
but this is taken care of in our calculation by including $H_{1}^{\dagger}$ 
also. The interaction of spin with the magnetic field in Roth's expression can 
be neglected since in the $y$-direction (small cyclotron mass direction) results 
in $\chi_{L}^{22}$ give the dominant contribution to $\chi_{\perp}$, the 
susceptibility with the magnetic field \textit{perpendicular} to the trigonal 
axis. This is also the direction where the simplified Lax two-band model is good 
for motion perpendicular to the magnetic field. One can neglect $\vec{\mu}_{0} 
\cdot \vec{B}$ in the effective Hamiltonian in this direction since the 
experimental $g$-factor due to \textit{pseudo-spin} being of the order of $100$ 
times that of free-electron spin moment. When the magnetic field is parallel to 
the trigonal axis, we denote the susceptibility as $\chi_{\parallel}$.

\subsection{Diamagnetism of Bismuth}

After consolidating various terms in the susceptibility expression similar to 
the one given by Wannier and Upadhyaya,\cite{wannierupadya} Buot and 
McClure\cite{buotMcClure} obtained a remarkably very simple expression for the 
most dominant contribution to $\chi_{\perp} \simeq \chi_{L}^{22}$ given by the 
expression
\begin{equation*}
 \chi_{L}^{22} = (6 \pi^{2} c^{2})^{-1} \bigg( \frac{Q_{3} Q_{1}}{Q_{2}} 
      \bigg) \int_{0}^{Z} d \eta \frac{f(\varepsilon) 
      - f(-\varepsilon)}{\varepsilon}
\end{equation*}
where
\begin{align*}
 \eta & = Q_{2} k_{y}\\
 \varepsilon & = \sqrt{\eta^{2} + \Delta^{2}}%
\end{align*}
We can write
\begin{align*}
 \chi_{L}^{22} & = (6 \pi^{2} c^{2})^{-1} \bigg( \frac{Q_{3} Q_{1}}{Q_{2}} 
      \bigg) \int_{0}^{Z} d \eta \bigg[ \frac{-1}{\varepsilon} + \bigg( 
      \frac{f(\varepsilon) + 1 - f(-\varepsilon)}{\varepsilon} \bigg) \bigg] \\
 & = \chi_{L,G}^{22} + \chi_{L,C}^{22}
\end{align*}
where $\chi_{L,G}^{22}$ is the large diamagentic background term, independent of 
Fermi level and temperature, $\chi_{L,C}^{22}$ is the carrier paramagnetism and 
depends on Fermi level and temperature. $\chi_{L,G}^{22}$ and $\chi_{L,C}^{22}$ 
both depend on the energy gap, $E_{g}$, at symmetry point $L$. The other 
diagonal components, $\chi_{L}^{11}$ and $\chi_{L}^{33}$ are obtain by 
permutation of the $Q_{i}$'s. These are less significant than $\chi_{L}^{22}$.

When the Fermi level lies in the forbidden gap and the temperature is low 
enough, then $\chi_{CP}^{22}$, $\chi_{LP}^{22}$, and $\chi_{L,C}^{22}$ are all 
zero and thus
\begin{equation*}
 \chi_{ID}^{22} = \chi_{L,G}^{22}
\end{equation*}
When the Fermi level is near the band edge at low temperature, then we have the 
following relations
\begin{align*}
 \chi_{LP}^{22} + \chi_{CP}^{22} & = \chi_{L,C}^{22}\\
 \chi_{ID}^{22} & = \chi_{L,G}^{22}\\
 \chi_{LP}^{22} & = -\frac{1}{3} \chi_{CP}^{22}
\end{align*}
The value of $\chi_{CP}$ is equal to the Pauli paramagnetism using the effective 
$g$-factor due to pseudospin moments. Similar relations hold for the other two 
principal directions of the magnetic field by simple rearrangement of the 
$Q_{i}$'s.

\subsection{Diamagnetism of Bi-Sb Alloys}

\begin{figure}[H]
 \centering
 \includegraphics[width=2.75in]{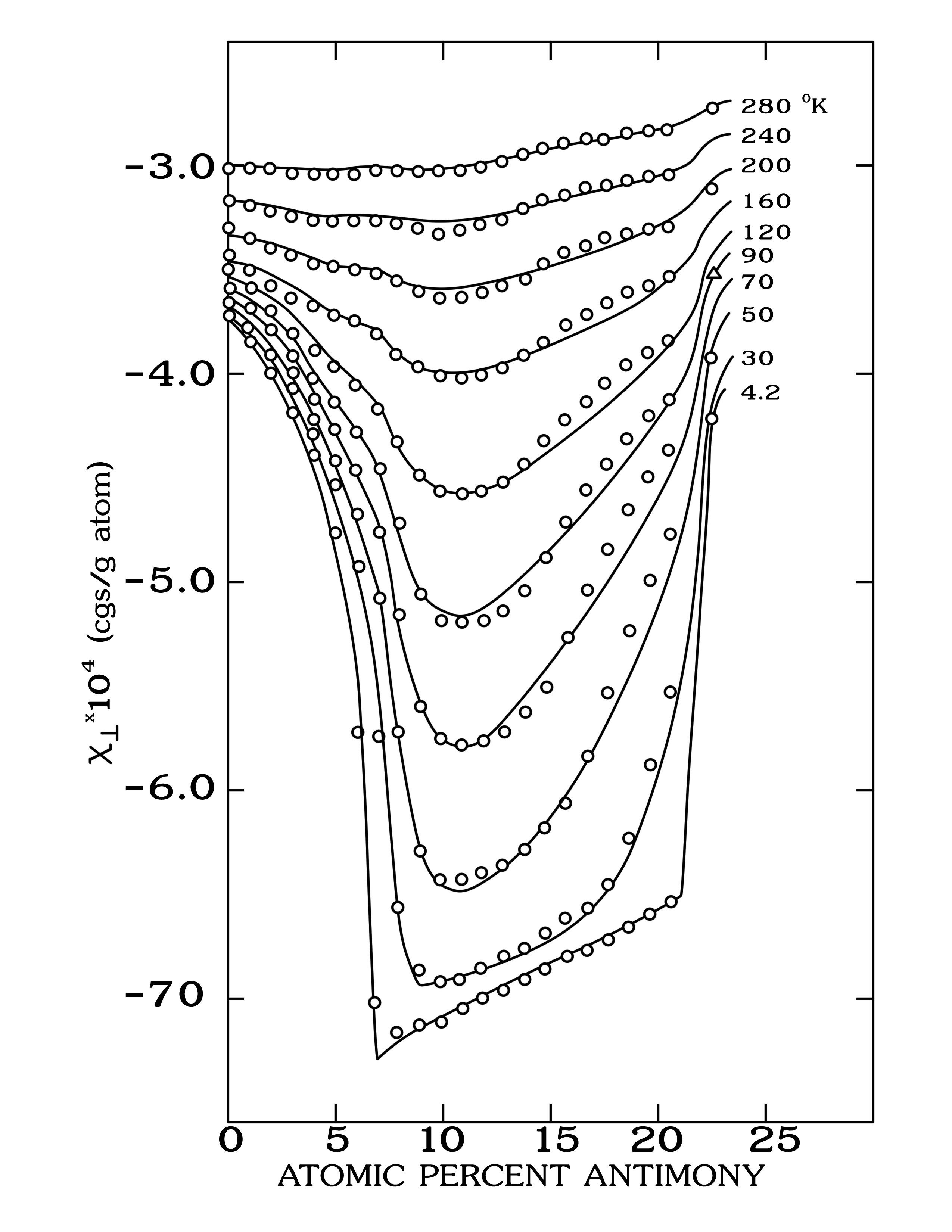}
 \caption{\small{Magnetic susceptibility, $\chi_{\perp}$, at different 
          temperatures perpendicular to the trigonal axis in Bi$_{1-x}$ - 
          Sb$_{x}$. The open circles are Wherli's experimental data for 
          $\chi_{\perp}$. The solid lines are the caculated data of 
          Buot\cite{supplement} using the Buot and McClure theory. 
          \cite{buotMcClure} [Reproduced from Ref.\cite{supplement}].}} 
          \label{fig1}
\end{figure}

Using the known energy band structure, band parameters, and matrix elements 
consistent with experimental data on bismuth and Bi-Sb thus \textit{implicitly 
including the spin-orbit coupling} in the $\vec{k} \cdot \vec{\pi}$ matrix 
elements, an outstanding fit of the calculated results with the detailed 
experimental data of Wherli on $\chi_{\perp}$ is obtained by Buot. 
\cite{supplement} This is shown in Fig. \ref{fig1}.

\begin{figure}[H]
 \centering
 \includegraphics[width=4in]{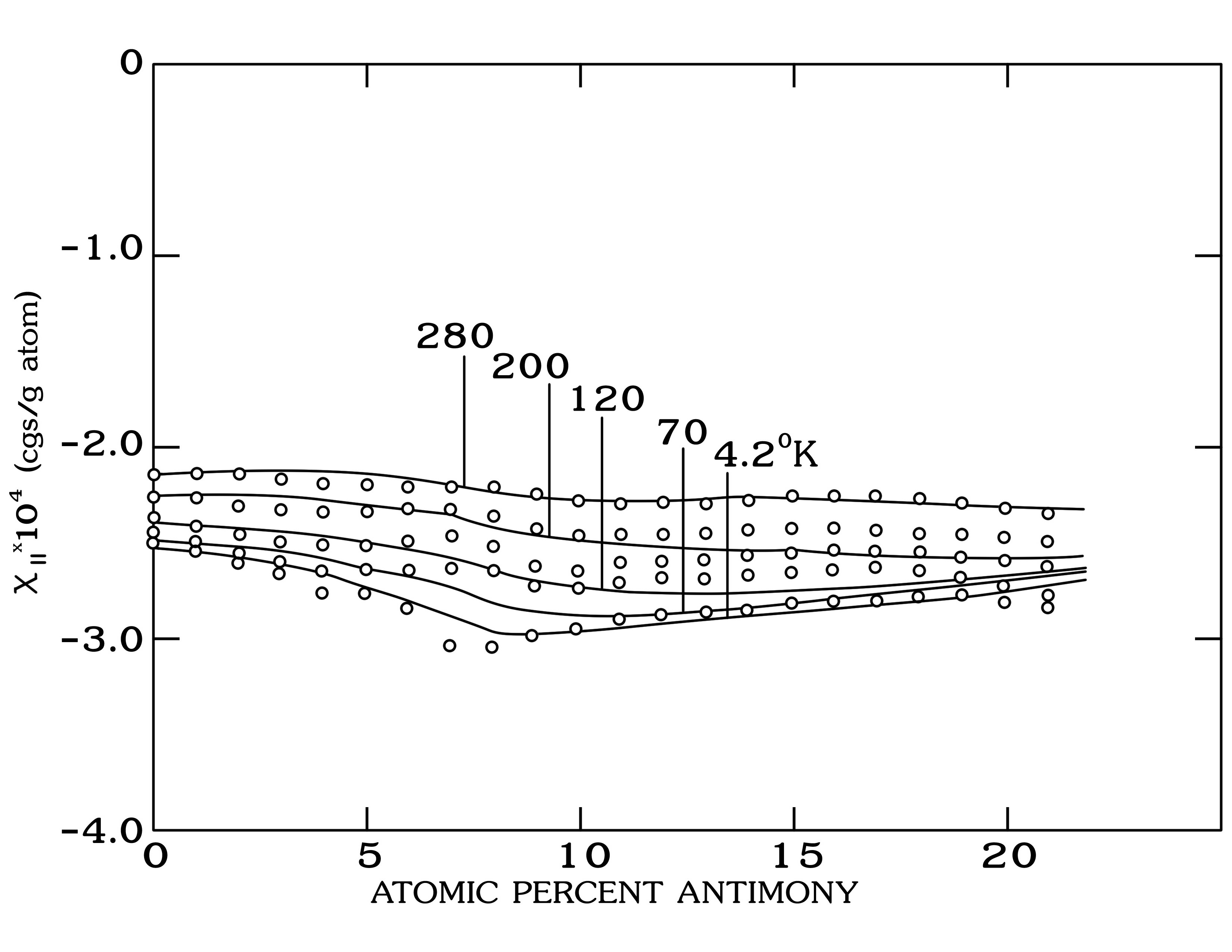}
 \caption{\small{Magnetic susceptibility parallel to the trigonal axis in 
          Bi$_{1-x}$ - Sb$_{x}$. The open circles are Wherli's experimental 
          data for $\chi_{\parallel}$. The solid lines represents the 
          calculated data using the Buot and McClure theory. [Reproduced from 
          Ref.\cite{supplement}]}} \label{fig2}
\end{figure}

The susceptibility when the magnetic field is parallel to the trigonal axis, 
$\chi_{\parallel}$, is also calculated. The main contribution comes from the 
$T$-point of the Brillouin zone. $\chi_{T,C}^{33}$ is calculated and 
$\chi_{T,G}^{33}$ adjusted to fit the experimental $\chi_{\parallel}$ data. 
$\chi_{T,G}^{33}$ is the contribution of the rest of the filled bands 
associated with symmetry point $T$ over and above the contribution at point $L$. 
The calculated result for $\chi_{\parallel}$ compared with the experimental data 
of Wherli is shown in Fig. \ref{fig2}.
The large diamagnetism of bismuth is only incidentally related to the spin-orbit 
coupling since the band dynamical effects dominate. In fact the same form of the 
Hamiltonian as in Eqs. (\ref{eq10}) and (\ref{eq10A}) applies at the $H$-point 
of graphite (without spin-orbit coupling) and also gives a large 
diamagnetism.\cite{McClure, buotMcClure}

\section{Band Theory of Magnetic Susceptibility of Relativistic Dirac Fermions}

In this section, we formulate the magnetic susceptibility of relativistic Dirac 
fermions analogous to energy-band dynamics of crystalline solids. The 
Hamiltonian of free relativistic Dirac fermions is of the form
\begin{equation}
 \mathcal{H} = \beta \Delta + c \vec{\alpha} \cdot \vec{P} \label{eq4}
\end{equation}
We designate quantum operators in capital letters and their corresponding 
eigenvalues in small letters. The equation for the eigenfunctions and 
eigenvalues is
\begin{equation}
 \mathcal{H} b_{\lambda} (x, p) = E_{\lambda}(p) b_{\lambda}(x, p) \label{eq7}
\end{equation}
where $E_{\lambda}(p) = \pm E(p)$, and $E_{\lambda}(\vec{q}^{\ \prime} - q) = 
\frac{1}{(2 \pi \hbar)^{3}} \int d \vec{p} \ e^{(\frac{i}{\hbar}) \vec{p} \cdot 
(\vec{q}^{\ \prime} - q)} E_{\lambda}(p)$, $\lambda$ labels the band index: 
$\pm$ 
spin band for positive energy states and $\pm$ spin band for negative energy 
states.
\begin{equation*}
 E_{\lambda}(p) = \pm \sqrt{(cp)^{2} + (mc^{2})^{2}}
\end{equation*}
The doubly degenerate bands is reminiscent of the Kramer conjugates in bismuth 
and Bi-Sb alloys. The localized function $a_{\lambda}(\vec{x} - 
\vec{q}^{\prime})$ is the `Wannier function' for relativistic Dirac fermions, 
defined below.\cite{buot9b}

In the absence of magnetic field we may define the Wannier function and Bloch
function of a relativistic Dirac fermions as
\begin{align*}
b_{\lambda}(x,p) & = \frac{1}{(2 \pi \hbar)^{\frac{3}{2}}} e^{(\frac{i}{\hbar}) 
                     \vec{p} \cdot \vec{x}} \ u_{\lambda}(\vec{p}) \\
a_{\lambda}(\vec{x} - \vec{q}) & = \frac{1}{(2 \pi \hbar)^{\frac{3}{2}}} \int 
                     d\vec{p} \ e^{(\frac{i}{\hbar}) \vec{p} \cdot \vec{q}} \ 
                     b_{\lambda}(x,p)
\end{align*}
where $b_{\lambda}(x,p)$ is the Bloch function, and $a_{\lambda (\vec{x} - 
\vec{q})}$ the corresponding Wannier function. $u_{\lambda}(\vec{p})$ is a 
four-component function. The $u_{\lambda}(\vec{p})$'s are related to the 
$u_{\lambda}(0)$'s by a unitary transformation, $S$, which also transforms the 
Dirac Hamiltonian into an \textit{even} form, i.e., no longer have interband 
terms. This is equivalent to the transformation from Kohn-Luttinger basis to 
Bloch functions in $\vec{k} \cdot \vec{p}$ theory. We have 
\begin{equation*}
 S = \frac{E + \beta \mathcal{H}}{\sqrt{2E(E + \Delta)}}
\end{equation*}
which can be written in matrix form as
\begin{equation*}
 S = 
      \begin{pmatrix}
       \sqrt{\frac{(E + \Delta)}{2E}} 
         & \frac{c \{\vec{\sigma} \cdot \vec{p}\}^{\ast}}{\sqrt{2E(E + 
         \Delta)}} \\
       -\frac{c \{ \vec{\sigma} \cdot \vec{p} \}^{\ast}}{\sqrt{2E(E + \Delta)}} 
         & \sqrt{\frac{(E + \Delta)}{2E}}
      \end{pmatrix}
\end{equation*}
where the entries are $2 \times 2$ matrices, $\Delta = mc^{2}$, and all matrix 
elements may be viewed as matrix elements of $S$ between the $u_{\lambda}(0)$'s, 
which are the spin functions in the Pauli representation. The transformed 
Hamiltonian is
\begin{equation}
 \mathcal{H} = S \mathcal{H} S^{\dagger} = \beta E \Big( \vec{P} \Big) 
               \label{eq5}
\end{equation}
The $a_{\lambda}(\vec{x} - \vec{q})$ is not a $\delta$-function because of the 
dependence of $u_{\lambda}(\vec{p})$ on $\vec{p}$; it is spread out over a 
region of the order of the Compton wavelength, $\frac{\hbar}{mc}$, of the 
electron and no smaller, as pointed out first by Newton and Wigner 
\cite{newtonwigner}, Foldy and Wouthuijsen\cite{foldy} and by Blount. 
\cite{blount}

The Weyl correspondence for the momentum and coordinate operator giving the 
correct dynamics of quasiparticles is given by the prescription that the 
momentum operator $\vec{P}$ and coordinate operator $\vec{Q}$ be defined with 
the aid of the Wannier function and the Bloch function as
\begin{align*}
 \vec{P} b_{\lambda}(x,p) & = \vec{p} b_{\lambda}(x,p) \\
 \vec{Q} a_{\lambda}(\vec{x}-\vec{q}) & = \vec{q} a_{\lambda}(\vec{x}-\vec{q})
\end{align*}
and the uncertainty relation follows in the formalism,
\begin{equation*}
 [Q_{i},P_{j}] = i \hbar \delta_{ij}
\end{equation*}
From Eq. (\ref{eq7}), we have
\begin{align*}
 \frac{1}{(2 \pi \hbar)^{\frac{3}{2}}} \int dq \ e^{(-\frac{i}{\hbar}) \vec{p} 
      \cdot \vec{q}} \mathcal{H} a_{\lambda}(\vec{x} - \vec{q}) & = 
      E_{\lambda}(p) \frac{1}{(2 \pi \hbar)^{\frac{3}{2}}} \int dq \ 
      e^{(-\frac{i}{\hbar}) \vec{p} \cdot \vec{q}} a_{\lambda}(\vec{x} 
      - \vec{q}) \\
 \mathcal{H} a_{\lambda}(\vec{x} - \vec{q}^{\ \prime}) & = \int dq \ 
     E_{\lambda} (\vec{q}^{\ \prime} - q) a_{\lambda}(\vec{x} - \vec{q})
\end{align*}
These relations allows us to transform the `bare' Hamiltonian operator to an 
`effective Hamiltonian' expressed in terms of the $\vec{P}$ operator and the 
$\vec{Q}$ operator. This is conveniently done by the use of the `lattice' Weyl 
transform\cite{buot4c} (`lattice' Weyl transform and Weyl transform will be used 
interchangeably for infinite translationally invariant system including 
crystaline solids). Thus, any operator $A(\vec{P},\vec{Q})$ which is a function 
of $\vec{P}$ and $\vec{Q}$ can be obtained from the matrix elements of the 
`bare' operator, $A_{op}^{b}$, between the Wannier functions or between the 
Bloch functions as,
\begin{align*}
 A(\vec{P},\vec{Q}) & = \sum_{\lambda \lambda^{\prime}} \int d \vec{v} \ d 
      \vec{u} \ a_{\lambda \lambda^{\prime}}(\vec{u},\vec{v}) \exp \Bigg[ 
      \bigg( -\frac{i}{\hbar} \bigg) \bigg(\vec{Q} \cdot \vec{u} + \vec{P} 
      \cdot \vec{v} \bigg) \Bigg] \Omega_{\lambda \lambda^{\prime}} \\
 a_{\lambda \lambda^{\prime}}(\vec{u},\vec{v}) & = h^{-8} \int d \vec{p} \ d 
      \vec{q} \ a_{\lambda \lambda^{\prime}}(\vec{p},\vec{q}) \exp \Bigg[ 
      \bigg( - \frac{i}{\hbar} \bigg) \bigg(\vec{q} \cdot \vec{u} + \vec{p} 
      \cdot \vec{v} \bigg) \Bigg] \\
 a_{\lambda \lambda^{\prime}}(\vec{p},\vec{q}) & = \int d \vec{v} 
      e^{\frac{i}{\hbar} \vec{p} \cdot \vec{v}} \bigg\langle \vec{q} - 
      \frac{1}{2} \vec{v}, \lambda \bigg| A_{op}^{b} \bigg| \vec{q} + 
      \frac{1}{2} \vec{v}, \lambda^{\prime} \bigg\rangle \\
 & = \int d \vec{u} e^{\frac{i}{\hbar} \vec{q} \cdot \vec{u}} \bigg\langle 
      \vec{p} + \frac{1}{2} \vec{u}, \lambda \bigg| A_{op}^{b} \bigg| 
      \vec{p} - \frac{1}{2} \vec{u},\lambda^{\prime} \bigg\rangle
\end{align*}
where $|\vec{p},\lambda \rangle$ and $|\vec {q},\lambda \rangle$ are the state 
vectors representing the Bloch functions and Wannier functions, respectively, 
and
\begin{align*}
 \Omega_{\lambda \lambda^{\prime}} & = \int d \vec{p} \ |\vec{p},\lambda \rangle 
     \langle \vec{p},\lambda^{\prime}| \\
 & = \int \ d \vec{q} |\vec{q}, \lambda \rangle \langle \vec{q}, 
     \lambda^{\prime}|
\end{align*}

\subsection{Canonical Conjugate Dynamical Variables in Band Quantum Dynamics}

A few more words about $\vec{Q}$ and $\vec{P}$. The use of $\vec{Q}$, conjugate 
to the operator $\vec{P}$ of the Hamiltonian in even form, is preferred in the 
band-dynamical formalism.\cite{wannierBT} The reason we now associate $\vec{Q}$ 
with the operator $\vec{P}$ of the Hamiltonian in even form is that this 
momentum operator now belongs to the respective bands (each of infinite width) 
of the \textit{decoupled} Dirac Hamiltonian. This operator is now analogous to 
the crystal momentum operator in crystalline solids. For the original Dirac 
Hamiltonian $\dot{x} = c$ [from Eq. (\ref{eq4})] leading to a complex 
\textit{zitterbewegung} motion in $x$-space, whereas for the Hamiltonian in 
even form $\dot{Q} = v$ [from Eq. (\ref{eq5})], $c$ is the speed of light and 
$v$ the velocity of a wave packet in the classical limit, and thus $Q$ is more 
closely related to the band dynamics of fermions than $x$. Moreover, on the 
cognizance that the continuum is the limit when the lattice constant of an 
array of lattice points goes to zero, there is a more compelling fundamental 
basis for using the lattice-position operator $Q$.\cite{buot9b} Since quantum 
mechanics is the mathematics of measurement processes,\cite{schwingerbook} the 
most probable measured values of the positions are the lattice-point 
coordinates. Indeed, these lattice points, or atomic sites, are where the 
electrons spend some time in crystalline solids. Therefore the \textit{lattice 
points} and \textit{crystal momentum} are clearly the \textit{observables} of 
the theory and $q$ and $p$ constitute the eigenvalues of the lattice-point 
position operator $Q$ and crystal momentum operator $P$, respectively. Thus, $Q$ 
is considered here as the generalized position operator in quantum theory for 
describing energy-band quantum dynamics, canonical conjugate to `crystal' 
momentum operator $\vec{P}$ of the Hamiltonian in even form. Although the `bare' 
operator $x$ can still be used as position operator it only unnecessarily 
renders very complicated and almost intractable resulting expressions, 
\cite{blount, suttorpdegroot} since this does not directly reflect the 
appropriate obsevables in band dynamics as first enunciated by Newton and 
Wigner\cite{newtonwigner} and by Wannier several decades ago.\cite{wannierBT} 
Thus, in understanding the dynamics of Dirac relativistic quantum mechanics 
succinctly, position space should be defined at discrete points $q$ which are 
eigenvalues of the operator $Q$.\cite{buot9b}

\subsection{The Even Form of Dirac Hamiltonian in a Uniform Magnetic Field}

The Dirac Hamiltonian for an electron with anomalous magnetic moment in a
magnetic field is
\begin{equation*}
 \mathcal{H}_{op} = \vec{\alpha} \cdot \vec{\Pi}_{op} + \beta mc^{2} 
      - \frac{1}{2} (g - 2) \mu_{B} \beta \vec{\sigma} \cdot \vec{B}
\end{equation*}
where
\begin{align*}
 \vec{\Pi}_{op} & = c \vec{P}_{op} - e \vec{A} \bigg(\vec{Q}_{op}\bigg) \\
 \mu_{B} & = \frac{e \hbar}{2mc}
\end{align*}

The transformed Hamiltonian in even form $\mathcal{H}_{B}^{\prime}$ is given by 
Ericksen and Kolsrud \cite{erikkols}
\begin{equation}
 \mathcal{H}_{B}^{\prime} = \beta \bigg[ m^{2} c^{4} + \Pi^{2} - e \hbar c 
      (1 + \lambda^{\prime}) \vec{\sigma} \cdot \vec{B} + \beta \bigg( 
      \frac{\lambda^{\prime} e \hbar}{2mc} \bigg) \sigma \cdot (B \times \Pi 
      - \Pi \times B) \bigg]^{\frac{1}{2}} \label{eq46-2}
\end{equation}
where $\lambda^{\prime} = \frac{1}{2} \ (g-2)$, and
\begin{align*}
 \tilde{\Pi} & = cP - eA(Q) - eA(r) \\
             & = cP - eA(Q + r) \\
 \\
 A(Q + r) & = \frac{1}{2} B \times (Q + r) \\
 r & = \beta \bigg( \frac{\lambda^{\prime} \hbar}{mc} \bigg) \sigma
\end{align*}
The above Hamiltonian can be written as
\begin{align}
 \mathcal{H}_{B}^{\prime} & = \beta \bigg[ m^{2} c^{4} + \Pi^{2} - e \hbar c 
      (1 + \lambda^{\prime}) \vec{\sigma} \cdot \vec{B} - 2 \bigg( \frac{1}{2} 
      B \times r \cdot \Pi \bigg) \bigg]^{\frac{1}{2}} \nonumber\\
 & = \beta \bigg[ m^{2} c^{4} + \Pi^{2} - e \hbar c (1 + \lambda^{\prime}) 
      \vec{\sigma} \cdot \vec{B} - 2A(r) \cdot \Pi \bigg]^{\frac{1}{2}} 
      \nonumber\\
 & = \beta \bigg[ m^{2} c^{4} + \tilde{\Pi}^{2} - e \hbar c(1 + 
      \lambda^{\prime}) \vec{\sigma} \cdot \vec{B} - A^{2}(r) 
      \bigg]^{\frac{1}{2}} \nonumber \\
 & = \beta \bigg[ m^{2} c^{4} + \tilde{\Pi}^{2} - e \hbar c(1 + 
      \lambda^{\prime}) \vec{\sigma} \cdot \vec{B} - \bigg( 
      \frac{\lambda^{\prime} e \hbar}{2mc} \bigg)^{2} B^{2} \bigg]^{\frac{1}{2}}
\end{align}

\subsection{\label{Tmag}Translation operator, $T_{M}(q)$, under uniform magnetic 
fields}

In the presence of a uniform magnetic field, magnetic Wannier Functions, 
$A_{\lambda}(x - q)$, and magnetic Bloch functions, $B_{\lambda}(x,p)$, exist. 
This is proved by using symmetry arguments. In general, these two basis 
functions are complete and span all the eigensolutions of the magnetic 
Hamiltonian belonging to a band index $\lambda$. The magnetic Wannier Functions 
$A_{\lambda}(x - q)$ and magnetic Bloch functions $B_{\lambda}(x,p)$ are related 
by similar unitary transformation in the absence of magnetic field, namely,
\begin{align*}
 B_{\lambda}(x,p) & = \frac{1}{(2 \pi \hbar)^{\frac{3}{2}}} 
      e^{(\frac{i}{\hbar}) \vec{p} \cdot \vec{x}} \ u_{\lambda}(\vec{p}) \\
 A_{\lambda (\vec{x} - \vec{q})} & = \frac{1}{(2 \pi \hbar)^{\frac{3}{2}}} 
      \int d \vec{p} \ e^{(\frac{i}{\hbar}) \vec{p} \cdot \vec{q}} \ 
      B_{\lambda}(x,p)
\end{align*}
where $\vec{p}$ and $\vec{q}$ are quantum labels.

Under a uniform magnetic fields, we have for a translation operator, $T_{M}(q)$, 
obeying the relation,
\begin{align}
 \nabla_{r} T_{M}(q) & = [P,T_{M}(q)] \nonumber\\
 & = \frac{ie}{\hbar c} A(q) T_{M}(q)  \label{eq6}
\end{align}
Therefore,
\begin{equation*}
 T_{M}(q) = \exp \bigg(\frac{-ie}{\hbar c} A(r) \cdot q \bigg) C(q)
\end{equation*}
where $C_{0}(q)$ is an operator which do not depend explicitly on $r$. Since 
$T_{M}(q)$ is a translation operator by amount $q$ leads us to write
\begin{equation*}
 C_{0}(q) = \exp (-q \cdot \nabla r), \qquad  \textrm{a pure displacement 
            operator by amount} - q
\end{equation*}
Equation (\ref{eq6}) means that $[P,T_{M}(q)]$ is diagonal if $T_{M}(q)$ is 
diagonal, and therefore they have the same eigenfunctions and the same quantum 
label. Therefore displacement operator in a translationally symmetric system 
under a uniform magnetic field acquire the so-called `\textit{Peierls phase 
factor}'.

Clearly, bringing the wavepacket or Wannier function around a closed loop, or 
around plaquette in the tight-binding limit, would acquire a phase equal to the 
magnetic flux through the area defined by the loop. This is the so-called 
Bohm-Aharonov effect or Berry phase. Thus, the concept of Berry phase has 
actually been floating around in the theory of band dynamics since the time of 
Peierls. Berry\cite{berry} has brilliantly generalized the concept to 
parameter-dependent Hamiltonians even in the absence of magnetic field through 
the so-called \textit{Berry connection}, \textit{Berry curvature}, and 
\textit{Berry flux}.

The magnetic translation operator generates all magnetic Wannier functions 
belonging to band index $\lambda$ from a given magnetic Wannier function 
centered at the origin, $A_{\lambda}^{0}(r - 0)$, as
\begin{align*}
 A_{\lambda}(r - q) & = T_{M}(q) A_{\lambda}^{0}(r - 0) \\
 & = \exp \bigg( \frac{-ie}{\hbar c} A(r) \cdot q \bigg) A_{\lambda}^{0} 
      (r - q)
\end{align*}
We also have the following relation,
\begin{equation*}
 T_{M}(q) T_{M}(\rho) = \exp \bigg( \frac{ie}{\hbar c} A(q) \cdot \rho \bigg) 
      T_{M}(q + \rho)
\end{equation*}
\begin{align*}
 [T_{M}(q), T_{M}(\rho)] & = \exp \bigg( \frac{ie}{\hbar c} A(q) \cdot \rho 
      \bigg) T_{M}(q + \rho) - \exp \bigg( \frac{ie}{\hbar c} A(\rho) \cdot q 
      \bigg) T_{M}(\rho + q) \\
 & = 2i \sin \bigg( \frac{e}{\hbar c} A(q) \cdot \rho \bigg) T_{M}(q + \rho)
\end{align*}
Moreover, we have,
\begin{align}
 \mathcal{H} B_{\lambda}(x,p) & = E_{\lambda} \Big( p - \frac{e}{c} A(q) \Big) 
                                  B_{\lambda}(x,p) \nonumber \\
 \mathcal{H}A_{\lambda}(\vec{x} - \vec{q}^{\ \prime}) & = \int dq \ e^{i 
      \frac{e}{c} A(q^{\ \prime}) \cdot q} E_{\lambda}(\vec{q}^{^{\prime}} - q) 
      A_{\lambda}(\vec{x} - \vec{q}) \label{eq25-2}
\end{align}
and the lattice Weyl transform of any operator, $A_{op}$, is
\begin{equation}
 a_{\lambda \lambda^{\prime}}(p,q) = \int d \vec{v} \ e^{\frac{i}{\hbar} 
      \vec{p} \cdot \vec{v}} \bigg\langle A_{\lambda} \bigg( \vec{q} - 
      \frac{1}{2} \vec{v} \bigg) \bigg| A_{op} \bigg| A_{\lambda^{\prime}} 
      \bigg( \vec{q} + \frac{1}{2} \vec{v} \bigg) \bigg\rangle \label{eq15}
\end{equation}
The Weyl transform of the Hamiltonian operator is easily calculated using Eq.
(\ref{eq25-2}) and Eq. (\ref{eq15}). The reader is referred to Ref. 
(\cite{buot3, buot4}) for details of the derivation. Applying Eq. (\ref{eq15}) 
to the even form of the Dirac Hamiltonian, we have
\begin{align*}
 h_{B}^{\prime}(\vec{p},\vec{q})_{\lambda \lambda^{\prime}} & = \int d\vec{v} \ 
      e^{\frac{i}{\hbar} \vec{p} \cdot \vec{v}} \bigg\langle A_{\lambda} \bigg( 
      \vec{q} - \frac{1}{2} \vec{v} \bigg) \bigg| \mathcal{H}_{B}^{\prime} 
      \bigg| A_{\lambda^{\prime}} \bigg( \vec{q} + \frac{1}{2} \vec{v} \bigg) 
      \bigg\rangle \\
 & = \int d \vec{v} \ \exp \bigg[ \frac{i}{\hbar} \bigg( p - \frac{e}{c} A(q)
      \bigg) \cdot v \bigg] \tilde{E}_{\lambda}(v; B) \delta_{\lambda 
      \lambda^{\prime}}\\
 & = E_{\lambda} \bigg( \vec{p} - \frac{e}{c} A(q); B \bigg) \delta_{\lambda 
      \lambda^{\prime}}
\end{align*}

\subsection{The function $E_{\lambda}(\vec{p} - \frac{e}{c} A(q); B) 
\delta_{\lambda \lambda^{\prime}}$}

The function $E_{\lambda}(\vec{p} - \frac{e}{c} A(q); B)$ is the Weyl transform 
of $\beta [\mathcal{H}^{2}]^{\frac{1}{2}}$, where the matrix $\beta$ served to 
designate the four bands. In order to calculate $\chi$ we only need the 
knowledge of $E_{\lambda}(\vec{p} - \frac{e}{c} A(q); B)$ as an expansion up to 
second order in the coupling constant $e$ and after a change of variable [this 
is effected by setting $A(q) = 0, p = \hbar k$ in the expansion], we obtain the
expression of $E_{\lambda}(\vec{p} - \frac{e}{c} A(q); B) |_{A(q) = 0}$, where 
the dependence in the field $B$ is beyond the vector potential,
\begin{equation*}
 E_{\lambda} \Big( \vec{k}; B \Big) = E_{\lambda} \Big( \vec{k}; B \Big) 
      + BE_{\lambda}^{(1)} \Big( \vec{k} \Big) + B^{2} E_{\lambda}^{(2)} \Big( 
      \vec{k} \Big) + \cdots
\end{equation*}
The function $E_{\lambda}(\vec{p} - \frac{e}{c} A(q); B)|_{A(q) = 0}$ which 
includes the anomalous magnetic moment of the electron is obtained as
\begin{align*}
 E_{\lambda}(k; B) = & \beta \Bigg\{ E - \frac{ec}{2E} \vec{L}_{c.m.} \cdot 
      \vec{B} - \frac{(1 + \lambda^{\prime})}{2E} \ e \hbar c \vec{\sigma} \cdot 
      \vec{B} - \frac{(1 + \lambda^{\prime})^{2}}{8E^{3}} \bigg( e \hbar c 
      \vec{\sigma} \cdot \vec{B} \bigg)^{2} \\
 & + \frac{(e \hbar c)^{2} \epsilon^{2}}{8E^{5}} B^{2} \bigg[ 1 + 
      \bigg(\frac{\lambda^{\prime}E}{mc^{2}} \bigg)^{2} \bigg] + O(e^{3}) 
      \Bigg\}
\end{align*}
where
\begin{align*}
 \vec{L}_{c.m.} & = \beta \bigg( \frac{\lambda^{\prime} \hbar}{mc} \bigg) 
      \vec{\sigma} \times \vec{p} \\
 \epsilon^{2} & = m^{2} c^{4} + c^{2} \hbar^{2} k_{z}^{2} \\
 E \Big( \vec{k} \Big) & = \Big[ m^{2} c^{4} + c^{2} \hbar^{2} k^{2} 
      \Big]^{\frac{1}{2}}
\end{align*}

The term, $\vec{L}_{c.m.}$, is a magnetodynamic effect, i.e., due to hidden
average angular momentum $\vec{L}_{c.m.}$ of a moving electron. Thus, the
introduction of the Pauli anomalous term in $\mathcal{H}$ at the outset endows
a rigid-body behavior to the electron, and its angular momentum about the
origin $\vec{L}_{0}$ is
\begin{equation*}
 \vec{L}_{0} = \vec{L}_{MO} + \vec{L}_{c.m.}
\end{equation*}
where $\vec{L}_{MO}$ is the angular momentum about the origin of the system of
charge concentrated as a point at the center of mass and $\vec{L}_{c.m.}$ is
the average angular momentumof the system, as a spread-out distribution of
charge about the center of mass. Thus,
\begin{align*}
 \vec{L}_{0} = & \vec{q} \times \vec{p} + \Bigg\langle \sum_{i} \vec{r}_{i} 
      \times \vec{p}_{i} \Bigg\rangle \\
 \Bigg\langle \sum_{i} \vec{r}_{i} \times \vec{p}_{i} \Bigg\rangle & = \beta 
      \bigg( \frac{\lambda^{\prime} \hbar}{mc} \bigg) \vec{\sigma} \times 
      \vec{p} 
\end{align*}
\begin{align}
 M & = -\Big[ 2E_{\lambda}^{(2)} \Big( \vec{k} \Big) B \Big]_{sp} \nonumber \\
   & = -\frac{(e \hbar c)^{2} \epsilon^{2}}{4 \Big[ E_{\lambda} \Big( \vec{k} 
       \Big) \Big]^{5}} \Bigg[ 1 + \bigg( \frac{\lambda^{\prime} E}{mc^{2}} 
       \bigg)^{2} \Bigg] B \label{eq50}%
\end{align}
The induced magnetic moment due to a distribution of electric charge is
\begin{equation}
 M = - \frac{Be^{2} \langle r^{2} \rangle}{4mc^{2}} \label{eq51}
\end{equation}
where $\langle r^{2} \rangle$ is the average of the square of the spatial spread 
of the distribution normal to the magnetic field. Equating Eqs. (\ref{eq50}) 
with (\ref{eq51}) we obtain
\begin{equation}
 \langle r^{2} \rangle = \frac{mc^{2}(\hbar c)^{2} 
      \epsilon^{2}}{[E_{\lambda}(k)]^{5}} \Bigg[ 1 + \bigg( 
      \frac{\lambda^{\prime} E}{mc^{2}} \bigg)^{2} \Bigg] \label{eq52}
\end{equation}
For positive energy states $E_{\lambda}(k) = (c^{2} \hbar^{2} k^{2} + m^{2} 
c^{4})^{\frac{1}{2}}$ and in the nonrelativistic limit, Eq. (\ref{eq52}) reduces 
to
\begin{equation*}
 \langle r^{2} \rangle = (1 + \lambda^{\prime 2}) \bigg( \frac{\hbar}{mc} 
      \bigg)^{2}
\end{equation*}
and thus the effective spread of the electron at rest, and for $\lambda^{\prime} 
= 0$, is precisely equal to the Compton wavelength.

\subsection{Magnetic Susceptibility of Dirac Fermions}

The magnetic susceptibility is given by
\begin{align*}
 \chi = & -\frac{1}{48 \pi^{3}} \Big( \frac{e}{\hbar c} \Big)^{2} 
          \sum_{\lambda} \int d\vec{k} \left\{ \frac{\partial^{2} E_{\lambda} 
          \Big( \vec{k}; 0 \Big)}{\partial k_{x}^{2}} \frac{\partial^{2} 
          E_{\lambda} \Big( \vec{k}; 0 \Big)}{\partial k_{y}^{2}} - \Bigg( 
          \frac{\partial^{2} E_{\lambda} \Big( \vec{k}; 0 \Big)}{\partial k_{x} 
          \partial k_{y}} \Bigg)^{2} \right\} \frac{\partial 
          f(E_{\lambda})}{\partial E_{\lambda}} \\
 & - \bigg( \frac{1}{2 \pi} \bigg)^{3} \sum_{\lambda} \int d\vec{k} \Big[ 
          E_{\lambda}^{(1)}(k) \Big]^{2} \frac{\partial f(E_{\lambda})}{\partial 
          E_{\lambda}} - \bigg( \frac{1}{2 \pi} \bigg)^{3} \sum_{\lambda} \int 
          d\vec{k} \ 2E_{\lambda}^{(2)}(k) f(E_{\lambda})
\end{align*}
Using the following change of variable of integration,
\begin{equation*}
 (\hbar c)^{3} \int d\vec{k} = \int_{-\infty}^{\infty} d\eta \int_{0}^{2 \pi} 
     d\phi \ E \Big( \vec{k} \Big) \ dE \Big( \vec{k} \Big)
\end{equation*}
where
\begin{equation*}
 \eta = \hbar ck_{z}
\end{equation*}
we obtain for the positive energy states the expression for $\chi$ which can be 
divided into more physically meaningful terms as
\begin{equation*}
 \chi = \chi_{_{LP}} + \chi_{_{P}} + \chi_{_{sp}} + \chi_{_{g}} + \chi_{_{MD}}
\end{equation*}
where
\begin{align}
 \chi_{_{LP}} = & \frac{1}{24 \pi^{3}} \bigg( \frac{e}{\hbar c} \bigg)^{2} 
                  \int_{-\infty}^{\infty} d\eta \int_{\epsilon}^{\infty} 
                  \frac{\epsilon^{2}}{E^{3}} \frac{\partial f(E)}{\partial E} 
                  dE  \label{chi1}\\
 \chi_{_{P}} = & -\frac{(1 + \lambda^{\prime})^{2}}{8 \pi^{2}} \bigg( 
                 \frac{e}{\hbar c} \bigg)^{2} \int_{-\infty}^{\infty} d\eta 
                 \int_{\epsilon}^{\infty} \frac{1}{E} \frac{\partial 
                 f(E)}{\partial E} dE \label{chi2}\\
 \chi_{_{sp}} = & -\frac{1}{8 \pi^{2}} \bigg( \frac{e^{2}}{\hbar c} \bigg) 
                  \int_{-\infty}^{\infty} d\eta \int_{\epsilon}^{\infty} 
                  \frac{\epsilon^{2}}{E^{4}} \Bigg[ 1 + \bigg( 
                  \frac{\lambda^{\prime} E}{mc^{2}} \bigg)^{2} \Bigg] f(E) dE 
                  \label{chi3}\\
 \chi_{_{g}} = & \frac{(1 + \lambda^{\prime})^2}{8 \pi^{2}} \bigg( 
                 \frac{e^{2}}{\hbar c} \bigg) \int_{-\infty}^{\infty} d\eta 
                 \int_{\epsilon}^{\infty} \frac{f(E)}{E^{2}} dE  \label{chi4}\\
 \chi_{_{MD}} = & -\frac{\lambda^{\prime2}}{8 \pi^{2}} \bigg( \frac{e^{2}}{\hbar 
                  c} \bigg) \int_{-\infty}^{\infty} d\eta 
                  \int_{\epsilon}^{\infty} \frac{(E^{2} - 
                  \epsilon^{2})}{(mc^{2})^{2} E} \frac{\partial f(E)}{\partial 
                  E} dE \label{chi5}
\end{align}
where
\begin{align*}
 \bigg( \frac{ec}{2E} \vec{L}_{c.m.} \bigg)_{z}^{2} & = \bigg(
     \frac{\lambda^{\prime} e \hbar c}{2mc^{2}} \bigg) \frac{(E^{2} -    
     \epsilon^{2})}{E^{2}}\\
 \vec{B} & = B \frac{\vec{z}}{|\vec{z}|}
\end{align*}
The total susceptibility for the positive energy states is
\begin{equation}
 \chi = \frac{1}{(2 \pi)^{2}} \bigg( \frac{e^{2}}{\hbar c} \bigg) \bigg[ (1 + 
      \lambda^{\prime})^{2} - \frac{1}{3} \bigg] \int_{0}^{\infty} d\eta \ 
      \frac{f(\epsilon)}{\epsilon} \frac{1}{(2 \pi)^{2}} \bigg( 
      \frac{e^{2}}{\hbar c} \bigg) \bigg( \frac{\lambda^{\prime}}{mc^{2}} 
      \bigg)^{2} \int_{0}^{\infty} d\eta \ G(\epsilon - \mu)  \label{eq62}
\end{equation}
where
\begin{align*}
 G(\epsilon - \mu) = & k_{B}T \ln \bigg\{ 1 + \exp \bigg[ - \frac{(\epsilon 
                       - \mu)}{k_{B}T} \bigg] \bigg\} \\
 & = \int_{\epsilon}^{\infty} f(E) dE
\end{align*}
The contributions of the holes is obtained by replacement of $f(\epsilon)$ and
$G(\epsilon - \mu)$ in Eq. (\ref{eq62}) by $(1 - f(-\epsilon))$ and $G(\epsilon 
+ \mu)$, respectively.

The relative importance of terms that made up $\chi$ at $T = 0$ of Dirac 
fermions, where $n$ is the electron density, $k_{F} = (3 \pi^{2} 
n)^{\frac{1}{3}}$, $\eta_{F} = \hbar ck_{F}$ and $E_{F} = (\Delta^{2} + 
\eta_{F}^{2})^{\frac{1}{2}}$, is summarized below.
\footnotesize
\begin{center}
 \begin{tabular}
  [c]{|l|l|l|}\hline
  Various Contributions to $\chi_{Dirac}$ at ${T = 0}$ & Nonrelativistic, 
  $\frac{\eta_{F}}{\Delta} \ll 1$ & Ultrarelativistic, $\frac{\eta_{F}}{\Delta} 
      \gg 1$ \\ \hline
  $\chi_{_{LP}} = -\frac{1}{12 \pi^{2}} \Big( \frac{e^{2}}{\hbar c} \Big) 
      \frac{1}{E_{F}^{3}} \Big( \frac{\eta_{F}^{3}}{3} + \Delta^{2} \eta_{F} 
      \Big)$ 
    & $-\frac{1}{12 \pi^{2}} \Big( \frac{e}{mc^{2}} \Big) k_{F}$ 
    & $-\frac{1}{12 \pi^{2}} \Big( \frac{e^{2}}{\hbar c} \Big) \frac{1}{3}$ \\ 
      \hline
  $\chi_{_{P}} = \frac{1}{4 \pi^{2}}(1 + \lambda^{\prime})^{2} \Big( 
      \frac{e^{2}}{\hbar c} \Big) \frac{\eta_{F}}{E_{F}}$ 
    & $\frac{1}{4 \pi^{2}}(1 + \lambda^{\prime})^{2} \Big( \frac{e}{mc^{2}} 
      \Big) k_{F}$ 
    & $\frac{1}{4 \pi^{2}}(1 + \lambda^{\prime})^{2} \Big( \frac{e^{2}}{\hbar c} 
      \Big)$ \\ \hline
  $\chi_{_{MD}} = -\frac{\lambda^{\prime 2}}{4 \pi^{2}} \Big( 
      \frac{e^{2}}{\hbar c} \Big) \frac{1}{\Delta^{2}} \bigg[ \frac{1}{E_{F}} 
      \Big( \frac{\eta_{F}^{3}}{3} + \Delta^{2} \eta_{F} \Big) - \eta_{F} 
      E_{F} \bigg]$ 
    & $\Longrightarrow 0$ 
    & $\frac{\lambda^{\prime 2}}{4 \pi^{2}} \Big( \frac{e^{2}}{\hbar c} \Big)  
      \frac{2}{3} \Big( \frac{\eta_{F}}{\Delta} \Big)^{2}$ \\ \hline
  $\chi_{_{spread}} = -\frac{1}{12 \pi^{2}} \Big( \frac{e^{2}}{\hbar c} \Big) 
      \sinh^{-1} \Big( \frac{\eta_{F}}{\Delta} \Big) - \chi_{_{LP}}$ 
    &  
    & $-\frac{1}{12 \pi^{2}} \Big( \frac{e^{2}}{\hbar c} \Big) \Big[ \ln \frac{2 
      \eta_{F}}{\Delta} - \frac{1}{3} \Big]$ \\ \hline
  \qquad $-\frac{\lambda^{\prime 2}}{4 \pi^{2}} \Big( \frac{e^{2}}{\hbar c} 
      \Big) \frac{1}{\Delta^{2}} \bigg[ \frac{\eta_{F}(\eta_{F}^{2} 
      + \Delta^{2})^{\frac{1}{2}}}{2} + \frac{\Delta^{2}}{2} \sinh^{-1} \Big( 
      \frac{\eta_{F}}{\Delta} \Big) \bigg]$ 
    &  
    & \\\hline
  \qquad $ + \frac{\lambda^{\prime 2}}{4 \pi^{2}} \Big( \frac{e^{2}}{\hbar c} 
      \Big) \frac{1}{E_{F}} \Big( \frac{\eta_{F}^{3}}{3} + \Delta^{2} \eta_{F} 
      \Big) \frac{1}{\Delta^{2}}$ 
    & $\Longrightarrow 0$ 
    & $-\frac{\lambda^{\prime 2}}{4 \pi^{2}} \Big( \frac{e^{2}}{\hbar c} \Big) 
      \Big[ \frac{1}{6} \big( \frac{\eta_{F}}{\Delta} \big)^{2} + \frac{1}{2} 
      \ln \frac{2 \eta_{F}}{\Delta} \Big]$ \\ \hline
  $\chi_{_{g}} = \frac{1}{4 \pi^{2}}(1 + \lambda^{\prime})^{2} \Big( 
      \frac{e^{2}}{\hbar c} \Big) \Big[ \sinh^{-1} \Big( \frac{\eta_{F}}{\Delta} 
      \Big) - \frac{\eta_{F}}{E_{F}} \Big]$ 
    & $\Longrightarrow 0$ 
    & $\frac{1}{4 \pi^{2}} (1 + \lambda^{\prime})^{2} \Big( \frac{e^{2}}{\hbar 
      c} \Big) \Big[ \ln \frac{2 \eta_{F}}{\Delta} - 1 \Big]$ \\ \hline
 \end{tabular}
\end{center}
\normalsize

\subsection{Displacement Operator under Uniform High External Electric Fields}

To complement Sec. \ref{Tmag}, we give the translation operator for uniform
electric field case, $\mathcal{H} = \mathcal{H} - e \vec{F} \cdot \vec{x}$. We 
have for the displacement operator, $T_{E}(q)$, obeying the relation,
\begin{align*}
 i \hbar \dot{T}_{E}(q) & = [T_{E}(q), \mathcal{H}] \\
 \dot{T}_{E}(q) & = \frac{ie}{\hbar} F \cdot q \ T_{E}(q)
\end{align*}
Therefore
\begin{equation*}
 T_{E}(q) = C_{0}(q, \tau) \exp \bigg( \frac{ie}{\hbar} \bigg) Ft \cdot q
\end{equation*}
where $C_{0}(q, \tau)$ is an operator which do not depend explicitly on time, 
$t$. $T_{E}(q)$, being a displacement operator in space and time lead us to 
write the operator 
\begin{equation*}
 C_{0}(q, \tau) = \exp \Bigg( q \cdot \frac{\partial}{\partial r} + \tau 
                  \frac{\partial}{\partial t} \Bigg)
\end{equation*}

$T_{E}(q)$ plays critical role similar to $T_{M}(q)$ for establishing the phase 
space quantum transport dynamics at very high electric fields, where we consider 
realistic transport problems as time-dependent many-body problems. For zero 
field case we are dealing with biorthogonal Wannier functions and Bloch 
functions because the Hamiltonian is no longer Hermetian due to the presence of 
energy variable, $z$, in the self-energy. This means that $[T_{E}(q), 
\mathcal{H}]$ is diagonal in the bilinear expansion if $T_{E}(q)$ is diagonal. 
The eigenfunction of the `lattice' translation operator $T_{E}(q)$ must then be 
labeled by a wavenumber $\vec{k}$ which is varying in time as
\begin{equation*}
 \vec{k} = \vec{k}_{0} + \frac{e \vec{F}}{\hbar} t
\end{equation*}
and $\mathcal{H}$ is also diagonal in $\vec{k}$. Similarly, the energy variable, 
$z$, in the Hamiltonian must also vary as
\begin{equation*}
 z = z_{0} + e \vec{F} \cdot \vec{q}
\end{equation*}

Similar developments for translationally invariant many-body system subjected to 
a uniform electric field allows us to define the corresponding electric Bloch 
functions and electric Wannier functions, in a unifying manner for both magnetic 
and electric fields. This electric-field version allows us to derive the quantum 
transport equation of the particle density at very high electric fields. This 
will be discussed in another communication dealing with quantum transport in 
many-body systems.

\section{Magnetic Susceptibility of Many-Body Systems in a Uniform Field}

Here, we shall see that symmetry arguments enable us to generalize, in a
unified manner, the derivation of $\chi$ for noninteracting to that of
interacting Fermi systems possessing translational symmetry.\cite{freeBloch}

The reduced one-particle Schrodinger equation of a many-body system in the
presence of a uniform magnetic field is defined by
\begin{equation}
 [\mathcal{H}_{0} + \Sigma(z)] \phi(z) = E(z) \phi(z), \label{eq2.1}
\end{equation}
where $\Sigma(z)$ is the nonlocal energy-dependent ($z$ is the energy variable) 
complex quantity called the self-energy operator. $\mathcal{H}_{0}$ is the 
non-interacting Hamiltonian in a magnetic field. In the absence of spin-orbit 
coupling, this is given by
\begin{equation}
 \mathcal{H}_{0} = \frac{1}{2m} \bigg( \frac{\hbar}{i} \nabla_{\vec{r}} - 
                   \frac{e}{c} \vec{A}(\vec{r}) \bigg)^{2} + V(\vec{r}) - 
                   g \mu_{B} S_{z} B  \label{eq2.2}
\end{equation}
$V(\vec{r}) + \Sigma(z)$ represents the effective potential which is a 
non-Hermitian operator leading to the use of biorthogonal eigenfunctions, with 
the dual sets obtained from the eigenfunctions of $\mathcal{H}_{0} + \Sigma(z)$ 
and its adjoint. In the presence of spin-orbit coupling, we have
\begin{equation}
 \mathcal{H}_{0} = \frac{1}{2m} \bigg( \frac{\hbar}{i} \nabla_{\vec{r}} - 
                   \frac{e}{c} \vec{A}(\vec{r}) - \frac{\mu_{B}}{c} \sigma 
                   \times \nabla V(r) \bigg)^{2} - g \mu_{B} \sigma \cdot B + 
                   V(r) \label{eqxtra}
\end{equation}
which gives the spin-orbit interaction correctly to order $\frac{e}{mc^{2}}$. 
\cite{roth} In the presence of inversion and time reversal symmetry the 
eigenfunctions are spinors and so are the magnetic Wannier function and magnetic 
Bloch function themselves.

\subsection{The Crystalline Effective Hamiltonian}

We will transform the many-body effective-Hamiltonian operator $\mathcal{H}_{0} 
+ \Sigma(z)$ to an effective Hamiltonian expressed in terms of the 
crystal-momentum operator, $\vec{P}$, and the lattice-position operator, 
$\vec{Q}$. This is done through the lattice Weyl-Wigner formalism of the quantum 
dynamics of solids.\cite{buot4c} We have,
\begin{align}
 \mathcal{H}_{eff} \Big( \vec{P}, \vec{Q}, z \Big) = & (N \hbar^{3})^{-2}
      \sum_{\vec{p},\vec{q},\lambda,\lambda^{\prime},\vec{u} \cdot \vec{v}} 
      H_{\lambda \lambda^{\prime}} \bigg( \vec{p} - \frac{e}{c} 
      \vec{A}(\vec{q}); B, z \bigg) \exp \Bigg( \frac{2i}{\hbar} \bigg( \vec{p} 
      - \vec{P} \bigg) \cdot \vec{v} \Bigg) \nonumber \\
 & \times \exp \Bigg( \frac{2i}{\hbar} \bigg( \vec{q} - \vec{Q} - \vec{v} 
      \bigg) \cdot \vec{u} \Bigg) \Omega_{\lambda \lambda^{\prime}}, 
      \label{eq2.3}
\end{align}
where
\begin{equation}
 H_{\lambda \lambda^{\prime}} \bigg( \vec{p} - \frac{e}{c} \vec{A}(\vec{q}); B, 
      z \bigg) = \sum_{\vec{v}} e^{(\frac{2i}{\hbar}) \vec{p} \cdot \vec{v}} 
      \Big\langle \vec{q} - \vec{v}, \lambda \Big| \mathcal{H}_{0} + \Sigma(z) 
      \Big| \vec{q} + \vec{v}, \lambda^{\prime} \Big\rangle \label{eq2.4}
\end{equation}
or in the presence of spin-orbit interaction, this is given by,
\begin{equation}
 H_{\lambda \lambda^{\prime}} \bigg( \vec{p} - \frac{e}{c} \vec{A}(\vec{q}) 
      - \frac{\mu_{B}}{c} \sigma \times \nabla V(q); B, z \bigg) = 
      \sum_{\vec{v}} e^{(\frac{2i}{\hbar}) \vec{p} \cdot \vec{v}} \Big\langle 
      \vec{q} - \vec{v}, \lambda \Big| \mathcal{H}_{0} + \Sigma(z) \Big| 
      \vec{q} + \vec{v}, \lambda^{\prime} \Big\rangle \label{eqxtra2}
\end{equation}
where
\begin{equation}
 \Omega_{\lambda \lambda^{\prime}} = \sum_{\vec{q}} \big| \vec{q}, \lambda 
      \big\rangle \big\langle \vec{q}, \lambda^{\prime} \big| = \sum_{\vec{p}} 
      \big| \vec{p}, \lambda \big\rangle \big\langle \vec{p}, \lambda^{\prime} 
      \big| \label{eq2.5}
\end{equation}
and the $|\vec{q},\lambda \rangle$ or $|\vec{p},\lambda \rangle$ are considered 
spinors for each band index $\lambda$ in the case with spin-orbit interaction. 
We expand the eigensolutions of (\ref{eq2.3}) in terms of the complete set of 
magnetic Wannier functions or of magnetic Bloch functions of non-interacting 
system, $\mathcal{H}_{0}$,
\begin{align}
 \phi(\vec{r}, z) = & \sum_{\vec{p},\lambda} f_{\lambda}(\vec{p},z) \big| 
                      \vec{p}, \lambda \big\rangle \label{eq2.6} \\
 \phi(\vec{r}, z) = & \sum_{\vec{q},\lambda} f_{\lambda}(\vec{q}, z) \big| 
                      \vec{q}, \lambda \big\rangle \label{eq2.7}
\end{align}
an equivalent eigenvalue problem is obtained for the coefficients $f_{\lambda} 
(\vec{q}, z)$. In $\vec{q}$-space this is,
\begin{equation}
 \sum_{\lambda^{\prime}} W_{\lambda \lambda^{\prime}}(\vec{\pi}; B, z)  
      f_{\lambda^{\prime}}(\vec{q}, z) = E(z) f_{\lambda}(\vec{q}, z) 
      \label{eq2.8}
\end{equation}
and the corresponding eigenvalue equation in $\vec{p}$-space
\begin{equation}
 \sum_{\lambda^{\prime}} W_{\lambda \lambda^{\prime}}(\vec{\pi}; B, z)  
      f_{\lambda^{\prime}}(\vec{p}, z) = E(z) f_{\lambda}(\vec{p}, z) 
      \label{eq2.9}
\end{equation}
where
\begin{equation}
 W_{\lambda \lambda^{\prime}}(\vec{\pi}; B, z) = (N \hbar^{3})^{-1} 
    \sum_{\vec{p}^{\ \prime},\vec{v}} H_{\lambda \lambda^{\prime}}(\vec{p}^{\ 
    \prime}; B, z) \exp \bigg( \frac{2i}{\hbar}(\vec{p}^{\ \prime} - \vec{\pi}) 
    \cdot \vec{v} \bigg)  \label{eq2.10} 
\end{equation}
\begin{equation}
 \vec{\pi} = \Bigg\{
 \begin{matrix}
  \frac{\hbar}{i} \nabla_{\vec{q}} - \frac{e}{c} \vec{A}(\vec{q}) 
      & \textrm{ in } \vec{q} \textrm{ space} \\
  \vec{p} + \frac{e}{c} \vec{A} \big( \frac{\hbar}{i} \nabla_{\vec{p}} \big)
      & \textrm{ in } \vec{p} \textrm{ space}
 \end{matrix}
 \label{eq2.11_12}
\end{equation}
Since $W_{\lambda \lambda^{\prime}}(\vec{\pi}; B, z)$ is a non-Hermitian 
operator, one also needs to solve the adjoint problem, either in $q$-space or 
$\vec{p}$-space,
\begin{eqnarray}
 \sum_{\lambda^{\prime}} W_{\lambda^{\prime} \lambda}^{\ast}(\vec{\pi}; B, z) 
      e_{\lambda^{\prime}}(\vec{q}, z) = E^{\ast}(z) e_{\lambda}(\vec{q}, z) 
      \label{eq2.13} \\
 \sum_{\lambda^{\prime}} W_{\lambda^{\prime} \lambda}^{\ast}(\vec{\pi}; B, z) 
      e_{\lambda^{\prime}}(\vec{p}, z) = E^{\ast}(z) e_{\lambda}(\vec{p}, z) 
      \label{eq2.14} 
\end{eqnarray}
$W_{\lambda \lambda^{\prime}}(\vec{\pi}; B, z)$ may be viewed as a generalized 
Hamiltonian of the Dirac type occurring in the relativistic quantum theory of 
electrons. Since we are using magnetic Wannier functions and magnetic Bloch 
functions of the noninteracting Bloch electrons in a uniform magnetic field as 
basis states, $\mathcal{H}_{0}$ is diagonal in band indices and we may write
\begin{equation}
 W_{\lambda \lambda^{\prime}}(\vec{\pi}; B, z) = W_{0}(\vec{\pi}; B, 
      z)_{\lambda} \delta_{\lambda \lambda^{\prime}} + \Sigma_{\lambda 
      \lambda^{\prime}}(\vec{\pi}; B, z), \label{eq2.15}%
\end{equation}
where $W_{0}(\vec{\pi}; B, z)_{\lambda}$ is the effective magnetic Hamiltonian, 
belonging to the band, $\lambda$, of noninteracting Bloch electrons in a uniform 
magnetic field. In the case with spin-orbit interaction, $W_{\lambda 
\lambda^{\prime}}(\vec{\pi}; B, z)$ is a spinor for each band index. Just like 
the reltivistic Dirac Hamiltonian, $W_{\lambda \lambda^{\prime}}(\vec{\pi}; B, 
z)$ can be transformed into an \textit{even form}, that is without any 
off-diagonal terms through the technique of successive transformation as defined 
below.

The eigenfunctions $f_{\lambda^{\prime}}(\vec{q}, z)$ of 
$W_{\lambda \lambda^{\prime}}(\vec{\pi}; 0, z) \equiv \mathcal{H}_{\lambda 
\lambda^{\prime}}^{(0)}$ and those of its adjoint define a similarity 
transformation which diagonalizes $\mathcal{H}_{\lambda 
\lambda^{\prime}}^{(0)}$. We have
\begin{equation}
 U^{-1} H^{0} U = \tilde{H}_{\lambda}^{0} \delta_{\lambda \lambda^{\prime}}
      \label{eq2.16}
\end{equation}
where the matrix of $U$ is given by $f_{ij}$, where $f_{ij}$ denotes the $i$th
component of an eigenvector belonging to the $j$th eigenvalue of the matrix
$\mathcal{H}_{\lambda \lambda^{\prime}}^{(0)}$. The matrix of $U^{-1}$ is the 
matrix formed by $e_{ji}^{\ast}$, where $e_{ji}$ is the $i$th component of the 
$j$th eigenvector of the adjoint matrix. $U$ and $U^{-1}$ also determine the 
transformation from the Wannier function and Bloch function of non-interacting 
Bloch electrons to the Wannier function and Bloch function of interacting Bloch 
electrons, which are, in general, energy dependent and biorthogonal. Denoting 
these by $|\vec{q}, \lambda, z\rangle$ and $|\vec{p}, \lambda, z\rangle$, we 
have
\begin{align}
 \Big|\vec{p}, \lambda, z \Big\rangle & = \sum_{i} f_{i \lambda}(\vec{p}, z) 
      \Big|\vec{p}, i \Big\rangle  \label{eq2.17} \\
 \Big\langle \vec{p}, \lambda, z\Big| & = \sum_{i} e_{\lambda i}^{\ast}(\vec{p}, 
      z) \Big\langle \vec{p}, i \Big|  \label{eq2.18}
\end{align}
and hence we have,
\begin{align}
 \Big|\vec{q}, \lambda, z \Big\rangle & = (N \hbar^{3})^{-\frac{1}{2}} 
      \sum_{\vec{p}} e^{\frac{i}{\hbar}(\vec{p} \cdot \vec{q})} |\vec {p}, 
      \lambda, z \Big\rangle \label{eq2.19}\\
 \Big\langle \vec{q}, \lambda, z \Big| & = \sum_{\vec{p}} 
      e^{\frac{i}{\hbar}(\vec{p} \cdot \vec{q})} \Big\langle \vec{p}, \lambda, 
      z \Big|  \label{eq2.20}
\end{align}
In terms of these basis states, $\tilde{H}_{\lambda}^{0}(\vec{p}, z) 
\delta_{\lambda \lambda^{\prime}}$, is given by
\begin{equation}
 \tilde{H}_{\lambda}^{0}(\vec{p}, z) = \sum_{\vec{v}} e^{\frac{2i}{\hbar} 
      \vec{p} \cdot \vec{v}} \Big\langle \vec{q} - \vec{v}, \lambda, 
      z \Big|\big[\mathcal{H}_{0} + \Sigma(z)\big] \Big|\vec{q} + \vec{v}, 
\lambda, z       \Big\rangle \label{eq2.21}
\end{equation}
or equivalently,
\begin{equation}
 \tilde{H}_{\lambda}^{0}(\vec{p}, z) = \sum_{\vec{u}} e^{\frac{2i}{\hbar} 
      \vec{q} \cdot \vec{u}} \Big\langle \vec{p} + \vec{u}, \lambda, 
      z \Big| \big[ \mathcal{H}_{0} + \Sigma(z) \big] \Big| \vec{p} - \vec{u}, 
      \lambda, z \Big\rangle 
      \label{eq2.22}
\end{equation}
The one-particle energy $z_{\lambda}$ belonging to the band index $\lambda$
is, in the quasiparticle picture given as usual by the solution of
\begin{equation}
 z_{\lambda} - \tilde{H}_{\lambda}^{0}(\vec{p}, z_{\lambda}) = 0 \label{eq2.23}
\end{equation}
which is doubly degenerate in the case with spin-orbit interaction.

\subsection{\label{chap3}Removal of Interband Terms of $\mathcal{H}_{eff}$ in
a Magnetic Field}

The basic idea is that instead of transforming the operator $W_{\lambda
\lambda^{\prime}}(\vec{\pi}; B, z)$ to an even form directly, one tries to 
transform the lattice Weyl transform of $\mathcal{H}_{eff} \big( \vec{P}, 
\vec{Q} \big)$ into an even form. The power and advantage of this approach lies 
in being able to deal with ordinary $c$-numbers instead of quantum mechanical 
operators. To diagonalize the operator $W_{\lambda \lambda^{\prime}}(\vec{\pi}; 
B, z)$ or transform to an \textit{even form}, we seek a transformation $S_{op}$ 
such that 
\begin{equation}
 S_{op}^{-1} \mathcal{H}_{eff} S_{op} \rightleftarrows \tilde{H}_{\lambda} 
      \bigg( \vec{p} - \frac{e}{c} \vec{A}(\vec{q}); B, z \bigg) \delta_{\lambda 
      \lambda^{\prime}} \label{eq3.1}
\end{equation}
where $\rightleftarrows$ indicates the one-to-one Weyl correspondence between
operator and its lattice Weyl transform. The $\delta_{\lambda \lambda^{\prime}}$ 
includes the diagonalization with rspect to the Kramer's conjugate or degenerate 
bands, analogous to the transformation of the Dirac relativistic Hamiltonian to 
an even form. The transformed effective Hamiltonian operator $\bar{W}_{\lambda} 
(\vec{\pi}; B, z)$ for each band index $\lambda$ is thus given using the 
diagonalized Weyl transform $\tilde{H}_{\lambda}$ as
\begin{equation}
 \bar{W}_{\lambda}(\vec{\pi}; B, z) = \big( N \hbar^{3} \big)^{-1} 
      \sum_{\vec{p}^{\ \prime}, \vec{v}} \tilde{H}_{\lambda}(\vec{p}^{\ 
      \prime}; B, z) \exp \bigg( \frac{2i}{\hbar} \big( \vec{p}^{\ \prime}      
      - \vec{\pi} \big) \cdot \vec{v} \bigg)  \label{eq3.2}
\end{equation}
The Weyl transform of a product of three operators is given by the following
expression,
\begin{align}
 S_{op}^{-1} & \mathcal{H}_{eff} S_{op} \nonumber\\
 \rightleftarrows & \exp \Bigg[  \frac{i\hbar eB}{2c} \Bigg( 
      \frac{\partial^{(a)}}{\partial \vec{k}_{x}} 
      \frac{\partial^{(b)}}{\partial \vec{k}_{y}} - 
      \frac{\partial^{(a)}}{\partial \vec{k}_{y}}
      \frac{\partial^{(b)}}{\partial \vec{k}_{x}} +
      \frac{\partial^{(a)}}{\partial \vec{k}_{x}}
      \frac{\partial^{(c)}}{\partial \vec{k}_{y}} - 
      \frac{\partial^{(a)}}{\partial \vec{k}_{y}}
      \frac{\partial^{(b)}}{\partial \vec{k}_{x}} +
      \frac{\partial^{(b)}}{\partial \vec{k}_{x}} 
      \frac{\partial^{(c)}}{\partial \vec{k}_{y}} -
      \frac{\partial^{(b)}}{\partial \vec{k}_{y}}
      \frac{\partial^{(c)}}{\partial \vec{k}_{x}}
      \Bigg) \Bigg] \nonumber \\
 & \times S^{-1(a)} \Big( \vec{k}, B, z \Big) H^{(b)} \Big( \vec{k}, B, z \Big)  
      S^{(c)} \Big( \vec {k}, B, z \Big)  \label{eq3.3}
\end{align}
The procedure is to diagonalize the lattice Weyl transform of $\mathcal{H}$,
by means of successive similarity transformations
\begin{equation}
 S_{op} = \prod_{i = 1}^{\infty} S_{op}^{0} e^{G_{op}^{(i)}} \label{eq3.4}
\end{equation}
To find $S_{op}^{0}$ we expand $H \big( \vec{k}, B, z \big)$ in powers of
$B$,
\begin{equation}
 H \big( \vec{k}, B, z \big) = H^{0} \big( \vec{k}, z \big) + BH^{(1)} \big( 
      \vec{k}, z \big) + \cdots \label{eq3.5}
\end{equation}
and require that the zero-order term on the right-hand side of Eq. (\ref{eq3.3}) 
be diagonal. Denoting the matrix which diagonalizes $H^{0} \Big( \vec{k}, z 
\big)$ by $U \big( \vec{k}, z \big)$ we have
\begin{equation}
 U^{-1} \big( \vec{k}, z \big) H^{0}\big( \vec{k}, z \big) U \big( \vec{k}, z 
      \big) = \tilde{H}_{\lambda}^{0} \big( \vec{k}, z \big) \delta_{\lambda 
      \lambda^{\prime}} \label{eq3.6}
\end{equation}
Equation (\ref{eq3.6}) is a pure matrix diagonalization problem and was already 
solved above for the zero-field case. There, we have assumed that the 
eigenvalues of $H^{0} \big( \vec{k}, z \big)$ are nondegenerate; the resulting 
eigenvectors of $H^{0} \big( \vec{k}, z \big)$ and those of its adjoint define 
a similarity transformation from Wannier function and Bloch function for $\Sigma 
= 0$ to the Wannier function and Bloch function for $\Sigma \neq 0$, which are, 
in general, energy dependent and biorthogonal.

\subsection{Iterative Solution for a Unitary $U(\vec{k}, z)$}

The operator corresponding to $U \big( \vec{k}, z \big)$, Eq. (\ref{eq3.6}), is, 
however, a similarity transformation only for the zero-field case. Thus, setting 
$\mathcal{H}_{eff} = 1$ in Eq. (\ref{eq3.3}), we have,
\begin{equation}
 \big\{ U_{op}^{-1} U_{op} \big\} \rightleftarrows \exp \Bigg[ \frac{i \hbar 
      eB}{2c} \Bigg( \frac{\partial^{(a)}}{\partial \vec{k}_{x}} 
      \frac{\partial^{(b)}}{\partial \vec{k}_{y}} - 
      \frac{\partial^{(a)}}{\partial \vec{k}_{y}} 
      \frac{\partial^{(b)}}{\partial \vec{k}_{x}} \Bigg) \Bigg] U_{op}^{-1(a)} 
      \big( \vec{k}, z \big) U_{op}^{(b)} \big( \vec{k}, z \big) \label{eq3.7}
\end{equation}
where $\big\{ U_{op}^{-1} U_{op} \big\}$ indicates that the product is not to be 
interpreted as exact product of an operator and its inverse. However, $\big\{ 
U_{op}^{-1} U_{op} \big\}$ can be made equal to unity up to an arbitrary order 
in the magnetic field strength $B$ by means of successive multiplication by 
exponential operators on the left- and right-hand sides. We have%
\begin{equation}
 U_{op}^{-1(n)} U_{op}^{(n)} = \prod_{i = n}^{1} e^{g_{op}^{(i)}} \big\{ 
      U_{op}^{-1} U_{op} \big\} \prod_{i = 1}^{n} e^{g_{op}^{(i)}} = 1 + 
      O(B^{n})  \label{eq3.8}
\end{equation}
where each successive $e^{g_{op}^{(i)}}$ is so chosen so as to make the product 
unity up to order $i$ in the magnetic field strength. 

To prove this, we need the expression for the lattice Weyl transform of an
arbitrary operator $A_{op}$ raised to any power $n$. For problems involving
uniform magnetic field and possessing translational symmetry, this can be
written analogous to Eq. (\ref{eq3.3}) as
\begin{equation}
 A_{op}^{n} \rightleftarrows \cos \Bigg[ \frac{e \hbar B}{2c} \sum_{\substack{j, 
      k=1 \\ j<k}}^{n} \Bigg( \frac{\partial^{(j)}}{\partial \vec{k}_{x}} 
      \frac{\partial^{(k)}}{\partial \vec{k}_{y}} - 
      \frac{\partial^{(j)}}{\partial \vec{k}_{y}} 
      \frac{\partial^{(k)}}{\partial \vec{k}_{x}} \Bigg) \Bigg] \frac{1}{2} 
      \Bigg( \prod_{l=1}^{n} A^{(l)} \bigg( \vec{k}; B \bigg) + \prod_{l=n}^{1} 
      A^{(l)} \bigg( \vec{k}; B \bigg) \Bigg)  \label{eq3.9}
\end{equation}
The lattice Weyl transform of an exponential operator $\exp \Big( 
g_{op}^{(i)} \Big)$ can therefore be expressed as
\begin{align}
 \exp \Big( g_{op}^{(i)} \Big) & \rightleftarrows \exp \big[ g^{(i)} \big( 
      \vec{k}; B \big) \big] + R \nonumber \\
 & = 1 + g^{(i)} \big( \vec{k}; B \big) + \cdots + R  \label{eq3.10}
\end{align}
where $R$ represents the remaining terms and $g_{op}^{(i)} \rightleftarrows 
g^{(i)} \big( \vec{k}; B \big)$. A complete iteration procedure for obtaining 
$S_{op}$ in Eq. (\ref{eq3.4}), up to an arbitrary order in $B$ can now be 
defined.

Let us write Eq. (\ref{eq3.7}) as
\begin{equation}
 \big\{ U_{op}^{-1} U_{op} \big\} \rightleftarrows 1 + BS^{(1)} \big( 
      \vec{k}, z \big) + B^{2} S^{(2)} \big( \vec{k}, z \big) + \cdots 
      \label{eq3.11}
\end{equation}
where the explicit dependence of $B$ comes from the exponential 
``Poisson-bracket operator'' in (\ref{eq3.7}). We choose $g_{op}^{(1)} 
\rightleftarrows -\frac{1}{2} BS^{(1)} \big( \vec{k}, z \big)$, obtaining
\begin{equation}
 e^{g_{op}^{(1)}} \big\{ U_{op}^{-1} U_{op} \big\} e^{g_{op}^{(1)}} 
      \rightleftarrows 1 + B^{2} S^{(2)} \big( \vec{k}, z \big) + 
      \Delta_{R}^{(1)} \label{eq3.12}%
\end{equation}
We next choose, $g_{op}^{(2)} = -\frac{1}{2} B^{2}S^{(2)} \big( \vec{k}, 
z \big)$ resulting in
\begin{equation}
 e^{g_{op}^{(2)}} e^{g_{op}^{(1)}} \big\{ U_{op}^{-1} U_{op} \big\} 
      e^{g_{op}^{(1)}} e^{g_{op}^{(2)}} \rightleftarrows 1 + B^{3} S^{(3)} 
      \big( \vec{k}, z \big) + \Delta_{R}^{(2)} \label{eq3.13}%
\end{equation}
In general order $n$, we have
\begin{equation}
 U_{op}^{-1(n)} U_{op}^{(n)} \rightleftarrows 1 + B^{n+1} S^{(n + 1)} \big( 
      \vec{k}, z \big) + \Delta_{R}^{(n)}  \label{eq3.14}
\end{equation}
and $g_{op}^{(n)}$ can be chosen such that $g_{op}^{(n)} = -\frac{1}{2} B^{n} 
S^{(n)} \big( \vec{k}, z \big)$. This completes the proof.

\subsection{Iterative Removal of Interband Terms of $\mathcal{H}_{eff}$}

We now proceed to the diagonalization of $\mathcal{H}_{eff}$, Eq. 
(\ref{eq3.1}). Let us write the expression containing the zero-order diagonal,
$\tilde{H}_{\lambda}^{0} \big( \vec{k}, z \big) \delta_{\lambda 
\lambda^{\prime}}$, as
\begin{equation}
 \Big( S_{op}^{0} \Big)^{-1} \mathcal{H}_{eff} S_{op}^{0} \rightleftarrows 
      \tilde{H}_{\lambda}^{0} \big( \vec{k}, z \big) \delta_{\lambda 
      \lambda^{\prime}} + BH^{(1)} \big( \vec{k}, z \big)_{\lambda 
      \lambda^{\prime}} + \Delta_{R} \label{eq3.15}
\end{equation}
Note that for even function of coordinates, linear momentum, the velocity, the
spin-orbit interaction, and a symmetrized product of an even number of linear
momenta do \textit{not} couple the Kramer's conjugates or doubly degenerate
states of the same energy. Thus, any other operators which couple the Kramer's
conjugate states must result in spin splitting of the doubly degenerate states
over the whole Brillouin zone. By virtue of time reversal and inversion 
symmetry, we assume that $\tilde{H}_{\lambda}^{0} \big( \vec{k}, z \big) 
\delta_{\lambda \lambda^{\prime}} \equiv \tilde{H}_{\lambda}^{0} \big( \vec{k}, 
z \big) \delta_{\lambda \lambda^{\prime}} \delta_{\lambda, \sigma 
\sigma^{\prime}} \delta_{\lambda^{\prime}, \sigma \sigma^{\prime}}$, indicating
that for $\Sigma\neq 0$, the Hamiltonian $\tilde{H}_{\lambda}^{0} \big( \vec{k}, 
z \big) \delta_{\lambda \lambda^{\prime}}$ in the absence of magnetic field is 
also brought to an even form in terms of the doubly degenerate states in the 
case with spin-orbit interaction. In the presence of magnetic field the doubly 
degenerate states are spin-split. Our task for $\Sigma\neq0$ is to diagonalize 
$\mathcal{H}_{eff}$ by method of successive similarity transformation starting 
with $S_{op}^{0}$, Eq. (\ref{eq3.4}). We will do this for a many-body system, 
$\Sigma \neq 0$, in a magnetic field.

Let us define ``odd'' and ``even'' operators and matrices. An even matrix is a 
diagonal matrix and the corresponding operator is called an even operator. An 
odd matrix and its corresponding operator is one where all diagonal (intraband) 
elements are zeros. Even operators and matrices commute, products of even 
matrices are even, whereas products of even and odd are odd. The zero-order term 
on the right-hand side of Eq. (\ref{eq3.15}) is even, the remaining terms may be 
written as a sum of even and odd matrices. Therefore an iterative procedure to 
diagonalize $\mathcal{H}_{eff}$ involves removing odd terms on the right-hand 
side of Eq. (\ref{eq3.15}) up to arbitrary orders in the magnetic field 
strength. Odd terms in Eq. (\ref{eq3.15}) correspond to the presence of 
interband terms (including the Kramer's conjugate bands in the case with 
spin-orbit interaction which are coupled in a magnetic field) in the effective 
Hamiltonian and its Weyl transform.

First we choose $G_{op}^{(1)} \rightleftarrows G^{(1)} \big( \vec{k}, B, z 
\big)$ such that
\begin{equation}
 \Big[ G^{(1)} \big( \vec{k}, B, z \big), \tilde{H}^{0} \big( \vec{k}, z 
      \big) \Big] = BH_{odd}^{(1)} \big( \vec{k}, z \big), \label{xtra3}
\end{equation}
where $H_{odd}^{(1)} \big( \vec{k}, z \big)$ is the odd part of $H^{(1)} \big( 
\vec{k}, z \big)$ in Eq. (\ref{eq3.15}). Then we have
\begin{equation}
 e^{-G_{op}^{(1)}} \Big[ \big( S_{op}^{0} \big)^{-1} \mathcal{H}_{eff} 
      S_{op}^{0} \Big] e^{G_{op}^{(1)}} \rightleftarrows \tilde{H}_{\lambda}^{0} 
      \big( \vec{k}, z \big) \delta_{\lambda \lambda^{\prime}} + BH_{even}^{(1)} 
      + B^{2} \Big[ H_{odd}^{(2)} + H_{even}^{(2)} \Big] + \Delta_{R}^{(1)}  
      \label{eq3.16}
\end{equation}
showing that the right-hand side of Eq. (\ref{eq3.16}) is even up to order $B$. 
Since $\tilde{H}^{0} \big( \vec{k}, z \big)$ is even, $G^{(1)} \big( \vec{k}, B, 
z \big)$ can be chosen odd. Its matrix elements is related to that of 
$H_{odd}^{(1)} \big( \vec{k}, z \big)$ and $\tilde{H}^{0} \big( \vec{k}, z 
\big)$. For $\tilde{H}_{j}^{0} \big( \vec{k}, z \big) \neq \tilde{H}_{i}^{0} 
\big( \vec{k}, z \big)$, this is given by the relation
\begin{equation}
 G_{ij}^{(1)} \big( \vec{k}, B, z \big) = \left\{
      \begin{array} [c]{ccc}
       & \ B \Big[ H_{odd}^{(1)} \big( \vec{k}, z \big) \Big]_{ij} \ \Big[ 
         \tilde{H}_{j}^{0} \big( \vec{k}, z \big) - \tilde{H}_{i}^{0} \big( 
         \vec{k}, z \big) \Big]^{-1} & \quad i \neq j\\
       & \ 0 & \quad i=j
      \end{array}
      \right. \qquad \label{eq3.17_18}
\end{equation}
For the Kramer's conjugate orthogonal states, $\tilde{H}_{i}^{0} \big( \vec{k}, 
z \big)_{\sigma} = \tilde{H}_{i}^{0} \big( \vec{k}, z \big)_{\sigma^{\prime}}$, 
we asume that the matrix $H_{odd}^{(1)} \big( \vec{k}, z \big)_{\sigma 
\sigma^{\prime}}$ for each band is of the form of $i\sigma_{y}$ where 
$\sigma_{y}$ is the Pauli matrix, $\sigma_{y} = 
\begin{pmatrix} 
 0 & -i \\
 i & 0
\end{pmatrix}
$. This form can be obtained by proper choice of the phase of one of the 
Kramer's conjugate orthogonal states. Then for the spinor we have from $\Big[ 
G_{i}^{(1)} \big( \vec{k}, B, z \big), \tilde{H}^{0} \big( \vec{k}, z \big) 
\Big] = BH_{odd}^{(1)} \big( \vec{k}, z \big)_{i}$, the following relation
\begin{equation}
 G_{i}^{(1)} \big( \vec{k}, B, z \big) = \frac{BH_{odd}^{(1)} \big( \vec{k}, z 
      \big)_{i}}{\tilde{H}_{i}^{0} \big( \vec{k}, z \big)} \label{eqxtra4}
\end{equation}
as the proper choice for the spinor $G_{i}^{(1)} \big( \vec{k}, B, z \big)$ for 
the each band index $i$. The procedure can now be reiterated, choosing 
$G_{op}^{(2)} \rightleftarrows G^{(2)} \big( \vec{k}, B, z \big)$ such that
\begin{equation}
 \Big[ G^{(2)} \big( \vec{k}, B, z \big), \tilde{H}^{0} \big( \vec{k}, z \big) 
      \Big] = B^{2 \ (1)} H_{odd}^{(2)} \big( \vec{k}, z \big) \label{eqxtra5}
\end{equation}
resulting in
\begin{align}
 e^{-G_{op}^{(2)}} e^{-G_{op}^{(1)}} \Big[ \big( S_{op}^{0} \big)^{-1} 
      \mathcal{H}_{eff} S_{op}^{0} \Big] e^{G_{op}^{(1)}} e^{G_{op}^{(2)}} 
      \rightleftarrows & \ \tilde{H}_{\lambda}^{0} \big( \vec{k}, z \big) 
      \delta_{\lambda \lambda^{\prime}} + BH_{even}^{(1)} + B^{2} 
      H_{even}^{(2)} \nonumber\\
 & \ + B^{3} \Big[ H_{even}^{(3)} + H_{odd}^{(3)} \Big] + \Delta_{R}^{(2)} 
      \label{eq3.19}
\end{align}
with $G^{(2)} \big( \vec{k}, B, z \big)$ given by
\begin{equation}
 G_{ij}^{(2)} \big( \vec{k}, B, z \big) = \left\{
 \begin{array} [c]{ccc}
  & \ B^{2} \Big[ H_{odd}^{(2)} \big( \vec{k}, z \big) \Big]_{ij} \ \Big[ 
      \tilde{H}_{j}^{0} \big( \vec{k}, z \big) - \tilde{H}_{i}^{0} \big(  
      \vec{k}, z \big) \Big]^{-1} & \quad i\neq j\\
  & \ 0 & \quad i=j
 \end{array}
 \right.  \label{eq3.20_21}
\end{equation}
with the spinor for each band $G_{i}^{(2)} \big( \vec{k}, B, z \big)_{\sigma 
\sigma^{\prime}}$ given by
\begin{equation}
 G_{i}^{(2)} \big( \vec{k}, B, z \big) = \frac{BH_{odd}^{(2)} \big( \vec{k}, z 
      \big)_{i}}{\tilde{H}_{i}^{0} \big( \vec{k}, z \big)} \label{eqxtra6}
\end{equation}

In general order $n$, we can choose $G_{op}^{(n + 1)} \rightleftarrows 
G^{(n + 1)} \big( \vec{k}, B, z \big)$ such that if
\begin{align}
 \prod_{i = n}^{1} e^{-G_{op}^{(i)}} \Big[ \big( S_{op}^{0} \big)^{-1} 
      \mathcal{H}_{eff} S_{op}^{0} \Big] \prod_{i = 1}^{n} e^{-G_{op}^{(i)}}  
      \rightleftarrows & \tilde{H}_{\lambda}^{0} \big( \vec{k}, z \big) 
      \delta_{\lambda \lambda^{\prime}} + \cdots B^{n} H_{even}^{(n)} \big(
      \vec{k}, z \big) \nonumber\\
 & + B^{n+1} \Big[ H_{odd}^{(n + 1)} \big( \vec{k}, z \big) + H_{even}^{(n + 1)} 
      \big( \vec{k}, z \big) \Big] + \Delta_{R}^{(n)} \label{eq3.22}
\end{align}
then we must have $G^{(n + 1)} \big( \vec{k}, B, z \big)$ given by
\begin{equation}
 G_{ij}^{(n + 1)} \big( \vec{k}, B, z \big) = \left\{
      \begin{array} [c]{ccc}
       & \ B^{n + 1} \Big[ H_{odd}^{(n + 1)} \big( \vec{k}, z \big) \Big]_{ij} 
         \ \Big[ \tilde{H}_{j}^{0} \big( \vec{k}, z \big) - \tilde{H}_{i}^{0} 
         \big( \vec{k}, z \big) \Big]^{-1} & \quad i\neq j\\
       & \ 0 & \quad i=j
      \end{array}
      \right.  \label{eq3.23_24}
\end{equation}
and for the spinor matrix for each band,
\begin{equation}
 G_{i}^{(n + 1)} \Big( \vec{k}, B, z \Big) = \frac{BH_{odd}^{(n + 1)} \big( 
      \vec{k}, z \big)_{i}}{\tilde{H}_{i}^{0} \big( \vec{k}, z \big)} 
      \label{eqxtra7}
\end{equation}

The above procedure can, of course, only be guaranteed to converge for very
small fields; for calculating the low-field susceptibility the removal of the
interband terms up to second order in $B$ is all that is required since
higher-order terms do not contribute.

\subsection{Direct Removal of Interband Terms of $\mathcal{H}$: Magnetic Basis
Functions}

A lattice Weyl transform of $\mathcal{H}$ which is free of interband terms
suggests the existence of magnetic Wannier function and magnetic Bloch
function of interacting Bloch electrons in a magnetic field. The starting
point in obtaining the lattice Weyl transform of $\mathcal{H}$, which is free
of interband terms, is the equation defining the magnetic Wannier function on
the corresponding equation defining the magnetic Bloch function. In the
magnetic Wannier function representation we have
\begin{equation}
 \mathcal{H} = \sum_{\lambda, \vec{q}^{\ \prime}, \vec{q}} \bigg[ \Big\langle 
      \lambda, \vec{q}^{\ \prime}, z, B \Big| \mathcal{H}_{0} + \Sigma(z) 
      \Big|\lambda, \vec{q}, z, B \Big\rangle \bigg] \ \Big| \lambda, 
      \vec{q}^{\ \prime}, z, B \Big\rangle \Big\langle \lambda, \vec{q}, z, B 
      \Big|   \label{eq3.25}
\end{equation}
where the matrix elements of $\mathcal{H}_{0} + \Sigma(z)$ in general have dual 
magnetic Wannier functions on the left- and right-hand sides instead of complex 
conjugate of the same wave function. The equation defining 
$|\lambda, \vec{q}, z, B \rangle$ becomes
\begin{equation}
 \mathcal{H} \ \Big|\lambda, \vec{q}, z, B \Big\rangle =  \sum_{\vec{q}^{\ 
      \prime}} \bigg[ \Big\langle \lambda, \vec{q}^{\ \prime}, z, B \Big| 
      \mathcal{H}_{0} + \Sigma(z) \Big| \lambda, \vec{q}, z, B \Big\rangle 
      \bigg] \ \Big| \lambda, \vec{q}^{\ \prime}, z, B \Big\rangle  
      \label{eq3.26}
\end{equation}
and that for $\langle \lambda, \vec{q}^{\ \prime}, z, B|$ is
\begin{equation}
 \Big\langle \lambda, \vec{q}^{\ \prime}, z, B \Big| \mathcal{H} = 
      \sum_{\vec{q}} \bigg[ \Big\langle \lambda, \vec{q}^{\ \prime}, z, B \Big| 
      \mathcal{H}_{0} + \Sigma(z) \Big| \lambda, \vec{q}, z, B \Big\rangle 
      \bigg] \ \Big\langle \lambda, \vec{q}, z, B \Big| \label{eq3.27}
\end{equation}
The form of the matrix elements is given in Appendix B of Ref. \cite{freeBloch}, 
which may be written as
\begin{equation}
 \Big\langle \lambda, \vec{q}^{\ \prime}, z, B \Big| \mathcal{H}_{0} + 
      \Sigma(z) \Big| \lambda, \vec{q}, z, B \Big\rangle = \exp \bigg[ \bigg( 
      \frac{ie}{\hbar c} \bigg) \vec{A}(\vec{q}) \cdot \vec{q} \bigg] \  
      H_{\lambda} \Big( \vec{q} - \vec{q}^{\ \prime}, B, z \Big) 
      \delta_{\lambda \lambda^{\prime}} \label{eq3.28}
\end{equation}
The magnetic translation operator is define by
\begin{align} 
 \Big| \lambda, \vec{q}, z, B \Big\rangle & = T(-\vec{q}) \Big| 
      \lambda, 0, z, B \Big\rangle_{0} 
  = \exp \bigg[ \bigg( \frac{-ie}{\hbar c} \bigg) \vec{A}(\vec{r}) \cdot 
      \vec{q} \bigg]  \Big| \lambda,- \vec{q}, z, B \Big\rangle_{0} 
      \label{eq3.29}
\end{align}
where $T(-\vec{q})$ is the magnetic translation operator and $w_{\lambda} 
\Big(\vec{r} - \vec{q}, z, B \Big) \equiv \Big| \lambda, -\vec{q}, z, B 
\Big\rangle_{o}$ is the modified Wannier function centered at the lattice point 
$\vec{q}$. The equation satisfied by $w_{\lambda} \Big( \vec{r} - 0, z, B \Big) 
\equiv w_{\lambda} \Big( \vec{r}, z, B \Big)$ 
can explicitly be written
\begin{equation}
 \mathcal{H}_{0} \ w_{\lambda}(\vec{r}, z, B) + \int d^{3} r^{\prime} 
      \Sigma(\vec{r}, \vec{r}^{\ \prime}, z, B) \ w_{\lambda} \Big( \vec{r}^{\ 
      \prime}, z, B \Big) = \sum_{\vec{q}^{\ \prime}} H_{\lambda} \Big( 
      -\vec{q}^{\ \prime}, z, B \Big) \ \Big| \lambda, \vec{q}^{\ \prime}, z, B 
      \Big\rangle \label{eq3.30}
\end{equation}
and hence,
\begin{equation}
 \mathcal{H} \big|\lambda, \vec{q}, z, B \big\rangle = T(-q) \ \mathcal{H} 
      \big|\lambda, 0, z, B \big\rangle \label{eq3.31}
\end{equation}
which can be written explicitly as
\small
\begin{align}
 \exp & \bigg[ \bigg( \frac{-ie}{\hbar c} \bigg) \vec{A}(\vec{r}) \cdot \vec{q} 
      \ \bigg] \mathcal{H}_{0}(\vec{r} - \vec{q}) \ w_{\lambda} \Big(\vec{r} - 
      \vec{q}, z, B \Big) \nonumber\\
 & + \exp \bigg[ \bigg( \frac{-ie}{\hbar c} \bigg) \vec{A}(\vec{r}) \cdot 
      \vec{q} \ \bigg] \int d^{3} r^{\prime} \Sigma \Big( \vec{r} - \vec{q}, 
      \vec{r}^{\ \prime}, z, B \Big) \ w_{\lambda} \Big( \vec{r}^{\ \prime} - 
      0, z, B \Big) \nonumber\\
 & = \sum_{\vec{q}^{\ \prime}} \exp \bigg[ \bigg( \frac{ie}{\hbar c} \bigg) 
      \vec{A}(\vec{q}) \cdot \vec{q}^{\ \prime} \bigg] H_{\lambda} \Big( 
      \vec{q} - \vec{q}^{\ \prime}, z, B \Big) \ \exp \bigg[ \bigg( 
      \frac{-ie}{\hbar c} \bigg) \vec{A}(\vec{r}) \cdot \vec{q}^{\ \prime} 
      \bigg] w_{\lambda} \Big( \vec{r} - \vec{q}^{\ \prime}, z, B \Big)  
      \label{eq3.32}
\end{align}
\normalsize
Changing the variable of integration $\vec{r}^{\ \prime}$ to $\vec{r}^{\ 
\prime} - q$, noting that by symmetry,
\begin{equation}
 \Sigma(\vec{r}, \vec{r}^{\ \prime}, z, B) = \exp \bigg[ \bigg( \frac{-ie}{\hbar 
      c} \bigg) \vec{A}(\vec{r}) \cdot \vec{r}^{\ \prime} \bigg] \tilde{\Sigma} 
      \Big( \vec{r}, \vec{r}^{\ \prime}, z, B \Big)  \label{eq3.33}
\end{equation}
where $\Sigma \Big( \vec{r} - \vec{q}, \vec{r}^{\ \prime} - q, z, B \Big) = 
\Sigma \Big( \vec{r}, \vec{r}^{\ \prime}, z, B \Big)$. Dividing both sides of 
the equation by $\exp \Big[ \Big( \frac{-ie}{\hbar c} \Big) \vec{A} (\vec{r}) 
\cdot \vec{q} \Big]$ and taking the lattice Fourier transform [i.e., multiply 
both sides by operation, $\Big( \frac{1}{N\hbar} \Big)^{3} \sum_{q} \exp \Big( 
\frac{i}{\hbar} \vec{p} \cdot \vec{q} \Big)$], we obtain
\begin{align}
 \mathcal{H}_{0} & \Big( \vec{p} - \frac{e}{c} \vec{A} \big( \vec{r} + i 
      \nabla_{\vec{k}} \big), \vec{r} \Big) \ b_{\lambda} \Big( \vec{r}, 
      \vec{k}, B, z \Big) \nonumber\\
 & + \int d^{3} r^{\prime} \exp \bigg[ \bigg( \frac{-ie}{\hbar c} \bigg) 
      \vec{A}(\vec{r}) \cdot \vec{r}^{\ \prime} \bigg] \tilde{\Sigma} 
      \Big( \vec{r}, \vec{r}^{\ \prime}, z, B \Big) \ b_{\lambda} \Big( 
      \vec{r}^{\ \prime}, \vec{k} + \frac{e}{\hbar c} \vec{A}(\vec{r} - 
      \vec{r}^{\ \prime}), B, z \Big)   \nonumber \\
 & = \sum_{\vec{q}} e^{i \vec{k} \cdot \vec{q}} H_{\lambda} \Big( \vec{q}, z, 
      B \Big) e^{\big( \frac{ie}{\hbar c} \big) \vec{A}(\vec{r}) \cdot \vec{q}} 
      \ b_{\lambda} \bigg( \vec{r}^{\ \prime}, \vec{k} + \frac{e}{\hbar c} 
      \vec{A}(\vec{q}); B, z \bigg)  \label{eq3.34}
\end{align}
where the modified Bloch function $b_{\lambda} \Big( \vec{r}^{\ \prime}, 
\vec{k} + \frac{e}{\hbar c} \vec{A}(\vec{q}); B, z \Big)$ is defined by
\begin{align}
 b_{\lambda} \Big( \vec{r}, \vec{k}, B, z \Big) & = \Big( N \hbar^{3} 
      \Big)^{-\frac{1}{2}} \sum_{\vec{q}} e^{i\vec{k} \cdot \vec{q}} 
      w_{\lambda} \Big( \vec{r} - \vec{q}, z, B \Big) \nonumber\\
 & \equiv e^{i \vec{k} \cdot \vec{r}} u_{\lambda} \Big( \vec{r}, \vec{k}, z, B 
      \Big)  \label{eq3.35}%
\end{align}
The equation satisfied by $u_{\lambda} \Big( \vec{r}, \vec{k}, z, B \Big)$ 
is
\begin{align}
 \mathcal{H}_{0} & \Big( \vec{p} + \hbar k - \frac{e}{c} \vec{A} \big( i 
      \nabla_{\vec{k}} \big); \vec{r} \Big) \ u_{\lambda} \Big( \vec{r}, 
      \vec{k}, z, B \Big) \nonumber\\
 & + \int d^{3} r^{\prime} e^{i \vec{k} \cdot (\vec{r}^{\ \prime} - \vec{r})} 
      \tilde{\Sigma} \big( \vec{r}, \vec{r}^{\ \prime}, z, B \big) \ u_{\lambda} 
      \Big( \vec{r}, \vec{k} + \frac{e}{\hbar c} \vec{A} (\vec{r} - 
      \vec{r}^{\ \prime}), z, B \Big) \nonumber\\
 & = \sum_{\vec{q}} e^{i \vec{k} \cdot \vec{q}} H_{\lambda}(\vec{q}, z, B) \ 
      u_{\lambda} \Big( \vec{r}, \vec{k} + \frac{e}{\hbar c} 
      \vec{A}(\vec{q}), z, B \Big)  \label{eq3.36}
\end{align}

\subsection{Expansions in Powers of $B$}

Let us expand $\mathcal{H}_{0}, u_{\lambda} \Big( \vec{r},\vec{k}, z, B \Big), 
\tilde{\Sigma} \Big( \vec{r}, \vec{r}^{\ \prime}, z, B \Big)$ in powers of the 
magnetic field, $B$, as well
\begin{equation*}
 u_{\lambda} \Big( \vec{r}, \vec{k} + \frac{e}{\hbar c} 
      \vec{A}(\vec{q}), z, B \Big) = \exp \Big[ \Big( \frac{e}{\hbar c} \Big) 
      A(q) \cdot \nabla_{k} \Big] u_{\lambda} \Big( \vec{r},\vec{k}, z, B \Big).
\end{equation*}
We expand both sides of Eq. (\ref{eq3.36}) up to second order in $B$ and
equate the coefficients on both sides. Multiplying both sides of the resulting
equations by the dual set of wave functions $\Big\langle u_{\lambda} \Big( 
\vec{r}, \vec{k}, z, B \Big) \Big|$, determined by Eq. (\ref{eq2.18})
biorthogonal to $\big|u_{\lambda} \big( \vec{r}, \vec{k}, z, B \big) 
\big\rangle$, and integrating, we obtain for $\delta = \lambda$ the expression 
for $H_{\lambda}^{(1)}(k, z)$ and $H_{\lambda}^{(2)}(k, z)$, where
\begin{equation}
 H_{\lambda}^{(l)} \big( \vec{k}, z \big) = \sum_{\vec{q}} e^{i \vec{k} \cdot 
      \vec{q}} H_{\lambda}^{(l)}(\vec{q}, z)  \label{eq3.41}
\end{equation}
We obtained the following expressions:
\small
\begin{align} 
 H_{\lambda}^{(1)}(k, z) = & \ \bigg\langle u_{\lambda}^{0} \ \bigg| 
      \ \mathcal{H}_{0}^{(1)} + \frac{e}{B \hbar c} \bigg[ \vec{A} \big( 
      \nabla_{\vec{k}} \big) \ \tilde{H}_{\lambda}^{0} \big( \vec{k}, z \big) 
      \bigg] \cdot \nabla_{\vec{k}} \ \bigg| \ u_{\lambda}^{0} \ \bigg\rangle 
      \nonumber \\
 & + \bigg\langle \ u_{\lambda}^{0} \big( \vec{r}, \vec{k}, z \big) \ \bigg| \ 
      \int d^{3} r^{\prime} e^{i \vec{k} \cdot \big(\vec{r}^{\ \prime} - \vec{r} 
      \big)} \ \tilde{\Sigma}^{0}(\vec{r}, \vec{r}^{\ \prime}, z) \ \frac{e}{B   
      \hbar c} \vec{A}(\vec{r} - \vec{r}^{\ \prime}) \cdot \nabla_{\vec{k}} \   
      \bigg| \ u_{\lambda}^{0} \big( \vec{r}^{\ \prime}, \vec{k}, z \big)      
      \bigg\rangle \nonumber\\
 & + \bigg\langle u_{\lambda}^{0} \big( \vec{r}, \vec{k}, z \big) \ \bigg| \int 
     d^{3} r^{\prime} e^{i \vec{k} \cdot \big( \vec{r}^{\ \prime} - \vec{r} 
     \big)} \tilde{\Sigma}^{(1)} (\vec{r}, \vec{r}^{\ \prime}, z) \ \bigg| 
     u_{\lambda}^{0} \big( \vec{r}^{\ \prime}, \vec{k}, z \big) \ \bigg\rangle 
     \label{eq3.42} \\
 H_{\lambda}^{(2)}(k, z) = & \ \Big\langle \ u_{\lambda}^{0} \ \Big| 
     \mathcal{H}_{0}^{(2)} \ \Big| \ u_{\lambda}^{0} \ \Big\rangle + 
     \Big\langle \ u_{\lambda}^{0} \ \Big| \ \mathcal{H}_{0}^{(1)} \ \Big| 
     u_{\lambda}^{(1)} \ \Big\rangle \nonumber \\
 & + \bigg\langle \ u_{\lambda}^{0} \ \bigg| \ \sum_{\vec{q}} e^{i \vec{k} \cdot 
     \vec{q}} \mathcal{\tilde{H}}_{\lambda}^{0}(\vec{q}, z) \frac{1}{2!} \bigg(  
     \frac{e}{B \hbar c} \vec{A}(q) \cdot \nabla_{\vec{k}} \bigg)^{2} \ \bigg| 
     \ u_{\lambda}^{0} \ \bigg\rangle \nonumber\\
 & + \bigg\langle \ u_{\lambda}^{0} \bigg| \ \frac{e}{B \hbar c} \vec{A}(q) 
     \cdot \nabla_{\vec{k}} \mathcal{\tilde{H}}_{\lambda}^{0}(\vec{q}, z) \Big( 
     \frac{e}{B \hbar c} \vec{A}(q) \cdot i \nabla_{\vec{k}} \Big) \ \bigg| \ 
     u_{\lambda}^{(1)} \bigg\rangle \nonumber\\
 & + \bigg\langle \ u_{\lambda}^{0} \bigg| \ \frac{e}{B \hbar c} 
     \vec{A}(q) \cdot \nabla_{\vec{k}} \mathcal{H}_{\lambda}^{(1)}(\vec{q}, z) 
     \Big( \frac{e}{B \hbar c} \vec{A}(q) \cdot i \nabla_{\vec{k}} \Big) \ 
     \bigg| \ u_{\lambda}^{0} \bigg\rangle \nonumber \\
 & - \mathcal{H}_{\lambda}^{(1)}(\vec{q}, z) \Big\langle \ u_{\lambda}^{0} 
     \Big| \Big| u_{\lambda}^{(1)} \Big\rangle \nonumber\\
 & + \bigg\langle \ u_{\lambda}^{0} \big( \vec{r}, \vec{k}, z \big) \bigg| \int 
     d^{3} r^{\prime} e^{i \vec{k} \cdot \big( \vec{r}^{\ \prime} - \vec{r} 
     \big)} \tilde{\Sigma}^{0}(\vec{r}, \vec{r}^{\ \prime}, z) \ \frac{1}{2!} 
     \bigg( \frac{e}{B \hbar c} \vec{A}(\vec{r} - \vec{r}^{\ \prime}) \cdot 
     \nabla_{\vec{k}} \bigg)^{2} \ \bigg| \ u_{\lambda}^{0} \big( \vec{r}, 
     \vec{k}, z \big) \bigg\rangle \nonumber\\
 & + \bigg\langle \ u_{\lambda}^{0} \big( \vec{r}, \vec{k}, z \big) \bigg| \int 
     d^{3} r^{\prime} e^{i \vec{k} \cdot \big( \vec{r}^{\ \prime} - \vec{r} 
     \big)} \tilde{\Sigma}^{0}(\vec{r}, \vec{r}^{\ \prime}, z) \ \bigg( 
     \frac{e}{B \hbar c} \vec{A}(\vec{r} - \vec{r}^{\ \prime}) \cdot 
     \nabla_{\vec{k}} \bigg) \ \bigg| \ u_{\lambda}^{1} \big( \vec{r}, 
     \vec{k}, z \big) \bigg\rangle \nonumber\\
 & + \bigg\langle \ u_{\lambda}^{0} \big( \vec{r}, \vec{k}, z \big) \bigg| \int 
     d^{3} r^{\prime} e^{i \vec{k} \cdot \big( \vec{r}^{\ \prime} - \vec{r} 
     \big)} \tilde{\Sigma}^{1}(\vec{r}, \vec{r}^{\ \prime}, z) \ \bigg( 
     \frac{e}{B \hbar c} \vec{A}(\vec{r} - \vec{r}^{\ \prime}) \cdot 
     \nabla_{\vec{k}} \bigg) \ \bigg| \ u_{\lambda}^{0} \big( \vec{r}, 
     \vec{k}, z \big) \bigg\rangle \nonumber\\
 & + \bigg\langle \ u_{\lambda}^{0} \big( \vec{r}, \vec{k}, z \big) \bigg| \int 
     d^{3} r^{\prime} e^{i \vec{k} \cdot \big( \vec{r}^{\ \prime} - \vec{r} 
     \big)} \tilde{\Sigma}^{1}(\vec{r}, \vec{r}^{\ \prime}, z) \ \bigg| \ 
     u_{\lambda}^{1} \big( \vec{r}, \vec{k}, z \big) \bigg\rangle \nonumber\\
 & + \bigg\langle \ u_{\lambda}^{0} \big( \vec{r}, \vec{k}, z \big) \bigg| \int 
     d^{3} r^{\prime} e^{i \vec{k} \cdot \big( \vec{r}^{\ \prime} - \vec{r} 
     \big)} \tilde{\Sigma}^{2}(\vec{r}, \vec{r}^{\ \prime}, z) \ \bigg| \ 
     u_{\lambda}^{0} \big( \vec{r}, \vec{k}, z \big) \bigg\rangle 
     \label{eq3.43}
\end{align}
\normalsize
Note that in the expression for $H_{\lambda}^{(2)}(k, z)$ we need $\big| 
u_{\lambda}^{(1)} \big( \vec{r}, \vec{k}, z \big) \big\rangle $, which can be 
written
\begin{equation}
 \Big| \ u_{\lambda}^{1} \big( \vec{r}, \vec{k}, z \big) \Big\rangle = 
      \sum_{\sigma} \beta_{\lambda \sigma} \Big| \ u_{\sigma}^{0} \big( 
      \vec{r}, \vec{k}, z \big) \Big\rangle \label{eq3.44}
\end{equation}
and for $\lambda \neq \delta, \beta_{\lambda \sigma}$ can de determined from
the same set of equations which determined $H_{\lambda}^{(i)}(k, z)$. It is 
given by
\begin{align}
 \Big\langle u_{\sigma}^{0} \big( \vec{r}, \vec{k}, z \big) \Big| 
      u_{\lambda}^{(1)}(\vec{r}, \vec{k}, z) \Big\rangle & = -\big[ 
      H_{\sigma}^{0}(k, z) - H_{\lambda}^{0}(k, z) \big]^{-1} \Big\langle 
      u_{\sigma}^{0} \big( \vec{r}, \vec{k}, z \big) \Big| H_{\lambda}^{(1)op} 
      \Big| \ u_{\lambda}^{0}\big(\vec{r}, \vec{k},z \big) \Big\rangle 
      \nonumber\\
 & = \beta_{\sigma \lambda} \label{eq3.45}
\end{align}
where $\big\langle \ u_{\sigma}^{0} \ \big| \ H_{\lambda}^{(1)op} \ \big| \ 
u_{\lambda}^{0} \big\rangle$ is given by Eq. (\ref{eq3.42}) with the band index 
$\lambda$ replaced by $\delta$ on the left-hand wave function. For $\lambda = 
\delta, \beta_{\lambda \lambda}$ can be obtained from the requirement that the 
magnetic Wannier functions are biorthogonal. This is expressed by the equation
\begin{equation}
 \int d^{3} r \exp \bigg[ \bigg( \frac{1}{2} \frac{ie}{\hbar c} \bigg) \Big( 
      \vec{B} \times \vec{r} \Big) \cdot \big( \vec{q}^{\ \prime} - \vec{q} 
      \big) \bigg] \Omega_{\lambda}^{\ast} \big( \vec{r} - \vec{q}^{\ \prime}, 
      z, B \big) w_{\lambda} \big( \vec{r} - \vec{q}, z, B \big) = 
      \delta_{\vec{q} \ \vec{q}^{\ \prime}} \label{eq3.48}
\end{equation}
where we have written
\begin{equation}
 \big\langle \lambda, \vec{q}^{\ \prime}, z, B \big| = \exp \bigg[ \bigg( 
      \frac{1}{2} \frac{ie}{\hbar c} \bigg) \Big( \vec{B} \times \vec{r} \Big) 
      \cdot \vec{q}^{\ \prime} \bigg] \Omega_{\lambda}^{\ast} \big( \vec{r} - 
      \vec{q}^{\ \prime}, z, B \big)  \label{eq3.49}
\end{equation}
Expanding the left-hand side of Eq. (\ref{eq3.48}) in powers of $B$ we obtain 
the following relations:
\begin{equation}
 \big\langle \lambda, \vec{q}^{\ \prime}, z \big| \lambda, 
\vec{q}, z \big\rangle = \delta_{\vec{q} \ \vec{q}^{\ \prime}} \label{eq3.50}
\end{equation}
\begin{align}
 \Big\langle \lambda, & \vec{q}^{\ \prime}, z \Big| w_{\lambda}^{(1)} \big( 
      \vec{r} - \vec{q}^{\ \prime}, z \big) \Big\rangle + \Big\langle 
      \Omega_{\lambda}^{(1)\ast} \big( \vec{r} - \vec{q}^{\ \prime}, z \big) 
      \Big| \lambda, \vec{q}, z \Big\rangle \nonumber \\
 & + \frac{1}{2} \bigg( \frac{ie}{\hbar c} \bigg) \Big\langle \big( \hat{z} 
      \times \vec{r} \big) \cdot \big( \vec{q}^{\ \prime} - \vec{q} \big) \ 
      \Omega_{\lambda}^{0\ast} \big( \vec{r} - \vec{q}^{\ \prime}, z \big) 
      w_{\lambda}^{0} \big( \vec{r} - \vec{q}, z \big) \Big\rangle = 0 
      \label{eq3.51}%
\end{align}
where $\Omega_{\lambda}^{0\ast} \big( \vec{r} - \vec{q}^{\ \prime},z \big) 
\Longrightarrow \big\langle \lambda,\vec{q},z \big|$ and $w_{\lambda}^{0} 
(\vec{r} - \vec{q}, z) \Longrightarrow \big| \lambda, \vec{q}, z \big\rangle$.

\subsection{Berry connection and Berry curvature}

By virtue of the identity
\begin{equation}
 \big( \hat{z} \times \vec{r} \big) \cdot \big( \vec{q}^{\ \prime} - 
      \vec{q} \big) = \hat{z} \cdot \Big[ \big( \vec{r} - \vec{q}^{\ \prime} 
      \big) \times \big( \vec{r} - \vec{q} \big) - \big( \vec{q}^{\ \prime} 
      \times \vec{q} \big) \Big]  \label{eqxtra8}
\end{equation}
and of Eq. (\ref{eq3.50}), we obtain from Eq. (\ref{eq3.51}) after lattice
Fourier transformation,
\begin{equation}
 \Big\langle u_{\lambda}^{0} \Big| u_{\lambda}^{(1)} \Big\rangle + \Big\langle 
      u_{\lambda}^{(1)} \Big| u_{\lambda}^{0} \Big\rangle = \frac{1}{2} \bigg(  
      \frac{e}{\hbar c} \bigg) \Bigg( \frac{\partial}{\partial \vec{k}_{y}} 
      X_{\lambda} - \frac{\partial}{\partial \vec{k}_{x}} Y_{\lambda} \Bigg) 
      \label{eq3.52}
\end{equation}
where
\begin{align}
 X_{\lambda} & = \bigg\langle u_{\lambda}^{0} \bigg| i \frac{\partial}{\partial 
      \vec{k}_{x}} u_{\lambda}^{0} \bigg\rangle \label{eq3.53}\\
 Y_{\lambda} & = \bigg\langle u_{\lambda}^{0} \bigg| i \frac{\partial}{\partial 
      \vec{k}_{y}} u_{\lambda}^{0} \bigg\rangle \label{eq3.54}
\end{align}
which resemble the \textit{Berry connections} in modern terminology, and thus
the terms in the parenthesis of Eq. (\ref{eq3.52}) resembles \textit{Berry
curvature}. Indeed, the Berry curvature of Bloch states physically represents
part of their perturbative response to uniform electromagnetic fields.
Equation (\ref{eq3.52}) yields the expression for $\beta_{\lambda \lambda}$,
\begin{equation}
 \beta_{\lambda \lambda} = \frac{1}{4} \bigg( \frac{e}{\hbar c} \bigg) \Bigg( 
      \frac{\partial}{\partial \vec{k}_{y}} X_{\lambda} - 
      \frac{\partial}{\partial \vec{k}_{x}} Y_{\lambda} \Bigg)  \label{eq3.55}
\end{equation}
Thus, $H_{\lambda}^{(1)}(k, z)$ and $H_{\lambda}^{(2)}(k, z)$ are completely 
determined in the expansion of $\tilde{H}_{\lambda} \Big( \vec{p} - \frac{e}{c} 
\vec{A} \big(\vec{q} \big); B, z \Big)$ in Eq. (\ref{eq3.1}). The lattice Weyl
transform, which is free of interband terms, is obtained by replacement of
$\hbar \vec{k}$ by $\Big[ \hbar \vec{k} - \frac{e}{c} A \big( \vec{q} \big) 
\Big]$ in $H_{\lambda}^{(1)}(k, z)$ and $H_{\lambda}^{(2)}(k,z)$. Equation 
(\ref{eq3.1}) written up to second order in its explicit dependence in $B$, 
beyond the vector potential, is thus given by
\begin{align}
 \tilde{H}_{\lambda} \bigg( \vec{p} - \frac{e}{c} \vec{A}\big( \vec{q} \big), B, 
      z \bigg) = & \ \tilde{H}_{\lambda}^{0} \bigg( \vec{p} - \frac{e}{c} 
      \vec{A} \big( \vec{q} \big), z \bigg) + BH_{\lambda}^{(1)} \bigg( \vec{p} 
      - \frac{e}{c} \vec{A} \big( \vec{q} \big), z \bigg) \nonumber\\
 & + B^{2} H_{\lambda}^{(2)} \bigg( \vec{p} - \frac{e}{c} \vec{A} \big( 
      \vec{q} \big), z \bigg) + \cdots  \label{eq3.56}
\end{align}

\subsection{Derivation of General Expression for Many-Body $\chi$}

In this section, will derive the most general expression for $\chi$ using the 
temperature Green's function formalism of Luttinger and Ward. 
\cite{luttingerward} The magnetic susceptibility for a system of volume $V$ is,
\begin{equation}
 \chi = \frac{1}{V} \lim_{B \Longrightarrow 0} \frac{\partial^{2}}{\partial 
        B^{2}} \bigg( \frac{1}{\beta} \ln Z \bigg)  \label{eq4.1}
\end{equation}
which at zero temperature can be expressed as,
\begin{equation}
 \chi = \frac{1}{V} \lim_{\substack{B \Longrightarrow0 \\ \beta \Longrightarrow 
        \infty}}\frac{\partial^{2}}{\partial B^{2}} \bigg( 
        \frac{\partial}{\partial \beta} \ln Z \bigg)  \label{eq4.2}
\end{equation}
The expression for $\ln Z$ as a functional of the temperature Green's function, 
$\mathcal{G}_{\zeta_{l}}$, is given by Luttinger and Ward,\cite{luttingerward}
\begin{equation}
 -\ln Z = \Phi(\mathcal{G}_{\zeta_{l}}) -Tr \ \Sigma( \mathcal{G}_{\zeta_{l}})  
          \mathcal{G}_{\zeta_{l}} + Tr \ln (-\mathcal{G}_{\zeta_{l}}) 
          \label{eq4.3}
\end{equation}
where the temperature Green's-function operator $\mathcal{G}_{\zeta_{l}}^{-1}$
is defined formally by
\begin{align}
 \mathcal{G}_{\zeta_{l}}^{-1} & = \zeta_{l} - \mathcal{H}_{0} 
      - \Sigma_{\zeta_{l}},  \label{eq4.4}\\
 \zeta_{l} & = (2l + 1) \frac{\pi i}{\beta} + \mu. \label{eq4.5}
\end{align}
$Tr$ is defined as $\sum_{l} \tilde{T}r$, where $\tilde{T}r$ refers to taking 
the trace in any convenient representation. The functional $\Phi 
\big( \mathcal{G}_{\zeta_{l}} \big)$ is defined as
\begin{equation}
 \Phi(\mathcal{G}_{\zeta_{l}}) = \lim_{\lambda \Longrightarrow 1} Tr 
      \sum_{n} \frac{\lambda^{n}}{2n} \Sigma^{(n)}(\mathcal{G}_{\zeta_{l}}) 
      \mathcal{G}_{\zeta_{l}} \label{eq4.6}
\end{equation}
$\Sigma^{(n)} \big( \mathcal{G}_{\zeta_{l}} \big)$ is the nth-order self-energy 
part where only th\symbol{126}interaction parameter $\lambda$ occurring 
explicitly in (\ref{eq4.6}) is used to determine the order.

It is convenient to work in the coordinate representation as a first step to
simplify the right-hand side of Eq. (\ref{eq4.1}). The total Hamiltonian in
this representation takes the form
\begin{align}
 \mathcal{H} = & \int d^{3}r \psi_{\alpha}^{\dagger} \big( \vec{r} \big) 
      \mathcal{H}_{0} \psi_{\alpha} \big( \vec{r} \big) + \frac{1}{2} \int 
      d^{3}r d^{3}r^{\prime} \psi_{\alpha}^{\dagger} \big( \vec{r} \big) 
      \psi_{\beta}^{\dagger} \big( \vec{r}^{\ \prime} \big) v_{\alpha \beta 
      \gamma \delta} \big( \vec{r}, \vec{r}^{\ \prime} \big) \psi_{\gamma} 
      \big( \vec{r}^{\ \prime} \big) \psi_{\delta} \big( \vec{r} \big)  
      \label{eq4.7}
\end{align}
where repeated spin indices are summed over, and for simplicity, we may take Eq, 
(\ref{eq2.2}) for $\mathcal{H}_{0}$. The term $v_{\alpha \beta \gamma \delta} 
\big( \vec{r}, \vec{r}^{\ \prime} \big)$ is the interaction between a pair of 
particles assumed to be velocity independent; this immediately implies that in 
coordinate representation the field dependence of $\ln Z$ in (\ref{eq4.3}) 
occurs only through the field dependence of $\mathcal{G}_{\zeta_{l}} \big( r, 
r^{\prime} \big)$. To take spin into account explicitly, both 
$\mathcal{G}_{\zeta_{l}} \big( r, r^{\prime} \big)$ and $\Sigma_{\zeta_{l}}$ 
must be considered as $2 \times 2$ matrices in spin indices. The form of 
$\mathcal{G}_{\zeta_{l}} \big( r, r^{\prime} \big)$ and $\Sigma_{\zeta_{l}}$ is 
given by Eq. (\ref{eq3.33}) by gauge invariance. It is convenient for our 
purpose to expand $\Sigma_{\zeta_{l}}$ in powers of its explicit dependence on 
the magnetic field (beyond the Peierls phase factor) and write
\begin{equation}
 \Sigma_{\zeta_{l}} \big( \vec{r}, \vec{r}^{\ \prime} \big) = 
      \Sigma_{\zeta_{l}}^{0} \big( \vec{r}, \vec{r}^{\ \prime} \big) + B 
      \Sigma_{\zeta_{l}}^{(1)} \big( \vec{r}, \vec{r}^{\ \prime} \big) + B^{2} 
      \Sigma_{\zeta_{l}}^{(2)} \big( \vec{r}, \vec{r}^{\ \prime} \big) + \cdots 
      \label{eq4.8}
\end{equation}
where the remaining field dependence of $\Sigma_{\zeta_{l}}^{0} \big( 
\vec{r}, \vec{r}^{\ \prime} \big), \Sigma_{\zeta_{l}}^{(1)} \big( \vec{r}, 
\vec{r}^{\ \prime} \big)$, and $\Sigma_{\zeta_{l}}^{(2)} \big( \vec{r}, 
\vec{r}^{\ \prime} \big)$ occurs only through the Peierls phase factor.

We have, using the definition of $\Phi \big( \mathcal{G}_{\zeta_{l}} \big)$, the 
following relations:
\scriptsize
\begin{align}
 \frac{\partial^{2} \Phi \big( \mathcal{G}_{\zeta_{l}} \big)}{\partial B^{2}} 
      \bigg|_{B \Rightarrow 0} = & \sum_{l} \int d^{3}r 
      d^{3}r^{\prime} \bigg( \tilde{\Sigma}_{\zeta_{l}}^{(1)} \big( \vec{r}, 
      \vec{r}^{\ \prime} \big) \frac{\partial \mathcal{\tilde{G}}_{\zeta_{l}} 
      \big( \vec{r}, \vec{r}^{\ \prime}, B \big)}{\partial B} \bigg|_{B 
      \Rightarrow 0} + \ \tilde{\Sigma}_{\zeta_{l}}^{0} \big( \vec{r}, 
      \vec{r}^{\ \prime} \big) \frac{\partial^{2} 
      \mathcal{\tilde{G}}_{\zeta_{l}} \big( \vec{r}, \vec{r}^{\ \prime}, B 
      \big)}{\partial B^{2}} \bigg|_{B \Rightarrow 0} \bigg) \label{eq4.9}      
      \\
 \frac{\partial^{2} Tr \big( \Sigma_{\zeta_{l}} \mathcal{G}_{\zeta_{l}} 
      \big)}{\partial B^{2}} \bigg|_{B \Rightarrow 0} = & - \sum_{l} \int 
      d^{3}r d^{3}r^{\prime} \Bigg( \tilde{\Sigma}_{\zeta_{l}}^{(1)} \big( 
      \vec{r}, \vec{r}^{\ \prime} \big) \mathcal{\tilde{G}}_{\zeta_{l}} \big( 
      \vec{r}, \vec{r}^{\ \prime}, B \big) \Big|_{B \Rightarrow 0} + 2 \ 
      \tilde{\Sigma}_{\zeta_{l}}^{(1)} \big( \vec{r}, \vec{r}^{\ \prime} \big) 
      \frac{\partial \mathcal{\tilde{G}}_{\zeta_{l}} \big( \vec{r}, \vec{r}^{\ 
      \prime}, B \big)}{\partial B} \Bigg|_{B \Rightarrow 0} \nonumber \\ 
 & + \tilde{\Sigma}_{\zeta_{l}}^{0} \big( \vec{r}, \vec{r}^{\ \prime} \big) 
      \frac{\partial^{2} \mathcal{\tilde{G}}_{\zeta_{l}} \big( \vec{r}, 
      \vec{r}^{\ \prime}, B \big)}{\partial B^{2}} \Bigg|_{B \Rightarrow 0} 
      \Bigg)  \label{eq4.10}
\end{align}
\normalsize
where $\tilde{\Sigma}_{\zeta_{l}}^{(i)} \big( \vec{r}, \vec{r}^{\ \prime} 
\big)$ and $\tilde{\Sigma}_{\zeta_{l}}^{0} \big( \vec{r}, \vec{r}^{\ \prime} 
\big)$ are field-independent quantities. The above relations lead to a more 
convenient expression for $\chi$,
\begin{equation}
 \chi = -\frac{1}{V} \bigg( \frac{\partial^{2}}{\partial B^{2}} 
      \frac{1}{\beta} Tr \ln \big( -\mathcal{G}_{\zeta_{l}} \big) 
      \bigg)_{B \Rightarrow 0} + \frac{1}{V} \bigg( \frac{1}{\beta} Tr \ 2 
      \tilde{\Sigma}_{\zeta_{l}}^{(2)} \mathcal{\tilde{G}}_{\zeta_{l}} 
      + \frac{1}{\beta} Tr \tilde{\Sigma}_{\zeta_{l}}^{(1)} 
      \frac{\partial \mathcal{\tilde{G}}_{\zeta_{l}}}{\partial B} \bigg| 
      \bigg)_{B \Rightarrow 0} \label{eq4.11}
\end{equation}
The first term in Eq. (\ref{eq4.11}) has exactly the same form as that of the
noninteracting Fermi systems, except for the replacement of the 
``noninteracting'' $\mathcal{G}_{\zeta_{l}}$ by the exact 
$\mathcal{G}_{\zeta_{l}}$ for the interacting, free, or Bloch, electrons. The 
second term can be immediately recognized as correction to the ``crystalline 
induced diamagnetism'' as calculated by the first term. The last term turns out 
to contain corrections to both the ``crystalline paramagnetism'' and 
``crystalline induced diamagnetism'' as calculated from the first term in Eq. 
(\ref{eq4.11}).

We are, in the present case, of course, interested in the effective 
one-particle Schrodinger Hamiltonian
\begin{equation}
 \mathcal{H}_{\zeta_{l}} = \mathcal{H}_{0} + \Sigma_{\zeta_{l}} \label{eq4.12}
\end{equation}
which is formally the same as that of Eq. (\ref{eq2.1}), with the replacement
$z \Rightarrow \zeta_{l}$ (we have chosen to indicate the discrete frequency
dependence of operators by a subscript). Therefore, all the results of Sec. 
\ref{chap3} can be formally carried over to apply to the effective Hamiltonian 
given above. The beauty and power in the use of the Weyl transform is that the 
Weyl transform of an operator is a physically meaningful quantity and faithfully 
corresponds to the original quantum-mechanical operator. Moreover, it provides a 
natural way of calculating the trace of any function of 
$\mathcal{H}_{\zeta_{l}}$, as a power-series expansion in $\hbar$, Planck's 
constant, which is equivalent to an expansion in the magnetic field strength for 
Fermi systems possessing translational symmetry.
\begin{align}
 \tilde{T}r F \big( \mathcal{H}_{\zeta_{l}} \big) = & \Bigg( \frac{1}{2 \pi} 
      \Bigg)^{3} \sum_{\lambda} \int d^{3}k \ d^{3}q \Bigg\{ F \Big( 
      \tilde{H}_{\lambda} \big( k, B, \zeta_{l} \big) \Big) \nonumber \\
 & - \frac{1}{24} \Bigg( \frac{eB}{\hbar c} \Bigg)^{2} F^{\prime \prime} \big( 
      \tilde{H}_{\lambda}^{0}(k, \zeta_{l}) \big) \Bigg[ \frac{\partial^{2} 
      \tilde{H}_{\lambda}^{0}}{\partial \vec{k}_{x}^{2}} \frac{\partial^{2} 
      \tilde{H}_{\lambda}^{0}}{\partial \vec{k}_{y}^{2}} - \Bigg( 
      \frac{\partial^{2} \tilde{H}_{\lambda}^{0}}{\partial \vec{k}_{x} \partial 
      \vec{k}_{y}} \Bigg)^{2} \Bigg] \Bigg\} + O(B^{4})  \label{eq4.13}
\end{align}
where, $\vec{p} - (e/c) A \big( \vec{q} \big)$ in (\ref{eq3.56}) is replaced
by $\hbar \vec{k}$ in Eq. (\ref{eq4.13}). In the above expression, it is assumed 
that $\tilde{H}_{\lambda}^{0}(k, \zeta_{l})$ is diagonal in spin indices; this 
is generally true for nonferromagnetic systems. The first term in Eq. 
(\ref{eq4.13}) can then be expanded up to second order in $B$ using the 
expansion given in Eq. (\ref{eq3.56}). Applying this result to the first term of 
Eq. (\ref{eq4.11}), we obtain
\begin{equation}
 - \frac{1}{V} \bigg( \frac{\partial^{2}}{\partial B^{2}} \frac{1}{\beta} Tr \ln 
      \big( - \mathcal{G}_{\zeta_{l}} \big) \bigg)_{B \Longrightarrow 0} = 
      \chi_{_{LP}} + \chi_{_{CP}} + \chi_{_{ID}} \label{eq4.14}
\end{equation}
where
\small
\begin{equation}
 \chi_{_{LP}} = \frac{1}{12} \bigg( \frac{e}{\hbar c} \bigg)^{2} 
      \sum_{\lambda} \bigg( \frac{1}{2 \pi} \bigg)^{3} \int d^{3}k \ k_{B} T 
      \sum_{l} \Bigg[ \frac{\partial^{2} \tilde{H}_{\lambda}^{0}}{\partial 
      \vec{k}_{x}^{2}} \frac{\partial^{2} \tilde{H}_{\lambda}^{0}}{\partial 
      \vec{k}_{y}^{ 2}} - \Bigg( \frac{\partial^{2} 
      \tilde{H}_{\lambda}^{0}}{\partial \vec{k}_{x} \partial \vec {k}_{y}} 
      \Bigg)^{2} \Bigg] \big( \mathcal{G}_{\zeta_{l}} \big( k, \zeta_{l} \big) 
      \big)^{2} \label{eq4.15}  
\end{equation}
\begin{equation}
 \chi_{_{CP}} = - \sum_{\lambda} \bigg( \frac{1}{2 \pi} \bigg)^{3} \int 
      d^{3}k \ k_{B} T \sum_{l} \Big[ \tilde{H}_{\lambda}^{(1)} \big( k, 
      \zeta_{l} \big) \Big]^{2} \Big( \mathcal{G}_{\zeta_{l}}(k, \zeta_{l}) 
      \Big)^{2} \label{eq4.16} 
\end{equation}
\begin{equation} 
 \chi_{_{ID}} = - \sum_{\lambda} \bigg( \frac{1}{2 \pi} \bigg)^{3} \int 
      d^{3}k \ k_{B} T \sum_{l} 2 \tilde{H}_{\lambda}^{(2)} \big( k, \zeta_{l} 
      \big) \mathcal{G}_{\zeta_{l}} \big( k, \zeta_{l} \big)  \label{eq4.17} 
\end{equation}
\begin{equation}  
 \mathcal{G}_{\zeta_{l}} \big( k, \zeta_{l} \big) = \Big[ \zeta_{l} - 
      \tilde{H}_{\lambda}^{0} \big( k, \zeta_{l} \big) \Big]^{-1} \label{eq4.18}
\end{equation}
\normalsize
In Eqs. (\ref{eq4.15}), (\ref{eq4.16}), and (\ref{eq4.17}), taking the trace 
over spin indices is implied. $\chi_{_{LP}}$ is a generalized Landau- Peierls 
formula for the orbital diamagnetism of free and Bloch electrons. It is for the 
case of an interacting free-electron gas that this term was derived by Fukuyama 
and McClure.\cite{fukuyamamcclure} In the limit of vanishing self-energy parts, 
Eq. (\ref{eq4.14}) exactly reproduces the expression for $\chi$ of Bloch 
electrons, both with or without spin-orbit coupling, as given by Roth, 
\cite{roth} and by Wannier and Upadhyaya,\cite{wannierupadya} respectively. 
Moreover, when the self-energy part is assumed to be independent of $\zeta_{l}$, 
which holds true in Hartree-Fock approximation, the form of Eq. (\ref{eq4.14}), 
after summation over $\zeta_{l}$, is exactly the same as that of the 
noninteracting case.

$\chi_{_{CP}}$, which includes the effect of free-electron spin and $g$-factor, 
will be referred to as the crystalline paramagnetism, and $\chi_{_{ID}}$ , the 
induced diamagnetism, although its sign cannot be determined \textit{a 
priori} even in the Hartree-Fock approximation and in the noninteracting case.

We consider the correction terms represented by the last two terms of Eq. 
(\ref{eq4.11}) explicitly from the self-energies, 
$\tilde{\Sigma}_{\zeta_{l}}^{(i)} \big( \vec{r}, \vec{r}^{\ \prime} \big)$. As 
we have mentioned earlier, these corrections only modify $\chi_{_{CP}}$ and 
$\chi_{_{ID}}$ but do not affect $\chi_{_{LP}}$. Let us recast the last two 
terms of Eq. (\ref{eq4.11}), which we now denote by $\chi_{cor}$ and write them 
as follows:
\begin{equation}
 \chi_{cor} = \frac{1}{V} \lim_{B \Longrightarrow 0} \frac{\partial}{\partial B} 
      Tr \frac{1}{\beta} \bigg( \tilde{\Sigma}_{\zeta_{l}}^{(1)} + 2B 
      \tilde{\Sigma}_{\zeta_{l}}^{(2)} \bigg) \mathcal{G}_{\zeta_{l}} 
      \label{eq4.19}
\end{equation}
Recall that in the coordinate representation, the Peierls phase factors
occurring in $\tilde{\Sigma}_{\zeta_{l}}^{(i)}$ and $\mathcal{G}_{\zeta_{l}}$ 
cancel. However, it is convenient to retain these phase factors in Eq. 
(\ref{eq4.19}) as the trace will now be taken using the biorthogonal magnetic 
function representation discussed in Sec. \ref{chap3}. The trace would then be 
expressed in terms of the Weyl transform, where indeed the Weyl transform of 
$\mathcal{G}_{\zeta_{l}}$ is diagonal in band indices, resulting in much 
simplification. We have
\footnotesize
\begin{equation}
 \chi_{cor} = \frac{1}{V} \lim_{B \Longrightarrow 0} \frac{\partial}{\partial B} 
      \sum_{\lambda} \bigg( \frac{1}{2 \pi} \bigg)^{3} \int d^{3}k \ k_{B} T 
      \sum_{l} \bigg[ \bigg( \tilde{\Sigma}_{\lambda \lambda}^{(1)} \Big( 
      \vec{k}, B, \zeta_{l} \Big) + 2B \tilde{\Sigma}_{\lambda \lambda}^{(2)} 
      \Big( \vec{k}, B, \zeta_{l} \Big) \bigg) \bigg] \mathcal{G}_{\lambda} 
      \Big( \vec{k}, B, \zeta_{l} \Big) \label{eq4.20}
\end{equation}
\normalsize
where in the last equation a familiar change of variable has been made,
$p - (e/c) A(q) \Longrightarrow \hbar k$, and from Eq. (\ref{eq3.9}) we have
\begin{equation}
 \mathcal{G}_{\lambda} \big( \vec{k}, B, \zeta_{l} \big) = \Big[ \zeta_{l} - 
      \tilde{H}_{\lambda} \big( \vec{k}, B, \zeta_{l} \big)  \Big]^{-1} + 
      O(B^{2})  \label{eq4.21}
\end{equation}
$\tilde{H}_{\lambda} \big( \vec{k}, B, \zeta_{l} \big)$ is of the form given by 
Eq. (\ref{eq3.56}) with the replacement $z \Longrightarrow \zeta_{l}$. Since we 
need $\tilde{\Sigma}_{\lambda \lambda}^{(2)} \big( \vec{k}, B, \zeta_{l} \big)$ 
only to zero order in the field, the calculation of the second term in Eq. 
(\ref{eq4.20}) is trivial. Denoting this contribution as $\chi_{cor}^{(2)}$ we 
have
\footnotesize
\begin{equation}
 \chi_{cor}^{(2)} = \sum_{\lambda} \Big( \frac{1}{2 \pi} \Big)^{3} \int d^{3}k 
      \ k_{B} T \sum_{l} 2 \Big\langle u_{\lambda}^{0} \big( \vec{r}, \vec{k}, 
      \zeta_{l} \big) \Big| \int d^{3} r^{\prime} e^{i \vec{k} \cdot \big( 
      \vec{r}^{\prime} - \vec{r} \big)} \tilde{\Sigma}_{\zeta_{l}}^{(2)} \big( 
      \vec{r}, \vec{r}^{\prime} \big) u_{\lambda}^{0} \big( \vec{r}^{\ \prime}, 
      \vec{k}, \zeta_{l} \big) \Big\rangle \mathcal{G}_{\lambda} \big( \vec{k}, 
      \zeta_{l} \big) \label{eq4.22}
\end{equation}
\normalsize
$\chi_{cor}^{(2)}$ is indeed a correction to $\chi_{ID}$ as can be seen from 
Eqs. (\ref{eq4.17}) and (\ref{eq3.43}).

To find $\tilde{\Sigma}_{\lambda \lambda}^{(1)} \big( \vec{k}, B, \zeta_{l} 
\big)$, we write down the effect of operating $\tilde{\Sigma}_{\zeta_{l}}^{(1)}$ 
on the magnetic Bloch function\cite{freeBloch} (same relation holds for 
$\tilde{\Sigma}_{\zeta_{l}}^{(2)}$)
\begin{equation}
 \Sigma_{\zeta_{l}}^{(1)} \ \Big| \vec{p}, \lambda, \zeta_{l}, B \Big\rangle 
      = \sum_{\vec{q}, \lambda^{\prime}} e^{\frac{i \vec{p} \cdot 
      \vec{q}}{\hbar}} \bigg[ \Sigma_{\zeta_{l}}^{(1)} \big( \vec{q}, B \big) 
      \bigg]_{\lambda \lambda^{\prime}} \ \Big| \vec{p} + \frac{e}{c} \vec{A} 
      \big( \vec{q} \big), \lambda^{\prime}, \zeta_{l}, B \Big\rangle 
      \label{eq4.23}
\end{equation}
The Weyl transform of $\Sigma_{\zeta_{l}}^{(1)}$ is
\begin{equation}
 \Sigma_{\lambda \lambda^{\prime}}^{(1)} \Big( \vec{p}, \vec{q}, \zeta_{l}, B 
      \Big) = \sum_{\vec{u}} e^{\frac{2i \vec{q} \cdot \vec{u}}{\hbar}} 
      \Big\langle \vec{p} + \vec{u}, \lambda, \zeta_{l}, B \Big| 
      \Sigma_{\zeta_{l}}^{(1)} \Big| \vec{p} - \vec{u}, \lambda^{\prime}, 
      \zeta_{l}, B \Big\rangle \label{eq4.24}
\end{equation}
and by virtue of Eq. (\ref{eq4.23}) we obtain
\begin{align}
 \Sigma_{\lambda \lambda^{\prime}}^{(1)} \big( \vec{p}, \vec{q}, \zeta_{l}, B 
      \big) & = \sum_{\vec{u}} e^{\frac{2i \vec{q} \cdot \vec{u}}{\hbar}} 
      \sum_{\vec{v}} \delta_{\vec{u}, \frac{e}{2c} \vec{A} \big( \vec{v} 
      \big)} e^{\frac{i \vec{p} \cdot \vec{v}}{\hbar}} \bigg[ 
      \Sigma_{\lambda \lambda^{\prime}}^{(1)} \Big( \vec{v}, \zeta_{l}, B \Big) 
      \bigg] \nonumber\\
 & = \sum_{\vec{v}} \exp \bigg[ \frac{1}{\hbar} \Big( \vec{p} - \frac{e}{c} 
      \vec{A} \big( \vec{q} \big) \Big) \cdot \vec{v} \bigg] \Sigma_{\lambda 
      \lambda^{\prime}}^{(1)} \big( \vec{v}, \zeta_{l}, B \big) \label{eq4.25}
\end{align}
which in turn yields
\begin{equation}
 \Sigma_{\lambda \lambda^{\prime}}^{(1)} \big( \vec{p}, \vec{q}, \zeta_{l}, B 
      \big) \bigg|_{\vec{p} - \frac{e}{c} \vec{A} \big( \vec{q} \big) 
      \Longrightarrow \hbar \vec{k}} = \sum_{\vec{v}} \exp \Bigg[ 
      \frac{1}{\hbar} \vec{k} \cdot \vec{v} \Bigg] \Sigma_{\lambda 
      \lambda^{\prime}}^{(1)} \big( \vec{v}, \zeta_{l}, B \big) \label{eq4.26}
\end{equation}
We are therefore interested in the right-hand side of Eq. (\ref{eq4.26}), to
obtain this we may proceed in a manner quite similar to that used in Sec. 
\ref{chap3}, i.e., Eqs. (\ref{eq3.30})-(\ref{eq3.43}). However, at this stage,
Eq. (\ref{eq4.23}) provides a very good starting point. The relation between
$\big| p, \lambda, \zeta_{l}, B \big\rangle$ and the modified Bloch function 
$b_{\lambda} \big( \vec{r}, \vec{k}, B, \zeta_{l} \big)$, used in the 
perturbation theory of Sec. \ref{chap3}, can be easily deduced from Eqs. 
(\ref{eq3.29}) and (\ref{eq3.35})
\begin{equation}
 \big| p, \lambda, \zeta_{l}, B \big\rangle = b_{\lambda} \Big( \vec{r}, \vec{k} 
      - \frac{e}{\hbar c} \vec{A} \big( \vec{r} \big), \zeta_{l}, B \big) 
      \label{eq4.27}
\end{equation}
Let us make the substitution $\vec{p} \Longrightarrow \vec{p}^{\ \prime} + 
(e/c) \vec{A}(\vec{r})$ in Eq. (\ref{eq4.23}) and obtain the relation
\begin{align}
 \int d^{3} r^{\prime} & \exp \bigg[ \bigg( \frac{-ie}{\hbar c} \bigg) 
      \vec{A} \big( \vec{r} \big) \cdot \vec{r}^{\ \prime} \bigg] 
      \tilde{\Sigma}_{\zeta_{l}}^{(1)} \big( \vec{r}, \vec{r}^{\ \prime} \big) 
      b_{\lambda} \Big( \vec{r}^{\ \prime}, \vec{p}^{\ \prime} + \frac{e}{c} 
      \vec{A} \big( \vec{r} - \vec{r}^{\ \prime} \big), \zeta_{l},B \Big) 
      \nonumber\\
 & = \sum_{\vec{q}, \lambda^{\prime}} \exp \bigg[ \bigg( \frac{i}{\hbar c} 
      \bigg) \Big( \vec{p}^{\prime} + \frac{e}{c} \vec{A} \big( \vec{r} \big)  
      \Big) \cdot \vec{q} \bigg] \tilde{\Sigma}_{\lambda \lambda^{\prime}}^{(1)} 
      \Big( \vec{q}, \zeta_{l}, B \Big) b_{\lambda^{\prime}} \bigg( \vec{r}, 
      \vec{p}^{\ \prime} + \frac{e}{c} \vec{A} \big( \vec{q} \big), \zeta_{l}, 
      B \bigg) \label{eq4.28}
\end{align}
The equation in terms of the modified periodic function $u_{\lambda^{\prime}} 
\big( \vec{r}, \vec{k}, \zeta_{l}, B \big)$ is therefore given by
\begin{align}
 & \int d^{3} r^{\prime} \exp \bigg[ i \vec{k} \cdot \Big( \vec{r} - \vec{r}^{\ 
      \prime} \Big) \bigg] \tilde{\Sigma}_{\zeta_{l}}^{(1)} \big( \vec{r}, 
      \vec{r}^{\ \prime} \big) u_{\lambda} \bigg( \vec{r}^{\ \prime}, \vec{k} + 
      \frac{e}{\hbar c} \vec{A} \big( \vec{r} - \vec{r}^{\ \prime} \big), 
      \zeta_{l}, B \bigg) \nonumber\\
 & = \sum_{\vec{q},\lambda^{\prime}} \exp \bigg[ i \vec{k} \cdot \vec{q} \bigg] 
      \tilde{\Sigma}_{\lambda \lambda^{\prime}}^{(1)} \big( \vec{q}, \zeta_{l}, 
      B \big) u_{\lambda^{\prime}} \bigg( \vec{r}, \vec{k} + \frac{e}{\hbar c} 
      \vec{A} \big( \vec{q} \big), \zeta_{l}, B \bigg)  \label{eq4.29}
\end{align}
Equation (\ref{eq4.29}) corresponds to Eq. (\ref{eq3.36}) of Sec. \ref{chap3}.
Perturbative treatment can then be carried out, using the expansion of 
$\tilde{\Sigma}_{\lambda \lambda^{\prime}}^{(1)} \big( \vec{q}, \zeta_{l}, B 
\big)$ and $u_{\lambda^{\prime}} \big( \vec{r}, \vec{k}, \zeta_{l}, B \big)$ in 
powers of $B$, and solution is obtained up to first order in $B$ for 
$\tilde{\Sigma}_{\lambda \lambda}^{(1)} \big( \vec{k}, \zeta_{l}, B \big)$. 
Writing
\begin{equation}
 \tilde{\Sigma}_{\lambda \lambda}^{(1)} \big( \vec{k}, \zeta_{l}, B \big) = 
      \tilde{\Sigma}_{\lambda \lambda}^{(1)0} \big( \vec{k}, \zeta_{l} \big) + B 
      \tilde{\Sigma}_{\lambda \lambda}^{(1)(1)} \big( \vec{k}, \zeta_{l} 
      \big) + \cdots  \label{eq4.30}
\end{equation}
we get, by equating the zero- and first-order coefficients of $B$, the following 
relations:
\small
\begin{equation}
 \int d^{3} r^{\prime} \exp \bigg[ i \vec{k} \cdot \big( \vec{r} - \vec{r}^{\ 
      \prime} \big) \bigg] \tilde{\Sigma}_{\zeta_{l}}^{(1)} \big( \vec{r}, 
      \vec{r}^{\ \prime} \big) u_{\lambda}^{0} \Big( \vec{r}^{\ \prime}, 
      \vec{k}, \zeta_{l} \Big) = \sum_{\vec{q}, \lambda^{\prime}} \exp \Big[ i 
      \vec{k} \cdot \vec{q} \Big] \tilde{\Sigma}_{\lambda 
      \lambda^{\prime}}^{(1)0} \big( \vec{q} \big) u_{\lambda^{\prime}}^{0} 
      \big( \vec{r}, \vec{k}, \zeta_{l} \big) \label{eq4.31} 
\end{equation}
\normalsize
\begin{align}
 \int & d^{3} r^{\prime} \exp Big[ i \vec{k} \cdot \big( \vec{r}^{\ \prime} 
      - \vec{r} \big) \Big] \tilde{\Sigma}_{\zeta_{l}}^{(1)} \big( \vec{r}, 
      \vec{r}^{\ \prime} \big) u_{\lambda}^{(1)} \big( \vec{r}^{\ \prime}, 
      \vec{k}, \zeta_{l} \big) \nonumber\\
 & \qquad \qquad + \int d^{3} r^{\prime} \exp \Big[ i \vec{k} \cdot \big( 
      \vec{r}^{\ \prime} - \vec{r} \big) \Big] \tilde{\Sigma}_{\zeta_{l}}^{(1)} 
      \big( \vec{r}, \vec{r}^{\ \prime} \big) \frac{e}{B \hbar c} A(\vec{r} - 
      \vec{r}^{\ \prime}) \cdot \nabla_{\vec{k}} \ u_{\lambda}^{0} \big( 
      \vec{r}^{\ \prime}, \vec{k}, \zeta_{l} \big) \nonumber\\
 & = \sum_{\vec{q}, \lambda^{\prime}} \exp \Big[  i \vec{k} \cdot 
      \vec{q} \Big] \bigg( \tilde{\Sigma}_{\lambda \lambda^{\prime}}^{(1)0} 
      \big( \vec{q}, \zeta_{l} \big) \frac{e}{B \hbar c} A(\vec{q}) \cdot 
      \nabla_{\vec{k}} \ u_{\lambda^{\prime}}^{0} \big( \vec{r}, \vec{k}, 
      \zeta_{l} \big) \nonumber \\
 & \qquad \qquad + \tilde{\Sigma}_{\lambda \lambda^{\prime}}^{(1)0} \big( 
      \vec{q}, \zeta_{l} \big) u_{\lambda^{\prime}}^{(1)} \big( \vec{r}^{\ 
      \prime}, \vec{k}, \zeta_{l} \big) + \tilde{\Sigma}_{\lambda 
      \lambda^{\prime}}^{(1)(1)} \big( \vec{q}, \zeta_{l} \big) 
      u_{\lambda^{\prime}}^{0} \big( \vec{r}, \vec{k}, \zeta_{l} \big) \bigg)  
      \label{eq4.32}
\end{align}
These relations yield for $\tilde{\Sigma}_{\lambda \lambda}^{(1)0} \big(
\vec{q}, \zeta_{l} \big)$ and $\tilde{\Sigma}_{\lambda \lambda}^{(1)(1)} \big( 
\vec{q}, \zeta_{l} \big)$ the following expressions:
\begin{align}
 \tilde{\Sigma}_{\lambda \lambda}^{(1)0} \big( \vec{q}, \zeta_{l} \big) = & \
      \Big\langle u_{\lambda}^{0} \big( \vec{r}, \vec{k}, \zeta_{l} \big) 
      \Big| \int d^{3} r^{\prime} e^{i \vec{k} \cdot \big( \vec{r}^{\ \prime} - 
      \vec{r} \big)} \tilde{\Sigma}_{\zeta_{l}}^{(1)} \big( \vec{r}, \vec{r}^{\ 
      \prime} \big) \ u_{\lambda}^{0} \big( \vec{r}^{\ \prime}, \vec{k}, 
      \zeta_{l} \big) \Big\rangle \label{eq4.33}\\
 \tilde{\Sigma}_{\lambda \lambda}^{(1)(1)} \big( \vec{q}, \zeta_{l} \big) = & \ 
      \Big\langle u_{\lambda}^{0} \big( \vec{r}, \vec{k}, \zeta_{l} \big) 
      \Big| \int d^{3} r^{\prime} e^{i \vec{k} \cdot \big( \vec{r}^{\ \prime} - 
      \vec{r} \big)} \tilde{\Sigma}_{\zeta_{l}}^{(1)} \big( \vec{r}, \vec{r}^{\ 
      \prime} \big) \ u_{\lambda}^{(1)} \big( \vec{r}^{\ \prime}, \vec{k}, 
      \zeta_{l} \big) \Big\rangle \nonumber\\
 & + \Big\langle u_{\lambda}^{0} \big( \vec{r}, \vec{k}, \zeta_{l} \big) \Big|
      \int d^{3} r^{\prime} e^{i \vec{k} \cdot \big( \vec{r}^{\ \prime} - 
      \vec{r} \big)} \tilde{\Sigma}_{\zeta_{l}}^{(1)} \big( \vec{r}, \vec{r}^{\ 
      \prime} \big) \frac{e}{B \hbar c} A \big( \vec{r} - \vec{r}^{\ \prime} 
      \big) \cdot \nabla_{\vec{k}} \ u_{\lambda}^{0} \big( \vec{r}^{\ \prime}, 
      \vec{k}, \zeta_{l} \big) \Big\rangle \nonumber\\
 & + \Big\langle u_{\lambda}^{0} \big( \vec{r}, \vec{k}, \zeta_{l} \big) \Big|
      \sum_{\lambda^{\prime}} \frac{e}{B \hbar c} A \big( \nabla_{\vec{k}} \big)
      \sum_{\vec{q}} e^{i \vec{k} \cdot \vec{q}} \ \tilde{\Sigma}_{\lambda 
      \lambda^{\prime}}^{(1)0} \big( q, \zeta_{l} \big) \cdot \ i 
      \nabla_{\vec{k}} u_{\lambda^{\prime}}^{0} \big( \vec{r}^{\ \prime}, 
      \vec{k}, \zeta_{l} \big) \Big\rangle \nonumber\\
 & - \sum_{\lambda^{\prime}} \ \tilde{\Sigma}_{\lambda \lambda^{\prime}}^{(1)0} 
      \big( q, \zeta_{l} \big) \Big\langle u_{\lambda}^{0} \big( \vec{r}^{\ 
      \prime}, \vec{k}, \zeta_{l} \big) \Big| u_{\lambda^{\prime}}^{(1)} \big( 
      \vec{r}^{\ \prime}, \vec{k}, \zeta_{l} \big) \Big\rangle \label{eq4.34}
\end{align}
The first and last terms of Eq. (\ref{eq4.34}) can be combined through the use
of Eqs. (\ref{eq3.44}), (\ref{eq3.45}), and (\ref{eq4.31}) to yield
\footnotesize
\begin{align}
 \Big\langle u_{\lambda}^{0} & \big( \vec{r}, \vec{k}, \zeta_{l} \big) \Big| 
      \int d^{3} r^{\prime} e^{i \vec{k} \cdot \big( \vec{r}^{\ \prime} - 
      \vec{r} \big)} \tilde{\Sigma}_{\zeta_{l}}^{(1)} \big( \vec{r}, \vec{r}^{\ 
      \prime} \big) u_{\lambda}^{(1)} \big( \vec{r}^{\ \prime}, \vec{k}, 
      \zeta_{l} \big) \Big\rangle - \sum_{\lambda^{\prime}} \Sigma_{\lambda 
      \lambda^{\prime}}^{(1)0} \big( \vec{k}, \zeta_{l} \big) \Big\langle 
      u_{\lambda}^{0} \big( \vec{r}^{\prime}, \vec{k}, \zeta_{l} \big) \Big| 
      u_{\lambda^{\prime}}^{(1)} \big( \vec{r}^{\ \prime}, \vec{k}, \zeta_{l}  
      \big) \Big\rangle \nonumber\\
 = & - \sum_{\lambda^{\prime} \neq \lambda} 2 \big( 
      \tilde{H}_{\lambda^{\prime}}^{0} - \tilde{H}_{\lambda}^{0} \big)^{-1} 
      \big\langle u_{\lambda}^{0} \big| \mathcal{H}_{\delta}^{(1)} \big| 
      u_{\lambda^{\prime}}^{0} \big\rangle \big\langle u_{\lambda^{\prime}}^{0} 
      \big| \mathcal{H}_{\delta}^{(1)} \big| u_{\lambda}^{0} \big\rangle 
      \nonumber\\
 & - \sum_{\lambda^{\prime} \neq \lambda} \big( 
      \tilde{H}_{\lambda^{\prime}}^{0} - \tilde{H}_{\lambda}^{0} \big)^{-1} 
      \Big( \big\langle u_{\lambda^{\prime}}^{0} \big| 
      \mathcal{H}_{\Delta}^{(1)} \big| u_{\lambda}^{0} \big\rangle \big\langle 
      u_{\lambda}^{0} \big| \mathcal{H}_{\delta}^{(1)} \big| 
      u_{\lambda^{\prime}}^{0} \big\rangle + \big\langle u_{\lambda}^{0} \big| 
      \mathcal{H}_{\Delta}^{(1)} \big| u_{\lambda^{\prime}}^{0} \big\rangle 
      \big\langle u_{\lambda^{\prime}}^{0} \big| \mathcal{H}_{\delta}^{(1)} 
      \big| u_{\lambda}^{0} \big\rangle \Big)  \label{eq4.35}
\end{align}
\normalsize
where the operators $\mathcal{H}_{\Delta}^{(1)}$ and 
$\mathcal{H}_{\delta}^{(1)}$ are defined such that $\big\langle 
u_{\lambda^{\prime}}^{0} \big| \mathcal{H}_{\Delta}^{(1)} \big| u_{\lambda}^{0} 
\big\rangle$ is given by the first two terms, and $\big\langle u_{\lambda}^{0} 
\big| \mathcal{H}_{\delta}^{(1)} \big| u_{\lambda}^{0} \big\rangle$ by the last 
term, of Eq. (\ref{eq3.42}) ($z \Longrightarrow \zeta_{l}$ and $\zeta_{l}$ occur 
as subscripts in $\tilde{\Sigma}_{\zeta_{l}}^{(1)}$). With the aid of Eq. 
(\ref{eq4.31}) and noting that the vector potential function used is in 
symmetric gauge, the second term of Eq. (\ref{eq4.34}) can be shown to be equal 
to the third term. Putting all these results together, Eqs. (\ref{eq4.22}), 
(\ref{eq4.30}), and (\ref{eq4.33})-(\ref{eq4.35}) in Eq. (\ref{eq4.20}), we 
obtain the total susceptibility correction $\chi_{cor}$ as
\footnotesize
\begin{align}
 \chi_{cor} & \nonumber\\
 = & \sum_{\lambda} \bigg( \frac{1}{2 \pi} \bigg)^{3} \int d^{3}k \ k_{B} T 
      \sum_{l} 2 \Big\langle u_{\lambda}^{0} \Big( \vec{r}, \vec{k}, \zeta_{l} 
      \Big) \Big| \int d^{3} r^{\prime} e^{i \vec{k} \cdot \big( \vec{r}^{\ 
      \prime} - \vec{r} \big)} \tilde{\Sigma}_{\zeta_{l}}^{(2)} \big( \vec{r}, 
      \vec{r}^{\ \prime} \big) \ u_{\lambda}^{0} \big( \vec{r}^{\ \prime}, 
      \vec{k}, \zeta_{l} \big) \Big\rangle \mathcal{G}_{\lambda} \big( \vec{k}, 
      \zeta_{l} \big) \nonumber\\
 & + \sum_{\lambda} \bigg( \frac{1}{2 \pi} \bigg)^{3} \int d^{3}k \ k_{B} T 
      \sum_{l} \Big\langle u_{\lambda}^{0} \Big| \mathcal{H}_{\delta}^{(1)} 
      \Big| u_{\lambda}^{0} \Big\rangle \tilde{H}_{\lambda}^{(1)} \big( 
      \vec{k}, \zeta_{l} \big) \Big[ \mathcal{G}_{\lambda} \big( \vec{k}, 
      \zeta_{l} \big) \Big]^{2} \nonumber\\
 & + \sum_{\lambda} \bigg( \frac{1}{2 \pi} \bigg)^{3} \int d^{3}k \ k_{B} T 
      \sum_{l} 2 \Big\langle u_{\lambda}^{0} \big( \vec{r}, \vec{k}, \zeta_{l} 
      \big) \Big| \int d^{3} r^{\prime} e^{i \vec{k} \cdot \big( \vec{r}^{\ 
      \prime} - \vec{r} \big)} \tilde{\Sigma}_{\zeta_{l}}^{(1)} \big( \vec{r}, 
      \vec{r}^{\ \prime} \big) \frac{e}{B \hbar c} A \big( \vec{r} - \vec{r}^{\ 
      \prime} \big) \cdot \nabla_{\vec{k}} \ u_{\lambda}^{0} \big( \vec{r}^{\ 
      \prime}, \vec{k}, \zeta_{l} \big) \Big\rangle \nonumber\\
 & \qquad \qquad \times \mathcal{G}_{\lambda} \big( \vec{k}, \zeta_{l} \big) 
      \nonumber\\
 & - \sum_{\lambda} \bigg( \frac{1}{2 \pi} \bigg)^{3} \int d^{3}k \ k_{B} T 
      \sum_{l} \sum_{\lambda^{\prime} \neq \lambda} 2 \big( 
      \tilde{H}_{\lambda^{\prime}}^{0} - \tilde{H}_{\lambda}^{0} \big)^{-1} 
      \Big\langle u_{\lambda^{\prime}}^{0} \Big| \mathcal{H}_{\delta}^{(1)}     
      \Big| u_{\lambda}^{0} \Big\rangle \Big\langle u_{\lambda}^{0} \Big| 
      \mathcal{H}_{\delta}^{(1)} \Big| u_{\lambda^{\prime}}^{0} \Big\rangle     
      \mathcal{G}_{\lambda} \big( \vec{k}, \zeta_{l} \big) \nonumber\\
 & - \sum_{\lambda} \bigg( \frac{1}{2 \pi} \bigg)^{3} \int d^{3}k \ k_{B} T 
      \sum_{l} \sum_{\lambda^{\prime} \neq \lambda} \big( 
      \tilde{H}_{\lambda^{\prime}}^{0} - \tilde{H}_{\lambda}^{0} \big)^{-1} 
      \Big( \Big\langle u_{\lambda^{\prime}}^{0} \Big| 
      \mathcal{H}_{\Delta}^{(1)} \Big| u_{\lambda}^{0} \Big\rangle \Big\langle 
      u_{\lambda}^{0} \Big| \mathcal{H}_{\delta}^{(1)} \Big| 
      u_{\lambda^{\prime}}^{0} \Big\rangle \nonumber \\
 & + \Big\langle u_{\lambda}^{0} \Big| \mathcal{H}_{\Delta}^{(1)} \Big| 
      u_{\lambda^{\prime}}^{0} \Big\rangle \Big\langle u_{\lambda^{\prime}}^{0} 
      \Big| \mathcal{H}_{\delta}^{(1)} \Big| u_{\lambda}^{0} \Big\rangle \bigg) 
      \mathcal{G}_{\lambda} \big( \vec{k},\zeta_{l} \big)  \label{eq4.36}
\end{align}
\normalsize
The second term gives a correction to $\chi_{CP}$ and the rest are corrections 
to $\chi_{ID}$. We shall see that these corrections to $\chi_{CP}$ and 
$\chi_{ID}$ lead, among other things, to the cancellation of the appearance of 
quadratic terms in $\tilde{\Sigma}_{\zeta_{l}}^{(1)}$ as well as the total 
cancellation of the appearance of $\tilde{\Sigma}_{\zeta_{l}}^{(2)}$. This 
important cancellation is expected and is in agreement with the work of 
Philippas and McClure.\cite{philipasmcclure} Using Eqs. (\ref{eq3.42}) and 
(\ref{eq3.43}) to write down $\chi_{CP}$ and $\chi_{ID}$ explicitly and denoting 
the corrected $\chi_{CP}$ and $\chi_{ID}$ by $\chi_{CP}^{\Sigma}$ and 
$\chi_{ID}^{\Sigma}$, respectively, we may write the total magnetic 
susceptibility of interacting free and Bloch electrons as
\begin{equation}
 \chi = \chi_{_{LP}} + \chi_{_{CP}}^{\Sigma} + \chi_{_{ID}}^{\Sigma} 
      \label{eq4.37}
\end{equation}
$\chi_{_{LP}}$ is given by Eq. (\ref{eq4.15}), $\chi_{_{CP}}^{\Sigma}$ and 
$\chi_{_{ID}}^{\Sigma}$ are given by the following relations:
\small
\begin{align}
 \chi_{_{CP}}^{\Sigma} & = - \sum_{\lambda} \bigg( \frac{1}{2 \pi} \bigg)^{3} 
      \int d^{3}k \ k_{B} T \sum_{l} \bigg\{ \Big[ \tilde{H}_{\Delta, 
      \lambda}^{(1)} \big( \vec{k}, \zeta_{l} \big) \Big]^{2} + 
      \tilde{H}_{\Delta, \lambda}^{(1)} \big( \vec{k}, \zeta_{l} \big) 
      \tilde{H}_{\delta, \lambda}^{(1)} \big( \vec{k}, \zeta_{l} \big) \bigg\} 
      \Big[ \mathcal{G}_{\lambda} \big( \vec{k}, \zeta_{l} \big) \Big]^{2} 
      \label{eq4.38}\\
 \chi_{_{ID}}^{\Sigma} & = - \sum_{\lambda} \bigg( \frac{1}{2 \pi} \bigg)^{3} 
      \int d^{3}k \ k_{B} T \sum_{l} 2W_{\lambda}^{(2)} \big( \vec{k}, 
      \zeta_{l} \big) \mathcal{G}_{\lambda} \big( \vec{k}, \zeta_{l} \big) 
      \label{eq4.39}
\end{align}
\normalsize
where
\footnotesize 
\begin{align}
 \tilde{H}_{\Delta, \lambda}^{(1)} \big( \vec{k}, \zeta_{l} \big) = & \ 
      \Big\langle u_{\lambda}^{0} \Big| \mathcal{H}_{0}^{(1)} \frac{e}{B 
      \hbar c} \Big[ A \big( \nabla_{\vec{k}} \big) \tilde{H}_{\lambda}^{0} 
      \big( \vec{k}, \zeta_{l} \big) \Big] \cdot \nabla_{\vec{k}} 
      u_{\lambda}^{0} \Big\rangle \nonumber\\
 & + \Big\langle u_{\lambda}^{0} \big( \vec{r}, \vec{k}, \zeta_{l} \big) \Big|
      \int d^{3} r^{\prime} e^{i \vec{k} \cdot \big( \vec{r}^{\ \prime} - 
      \vec{r} \big)} \tilde{\Sigma}_{\zeta_{l}}^{0} \big( \vec{r}, \vec{r}^{\ 
      \prime} \big) \frac{e}{B \hbar c} A \big( \vec{r} - \vec{r}^{\ \prime} 
      \big) \cdot \nabla_{\vec{k}} \ u_{\lambda}^{0} \big( \vec{r}^{\ \prime}, 
      \vec{k}, \zeta_{l} \big) \Big\rangle \label{eq4.40}\\
 \tilde{H}_{\delta, \lambda}^{(1)} \big( \vec{k}, \zeta_{l} \big) = & \ 
      \Big\langle u_{\lambda}^{0} \big( \vec{r}, \vec{k}, \zeta_{l} \big) \Big| 
      \int d^{3} r^{\prime} e^{i \vec{k} \cdot \big( \vec{r}^{\ \prime} - 
      \vec{r} \big)} \tilde{\Sigma}_{\zeta_{l}}^{(1)} \big( \vec{r}, \vec{r}^{\ 
      \prime} \big) \ u_{\lambda}^{0} \big( \vec{r}^{\ \prime}, \vec{k}, 
      \zeta_{l} \big) \Big\rangle = \Sigma_{\lambda \lambda}^{(1)0} \big( 
      \vec{k}, \zeta_{l} \big)  \label{eq4.41} \\
 W_{\lambda}^{(2)} \big( \vec{k}, \zeta_{l} \big) = & \ \Big\langle 
      u_{\lambda}^{0} \Big| \mathcal{H}_{0}^{(2)} \Big| u_{\lambda}^{0} 
      \Big\rangle + \Big\langle u_{\lambda}^{0} \Big| \mathcal{H}_{0}^{(1)} 
      \Big| u_{\lambda}^{(1)} \Big\rangle - \Big\langle 
      u_{\lambda}^{0} \Big| \sum_{\vec{q}} e^{ik \cdot q} 
      \tilde{H}_{\lambda}^{0} \big( q, \zeta_{l} \big) \frac{1}{2!} \Big( 
      \frac{e}{B \hbar c} A(q) \cdot \nabla_{\vec{k}} \Big)^{2} \Big| 
      u_{\lambda}^{0} \Big\rangle \nonumber\\
 & + \Big\langle u_{\lambda}^{0} \Big| \frac{e}{B \hbar c} A \big( 
      \nabla_{\vec{k}} \big) \tilde{H}_{\lambda}^{0} \big( k, \zeta_{l} \big) 
      \cdot i \nabla_{\vec{k}} \Big| u_{\lambda}^{(1)} \Big\rangle + 
      \Big\langle u_{\lambda}^{0} \Big| \frac{e}{B \hbar c} A \big( 
      \nabla_{\vec{k}} \big) \tilde{H}_{\lambda}^{(1)} \big( k, \zeta_{l} \big) 
      \cdot i \nabla_{\vec{k}} \Big| u_{\lambda}^{0} \Big\rangle \nonumber\\
 & - \tilde{H}_{\Delta, \lambda}^{(1)} \big( k, \zeta_{l} \big) \frac{1}{4} 
      \frac{e}{\hbar c} \bigg( \frac{\partial}{\partial k_{y}} \Big\langle 
      u_{\lambda}^{0} \Big| i \frac{\partial}{\partial k_{x}} u_{\lambda}^{0} 
      \Big\rangle -\frac{\partial}{\partial k_{x}} \Big\langle u_{\lambda}^{0} 
      \Big| i \frac{\partial}{\partial k_{y}} u_{\lambda}^{0} \Big\rangle 
      \bigg) \nonumber\\
 & + \Big\langle u_{\lambda}^{0} \big( \vec{r}, \vec{k}, \zeta_{l} \big) \Big| 
      \int d^{3} r^{\prime} e^{i \vec{k} \cdot \big( \vec{r}^{\ \prime} - 
      \vec{r} \big)} \tilde{\Sigma}_{\zeta_{l}}^{0} \big( \vec{r}, \vec{r}^{\ 
      \prime} \big) \frac{1}{2!} \bigg( \frac{e}{B \hbar c} A \big( \vec{r} - 
      \vec{r}^{\ \prime} \big) \cdot \nabla_{\vec{k}} \bigg)^{2} \ 
      u_{\lambda}^{0} \big( \vec{r}^{\ \prime}, \vec{k}, \zeta_{l} \big) 
      \Big\rangle \nonumber\\
 & + \Big\langle u_{\lambda}^{0} \big( \vec{r}, \vec{k}, \zeta_{l} \big) \Big|
      \int d^{3} r^{\prime} e^{i \vec{k} \cdot \big( \vec{r}^{\ \prime} - 
      \vec{r} \big)} \tilde{\Sigma}_{\zeta_{l}}^{0} \big( \vec{r}, \vec{r}^{\ 
      \prime} \big) \bigg( \frac{e}{B \hbar c} A \big( \vec{r} - \vec{r}^{\ 
      \prime} \bigg) \cdot \nabla_{\vec{k}} \bigg) \ u_{\lambda}^{(1)} \big( 
      \vec{r}^{\ \prime}, \vec{k}, \zeta_{l} \big) \Big\rangle \nonumber\\
 & - \frac{1}{2} \sum_{\lambda^{\prime} \neq \lambda} \Big( 
      \tilde{H}_{\lambda^{\prime}}^{0} - \tilde{H}_{\lambda}^{0} \Big)^{-1} 
      \bigg( \Big\langle u_{\lambda^{\prime}}^{0} \Big| 
      \mathcal{H}_{\Delta}^{(1)} \Big| u_{\lambda}^{0} \Big\rangle \Big\langle 
      u_{\lambda}^{0} \Big| \mathcal{H}_{\delta}^{(1)} \Big| 
      u_{\lambda^{\prime}}^{0} \Big\rangle - \Big\langle u_{\lambda}^{0} \Big| 
      \mathcal{H}_{\Delta}^{(1)} \Big| u_{\lambda^{\prime}}^{0} \Big\rangle 
      \Big\langle u_{\lambda^{\prime}}^{0} \Big| \mathcal{H}_{\delta}^{(1)} 
      \Big| u_{\lambda}^{0} \Big\rangle \bigg)  \label{eq4.42}
\end{align}
\normalsize
Indeed, $\chi$ is a linear function of the operator 
$\tilde{\Sigma}_{\zeta_{l}}^{(1)}$ and is independent of 
$\tilde{\Sigma}_{\zeta_{l}}^{(2)}$ For reasons which maybe clarified in some 
well-known cases, we will refer to the $\tilde{\Sigma}_{\zeta_{l}}^{(1)}$ term 
in $\chi_{CP}^{\Sigma}$ as the ``enhancement term''. Consequently, we will
also refer to the $\tilde{\Sigma}_{\zeta_{l}}^{(1)}$ terms in 
$\chi_{ID}^{\Sigma}$ as the ``second-order effect of the enhancement''.

\subsection{Application of $\chi$ to Some Many-Body Systems}

The general formula will be applied to $(a)$ a Fermi liquid and to $(b)$
correlated electrons represented by the Hubbard model,\cite{white} in
Hartree-Fock approximation for simplicity. In what follows, electric charge
$e \Longrightarrow -e$.

\subsubsection{Fermi liquid}

Since the periodic wave function $u_{\lambda}^{0}(r, k, \zeta_{l})$ occurring in
Eqs. (\ref{eq4.40}) and (\ref{eq4.42}) is a constant quantity for Fermi liquids, 
we can immediately write down the magnetic susceptibility of the quasiparticles 
as
\begin{equation}
 \chi = \chi_{_{LP}} + \chi_{_{CP}}^{\Sigma} \label{eq5.1}
\end{equation}
We obtain using Eqs. (4.62), (4.67), and (4.69) of Ref. \cite{white}, the total 
quasiparticle energy [more appropriately the Weyl transform with $\vec{p} + 
(e/c) \vec{A}(\vec{q}) \Longrightarrow \hbar \vec{k}$] in a magnetic field as
\begin{equation}
 \tilde{H}_{\lambda} \big( \vec{k}, B, \zeta_{l} \big) = e^{0} \big( \vec{k} 
      \big) 
      \begin{pmatrix}
       1 & 0\\
       0 & 1
      \end{pmatrix}
 + B \mu_{B} 
      \begin{pmatrix}
       1 & 0\\
       0 & -1
      \end{pmatrix}
 + \frac{B \chi_{_{CP}}^{\Sigma}}{\mu_{B}} \frac{\pi^{2} \hbar^{2}}{m^{\ast} 
      k_{F}} B_{0} 
      \begin{pmatrix}
       -1 & 0\\
       0 & 1
      \end{pmatrix}
      \label{eq5.2}
\end{equation}
where $m^{\ast} = (1 + \frac{1}{3} A_{1})m; A_{1}$ and $B_{0}$ are well-known
Fermi-liquid parameters. We immediately identify, upon examination of Eqs.
(\ref{eq3.56}), (\ref{eq4.40}), and (\ref{eq4.41}), the following relations:
\begin{align}
 \tilde{H}_{\lambda}^{0} \big( k, \zeta_{l} \big) & = e^{0}(k)
      \begin{pmatrix}
       1 & 0\\
       0 & 1
      \end{pmatrix}
      \label{eq5.3}\\
 \tilde{H}_{\Delta, \lambda}^{(1)} \big( k, \zeta_{l} \big) & = \mu_{B}
      \begin{pmatrix}
       1 & 0\\
       0 & -1
      \end{pmatrix}
      \label{eq5.4}\\
 \tilde{H}_{\delta, \lambda}^{(1)} \big( k, \zeta_{l} \big) & = 
      \frac{\chi_{_{CP}}^{\Sigma}}{\mu_{B}} \frac{\pi^{2} \hbar^{2}}{m^{\ast} 
      k_{F}} B_{0} 
      \begin{pmatrix}
       -1 & 0\\
       0 & 1
      \end{pmatrix}
      \label{eq5.5}
\end{align}
Substituting these quantities in $\chi_{_{CP}}^{\Sigma}$, Eq. (\ref{eq4.38}), we
get
\begin{equation}
 \chi_{_{CP}}^{\Sigma} = \bigg[ \bigg( \frac{1 + \frac{1}{3} A_{1}}{1 + B_{0}} 
      \bigg) \bigg] \chi_{_{P}}^{0} \label{eq5.6}
\end{equation}
where $\chi_{_{P}}^{0}$ is the Pauli spin susceptibility for a noninteracting
electron gas. The calculation of $\chi_{_{LP}}$ is very elementary and the total
$\chi$ is thus given by
\begin{equation}
 \chi = \bigg( 1 + \frac{1}{3} A_{1} \bigg)^{-1} \chi_{_{LP}}^{0} + \bigg[ 
      \bigg( \frac{1 + \frac{1}{3} A_{1}}{1 + B_{0}} \bigg) \bigg] 
      \chi_{_{P}}^{0} \label{eq5.7}
\end{equation}
This is a very well-known result for the orbital and spin susceptibility of
Fermi liquids. Note that a small effective mass enhances $\chi_{LP}^{0}$.

\subsubsection{Hubbard model in Hartree-Fock approximation}

The model under consideration assumes that there is only one band of interest
energetically far removed from the other bands. For a very narrow band we may
write
\begin{equation}
 \chi \simeq \chi_{_{CP}}^{\Sigma} + \chi_{ID}^{\Sigma} \label{eq5.8}
\end{equation}
Upon transforming Eqs. (4.75) and (4.76) of Ref. \cite{white} to $k$ space, we
have for the expression of the total Hubbard Hamiltonian in a magnetic field
in the Hartree-Fock approximation as
\small
\begin{equation}
 H = \sum_{k, \sigma} e \big( \vec{k} \big) n \big( \vec{k}, \sigma \big) + 
      \sum_{k, \sigma} I \Big\langle n \big( \vec{k}, \sigma \big) \Big\rangle 
      n \big( \vec{k}, -\sigma \big) + \frac{1}{2} g \mu_{B} B \sum_{k} \Big[ n 
      \big( \vec{k}, \uparrow \big) - n \big( \vec{k}, \downarrow \big) \Big] 
      \label{eq5.9}
\end{equation}
Therefore, $\tilde{H}_{\lambda} \big( \vec{k}, \zeta_{l}, B \big)$ is given by
\begin{align}
 \tilde{H}_{\lambda} \big( \vec{k}, \zeta_{l}, B \big) = & e \big( \vec{k} 
      \big) 
      \begin{pmatrix}
       1 & 0\\
       0 & 1
      \end{pmatrix}
      + I 
      \begin{pmatrix}
       \big\langle n \big( \vec{k}, \downarrow \big) \big\rangle & 0\\
       0 & \big\langle n \big( \vec{k}, \uparrow \big) \big\rangle
      \end{pmatrix}
 + \frac{1}{2} g \mu_{B} B
      \begin{pmatrix}
       1 & 0\\
       0 & -1
      \end{pmatrix}
      \label{eq5.10}
\end{align}
In view of the fact that $\big\langle n \big( \vec{k}, \downarrow \big) 
\big\rangle$ is greater than $\big\langle n \big( \vec{k}, \uparrow \big) 
\big\rangle$, we may write
\begin{align}
 \Big\langle n \big( \vec{k}, \downarrow \big) \Big\rangle & = n + \delta n 
      \label{eq5.11}\\
 \Big\langle n \big( \vec{k}, \uparrow \big) \Big\rangle & = n - \delta n 
      \label{eq5.12}\\
 \frac{2N \delta n}{V} \frac{1}{2} g \mu_{B} & = \chi_{_{CP}} B \label{eq5.13}
\end{align}
and readily obtain
\begin{align}
 \tilde{H}_{\lambda}^{0} \big( k, \zeta_{l} \big) & = \Big[ e^{0}(k) + In \Big] 
      \begin{pmatrix}
       1 & 0\\
       0 & 1
      \end{pmatrix}
      \label{eq5.14}\\
 \tilde{H}_{\Delta, \lambda}^{(1)} \big( k, \zeta_{l} \big) & = \frac{1}{2} g 
      \mu_{B}
      \begin{pmatrix}
       1 & 0\\
       0 & -1
      \end{pmatrix}
      \label{eq5.15}\\
 \tilde{H}_{\delta, \lambda}^{(1)} \big( k, \zeta_{l} \big) & = I \Big( 
      \frac{V}{N} \Big) \frac{\chi_{_{CP}}}{g \mu_{B}}
      \begin{pmatrix}
       1 & 0\\
       0 & -1
      \end{pmatrix}
      \label{eq5.16}
\end{align}
Upon substitution of these quantities in Eq. (\ref{eq4.38}), we obtain%
\begin{equation}
 \chi_{_{CP}}^{\Sigma} = \chi_{_{0}} \bigg( 1 - 2I \bigg( \frac{V}{N} \bigg) 
      \frac{\chi_{_{0}}}{(g \mu_{B})^{2}} \bigg)^{-1} \label{eq5.17}
\end{equation}
leading to the Stoner criterion for the appearance of ferromagnetism. 
\cite{white}

To obtain $\chi_{_{ID}}^{\Sigma}$, we note that in Eq. (\ref{eq4.42})
\begin{equation}
 W_{\lambda}^{(2)} \big( \vec{k}, \zeta_{l} \big) \simeq \big\langle 
      u_{\lambda}^{0} \big( \vec{r}, \vec{k} \big) \big| \mathcal{H}_{0}^{(2)} 
      \big| u_{\lambda}^{0} \big( \vec{r}, \vec{k} \big) \big\rangle. 
      \label{eq5.18}
\end{equation}
The second term, representing a Van Vleck paramagnetism, and last term of Eq.
(\ref{eq4.42}) are neglected since the band of interest is energetically far
removed from other bands. The third up to sixth term, inclusive, are neglected
by the assumption of a very narrow band and the rest of Eq. (\ref{eq4.42}) is
neglected due to the $\delta$-function locality of 
$\tilde{\Sigma}_{\zeta_{l}}^{0} \big( \vec{r}, \vec{r}^{\ \prime} \big)$.  
Expressing $e^{ik \cdot r} u_{\lambda}(r, k, \zeta_{l})$ as a linear combination 
of atomic orbitals we obtain, upon substitution in Eq. (\ref{eq4.39}), a 
familiar ``atomic diamagnetism'' multiplied by the total number of electrons 
$N$ in the band
\begin{equation}
 \chi_{_{ID}}^{\Sigma} = - \bigg( \frac{Ne^{2}}{4mc^{2}} \bigg) \big\langle
      \phi_{\lambda}(r) \big| x^{2} + y^{2} \big| \phi_{\lambda}(r) \big\rangle 
      \label{eq5.19}
\end{equation}
where $\phi_{\lambda}(r)$ is the atomic orbital of the band. For most purposes 
$\chi_{ID}^{\Sigma}$ is neglected and $\chi \simeq \chi_{CP}^{\Sigma}$.

\section{Magnetic Susceptibility of Dilute Nonmagnetic Alloys}

Theoretical efforts toward giving a general expression ofr the magnetic
suseptibility $\chi$ for solids with nonmagnetic impurities were initiated by
Kohn and Luming\cite{kohnluming} by considering an idealized model of
free-electron band. Other attempts to give $\chi$ for general Bloch bands can
at best proceed only as a power-series expansion in the strength of the
impurity potential. However, experimental data indicate the need for a better
understanding and a more complete theory that incorporates the band-structure
effects of the host lattice.

\subsection{Lattice Weyl-Wigner Formalism Approach}

We approach the problem by the use of the lattice Weyl-Wigner formalism of 
quantum theory, not very widely known in solid-state physics, although its 
embryonic and disguised form is already apparent in the operator method of 
Roth\cite{roth} and Blount\cite{blount} and in the formalism of the dynamics of 
band electrons by Wannier\cite{wannierBT}. The result for $\chi$ is given to 
order $\hbar^{2}$ valid for general nondegenerate Bloch bands and to all orders 
in the impurity potential. The effect of Bloch-electron interaction can, in 
principle, be incorporated by the use of a screened impurity potential. The 
expression for $\chi$ reduces to all well-known limiting cases. It is applied to 
the free-electron-band model of dilute alloys of copper. The result gives a firm 
theoretical foundation to the empirical theory of Henry and Rogers, which 
accounts quite well of their experimental results. 

The final result for $\chi$, the change of the magnetic susceptibility of the
crystalline solid due to the presence of impurity centers, may be written as,
\begin{equation*}
 \Delta \chi = N_{I}(\chi - \chi_{0})
\end{equation*}
where $N_{I}$ is the number of impurity centers, $\chi_{0}$ is the magnetic 
susceptibility of the pure crystal host, and $\chi$ is given by the following 
formula:
\footnotesize
\begin{equation}
 \chi = -\frac{1}{V} \bigg( \frac{1}{h} \bigg)^{3} Tr \int d^{3}p d^{3}q \left(
    \begin{array} [c]{c}
     \frac{\partial f \Big( \Sigma^{0} \Big)}{\partial \Sigma^{0}} \big( 
        \Sigma^{(1)} \big)^{2} + f \big( \Sigma^{0} \big) \Sigma^{(2)}\\
        -\frac{\hbar^{2}}{48} \left\{
        \begin{array} [c]{c}
         \frac{\partial f \big( \Sigma^{0} \big)}{\partial \Sigma^{0}} \left(
            \begin{array} [c]{c}
             \left[ \frac{\partial^{2} \Sigma^{(2)}}{\partial \vec{p} 
                \partial \vec{p}}; \frac{\partial^{2} V^{0}}{\partial \vec{q} 
                \partial \vec{q}} \right] + 2 \left[ \frac{\partial^{2} 
                \Sigma^{(1)}} {\partial \vec{p} \partial \vec{p}}; 
                \frac{\partial^{2} V^{(1)}}{\partial \vec{q} \partial \vec{q}} 
                \right] \\
             + \left[ \frac{\partial^{2} \Sigma^{0}}{\partial \vec{p} \partial 
                \vec{p}}; \frac{\partial^{2} \Sigma^{(2)}}{\partial \vec{q} 
                \partial \vec{q}} \right] - \left[ \frac{\partial^{2} 
                \Sigma^{(2)}} {\partial \vec{p} \partial \vec{q}}; 
                \frac{\partial^{2} V^{0}}{\partial \vec{q} \partial \vec{p}} 
                \right] \\
             - 2 \left[ \frac{\partial^{2} \Sigma^{(1)}}{\partial \vec{p} 
                \partial \vec{q}}; \frac{\partial^{2} \Sigma^{(1)}}{\partial 
                \vec{q} \partial \vec{p}} \right] - \left[ \frac{\partial^{2} 
                V^{0}}{\partial \vec{p} \partial \vec{q}}; \frac{\partial^{2} 
                \Sigma^{(2)}}{\partial \vec{q} \partial \vec{p}} \right]
            \end{array}
         \right) \\
         \Sigma^{(2)} \frac{\partial^{2} f \left( \Sigma^{0} \right)}{\left(  
            \partial \Sigma^{0} \right)^{2}} \left( \left[ \frac{\partial^{2} 
            \Sigma^{0}}{\partial \vec{p} \partial \vec{p}}; \frac{\partial^{2} 
            V^{0}}{\partial \vec{q} \partial \vec{q}} \right] - \left[ 
            \frac{\partial^{2} V^{0}}{\partial \vec{p} \partial \vec{q}}; 
            \frac{\partial^{2} V^{o}}{\partial \vec{q} \partial \vec{p}} 
            \right] \right) \\
         + \left( \Sigma^{(1)} \right)^{2} \frac{\partial^{3} f \left( 
            \Sigma^{0} \right)}{\left( \partial \Sigma^{0} \right)^{3}} \left( 
            \left[ \frac{\partial^{2} \Sigma^{0}}{\partial \vec{p} \partial 
            \vec{p}}; \frac{\partial^{2} V^{0}}{\partial \vec{q} \partial 
            \vec{q}} \right] - \left[ \frac{\partial^{2} V^{0}}{\partial 
            \vec{p} \partial \vec{q}}; \frac{\partial^{2} V^{0}}{\partial 
            \vec{q} \partial \vec{p}} \right] \right) \\
         + 2 \Sigma^{(1)} \frac{\partial^{2} f \left( \Sigma^{0} \right)}{\left( 
            \partial \Sigma^{0} \right)^{2}} \left( 
            \begin{array} [c]{c}
             \left[ \frac{\partial^{2} \Sigma^{(1)}}{\partial \vec{p} \partial 
                \vec{p}}; \frac{\partial^{2} V^{0}}{\partial \vec{q} \partial 
                \vec{q}} \right] + \left[ \frac{\partial^{2} 
                \Sigma^{0}}{\partial \vec{p} \partial \vec{p}}; 
                \frac{\partial^{2} V^{(1)}}{\partial \vec{q} \partial \vec{q}} 
                \right] \\
             - \left[ \frac{\partial^{2} \Sigma^{(1)}}{\partial \vec{p} \partial 
                \vec{q}}; \frac{\partial^{2} V^{0}}{\partial \vec{q} \partial 
                \vec{p}} \right]  -\left[ \frac{\partial^{2} V^{0}}{\partial 
                \vec{p} \partial \vec{q}}; \frac{\partial^{2} 
                \Sigma^{(1)}}{\partial \vec{q} \partial \vec{p}}\right]
            \end{array}
         \right)
        \end{array}
     \right\}
    \end{array}
 \right)  \label{eq37alloy}
\end{equation}
\normalsize
For magnetic field in the $z$-direction using a symmetric gauge, the various
quantities entering in the above expression are defined as follows: $f(x)$ is 
the Fermi-dirac distriution function, and
\begin{align}
 \Sigma^{0} = & W^{0} \big( \vec{p} \big) + V^{0} \big( \vec{p}, \vec{q} \big), 
       \label{eq38alloy}\\
 \Sigma^{(1)} = & \bigg( \frac{e}{2c} \bigg) \big( \vec{q} \times 
      \nabla_{\vec{p}} \big)_{z} \big( \vec{q} \times \nabla_{\vec{p}} 
      \big)_{z} W^{0} \big( \vec{p} \big) + W^{(1)} \big( \vec{p} \big) + 
      V^{(1)} \big( \vec{p}, \vec{q} \big),  \label{eq39alloy}\\
 \Sigma^{(2)} = & \bigg( \frac{e}{2c} \bigg)^{2} \big( \vec{q} \times 
      \nabla_{\vec{p}} \big)_{z} \big( \vec{q} \times \nabla_{\vec{p}} 
      \big)_{z} W^{0} \big( \vec{p} \big) + \bigg( \frac{e}{c} \bigg) \big( 
      \vec{q} \times \nabla_{\vec{p}} \big)_{z} W^{(1)} \big( \vec{p} \big)  
      \nonumber\\
 & + 2W^{(2)} \big( \vec{p} \big) + 2V^{(2)} \big( \vec{p}, \vec{q} \big) 
      \label{eq40alloy}
\end{align}
For simplicity one may take $V^{0} \big( \vec{p}, \vec{q} 
\big)_{\lambda \lambda^{\prime}} = V_{\lambda}^{0} \big( \vec{p},\vec{q} 
\big) \delta_{\lambda \lambda^{\prime}}$, making $\Sigma^{0}$ diagonal in 
bands. Equation (\ref{eq37alloy}) has the novel features of being transparent 
and of being valid to all orders in the impurity potential for general 
nondegenerate Bloch bands.

\subsection{Application to Dilute Alloys of Copper}

We will show that in the free-electron-band model Eq. (\ref{eq37alloy}) gives a 
firm theoretical foundation of the empirical theory of Henry and Rogers 
\cite{henryrogers} which accounts quite well for their experimental results for 
various solutes in copper. This is in marked contrast to the theory of Kohn and 
Luming\cite{kohnluming} which fails to justify the formula for $\Delta \chi$ 
per solute atom used by Henry and Rogers.

For the free-electrons in cooper, we have
\begin{align}
 \Sigma^{0} & = \frac{p^{2}}{2m} + V_{I} \big( \vec{q} \big),\label{eq41alloy}\\
 \Sigma^{(1)} & = \bigg( \frac{e}{2mc} \bigg) \big( \vec{q} \times 
      \vec{p} \big)_{z} + \mu_{B} 
      \begin{pmatrix}
       1 & 0\\
       0 & -1
      \end{pmatrix}
      ,\label{eq42alloy}\\
 \Sigma^{(2)} & = \bigg( \frac{e}{2c} \bigg)^{2} \frac{1}{m} \Big( q_{x}^{2} + 
      q_{y}^{2} \Big). \label{eq43alloy}
\end{align}
Thus, Eq. (\ref{eq37alloy}) reduces to
\begin{equation}
 \chi = \chi_{_{1}} + \chi_{_{2}} + \chi_{_{3}} + \chi_{_{4}} + \chi_{_{5}} 
      \label{eq44alloy}
\end{equation}
where
\begin{align}
 \chi_{1} & = -2 \frac{1}{Vh^{3}} \int d^{3}p \ d^{3}q \ f^{\prime} \Big( 
      \Sigma^{0} \Big) \bigg[ \bigg( \frac{e}{2mc} L_{z} \bigg)^{2} + 
      \mu_{B}^{2} \bigg],  \label{eq45alloy}\\
 \chi_{2} & = -2 \frac{1}{Vh^{3}} \int d^{3}p \ d^{3}q \ f \Big( \Sigma^{0} 
      \Big) \frac{e^{2}}{4mc^{2}} \Big( q_{x}^{2} + q_{y}^{2} \Big),  
      \label{eq46alloy}\\
 \chi_{3} & = 2 \frac{1}{Vh^{3}} \int d^{3}p \ d^{3}q \ f^{\prime} \Big( 
      \Sigma^{0} \Big) \mu_{B}^{2},   \label{eq47alloy}\\
 \chi_{4} & = 2 \frac{1}{24Vh^{3}} \int d^{3}p \ d^{3}q \ f^{\prime \prime 
      \prime} \Big( \Sigma^{0} \Big) \frac{\hbar^{2}}{m} \nabla^{2} V_{I}(q) 
      \bigg[ \bigg( \frac{e}{2mc} L_{z} \bigg)^{2} + \mu_{B}^{2} \bigg], 
      \label{eq48alloy}\\
 \chi_{5} & = 2 \frac{1}{24Vh^{3}} \int d^{3}p \ d^{3}q \ f^{\prime \prime} 
      \Big( \Sigma^{0} \Big) \frac{\hbar^{2}}{m} \nabla^{2} V_{I}(q) 
      \frac{e^{2}}{4mc^{2}} \Big( q_{x}^{2} + q_{y}^{2} \Big)  \label{eq49alloy}
\end{align}
where factors of $2$ in front of integrals account for the $\pm$ spin band. We
note that for $V_{I}(q) = 0$, we have
\begin{equation*}
 \chi = \chi^{0} = \chi_{spin}^{0} + \chi_{orb}^{0}
\end{equation*}
where
\begin{align}
 \chi_{orb}^{0} & = \frac{2}{3Vh^{3}} \int d^{3}p \ d^{3}q \ f^{\prime} \bigg( 
      \frac{p^{2}}{2m} \bigg) \mu_{B}^{2}, \label{eq50alloy}\\
 \chi_{spin}^{0} & = -\frac{2}{Vh^{3}} \int d^{3}p \ d^{3}q \ f^{\prime} \bigg( 
      \frac{p^{2}}{2m} \bigg) \mu_{B}^{2}. \label{eq51alloy}
\end{align}
In view of Eqs. (\ref{eq45alloy}), (\ref{eq47alloy}), and (\ref{eq48alloy}), and 
Eqs. (\ref{eq50alloy}) and (\ref{eq51alloy}), we can write for an arbitrary 
strength of $V_{I}(q)$,
\begin{align}
 \Delta \chi_{orb} & = \chi_{orb}^{0} \frac{\Delta g_{1}}{g}, \label{eq52alloy} 
      \\
 \Delta\chi_{spin} & = \chi_{spin}^{0} \frac{\Delta g_{2}}{g}, \label{eq53alloy}
\end{align}
where
\begin{align}
 g & = \frac{2}{Vh^{3}} \int d^{3}p \ d^{3}q \ f^{\prime} \bigg( 
      \frac{p^{2}}{2m} \bigg),  \label{eq54alloy}\\
 \Delta g_{1} & = \frac{2}{Vh^{3}} \int d^{3}p \ d^{3}q \bigg[ f^{\prime} \Big( 
      \Sigma^{0} \Big) - \ f^{\prime} \bigg( \frac{p^{2}}{2m} \bigg) \bigg],  
      \label{eq55alloy}\\
 \Delta g_{2} & = \Delta g_{1} + \frac{2}{24Vh^{3}} \int d^{3}p \ d^{3}q 
      f^{\prime \prime \prime} \Big( \Sigma^{0} \Big) \frac{\hbar^{2}}{m} 
      \nabla^{2} V_{I}(q)  \label{eq56alloy}%
\end{align}
By writing $\big( L_{z} \big)^{2} = (q \times p)_{z}^{2} = q_{x}^{2} p_{y}^{2} - 
2q_{x} q_{y} p_{x} p_{y} + q_{y}^{2} p_{x}^{2}$ and integrating with respect to 
$\vec{p}$, the first terms of $\chi_{1}$, and $\chi_{4}$ can be combined with 
$\chi_{2}$ and $\chi_{5}$, resulting in the expression for $\Delta\chi$ per 
solute atom as
\begin{equation}
 \Delta \chi = -\frac{e^{2}}{6mc^{2}} \int d^{3}q \ \Delta \rho \big( \vec{q} 
      \big) \ |q|^{2} + \chi_{orb}^{0} \frac{\Delta g_{1}}{g} + \chi_{spin}^{0} 
      \frac{\Delta g_{2}}{g}, \label{eq57alloy}
\end{equation}
where
\begin{align}
 \Delta \rho \big( \vec{q} \big) = & \frac{2}{Vh^{3}} \int d^{3} p \ \bigg\{ 
      \bigg[ f \Big( \Sigma^{0} \Big) + \frac{1}{24} f^{\prime \prime} \Big( 
      \Sigma^{0} \Big) \frac{\hbar^{2}}{m} \nabla^{2} V_{I}(q) - f \Big( 
      \frac{p^{2}}{2m} \Big) \bigg] + \bigg[ f^{\prime} \Big( \Sigma^{0} 
      \Big) \nonumber \\
 & + \frac{1}{24} f^{\prime \prime \prime} \Big( \Sigma^{0} \Big) 
      \frac{\hbar^{2}}{m} \nabla^{2} V_{I}(q) - f^{\prime} \Big( 
      \frac{p^{2}}{2m} \Big) \bigg] \frac{p^{2}}{3m} \bigg\} \label{eq58alloy}
\end{align}
Equation (\ref{eq57alloy}), with $\Delta g_{1} = \Delta g_{2}$ and a similar
consistent approximation for $\rho\left(  \vec{q}\right)  $, is exactly the
expression used by Henry and Rogers,\cite{henryrogers} as pointed out by Kohn
and Luming\cite{kohnluming}, in analyzing their data on dilute alloys of Zn,
Ga, Ge, and As with Cu which accounts quite well for their experimental results. 
Thus the use of Eq. (\ref{eq57alloy}) by Henry and Rogers, as pointed out by 
Kohn and Luming, is given a firm theoretical foundation. Here lies the essential 
discrepancy between Eq. (\ref{eq57alloy}) and the theory of Kohn and Luming. We 
believe that the copper conduction electron can be approximately described by a 
free electron-band model and Eq. (\ref{eq37alloy}) should provide a good 
approximation for copper as used by Henry and Rogers. On the other hand, the 
theory presented by Kohn and Luming does not contain the entire emperical 
expression for $\Delta \chi$ per solute atom used by Henry and Rogers. 
\cite{henryrogers}

\section{Magnetic Susceptibility of Dilute Magnetic Alloys}

For completeness, we will briefly treat the magnetic susceptibility of dilute 
magnetic alloys. The study of dilute magnetic alloys evolves with the general 
problem of how magnetism develops in magnets. As it turns out the study of 
dilute magnetic alloys has stood out as a complex many-body problem dealing
with self-consistent fluctuating scattering potential, and hence becomes an
important physics problem in its own right. Self-consistency in the sense that
the electrons interacting with the impurity themselves create the fluctuating
potential, signaling a \textit{bonafide} many-body and perhaps a time-dependent 
and/or highly nonlinear problem. Roughly speaking, each electron as it passes 
the impurity influences the state of the impurity and is influenced by the 
impurity. Therefore, the state of the ion which a given electron sees is 
determined by all previous electron-impurity encounters. The problem is 
essentially a many-body problem with nonstationary impurity potential.

The increase in resistance due to strong fluctuations at low temperature is
reminiscent of the Anderson localization problem with random scattering 
potential. Here, the magnetic impurities are represented by localized spins that 
couple to the conduction-band electrons of the nonmagnetic host metal via a 
spin-exchange interaction, in particular via anti-ferromagnetic coupling. 
Whether the magnetic moment of the impurity persists down to zero temperature is 
not very well understood and this has given way to the so-called Kondo problem, 
where one studies the low-temperature behavior of a system. 

\begin{figure}[H]
 \centering
 \includegraphics[width=4in]{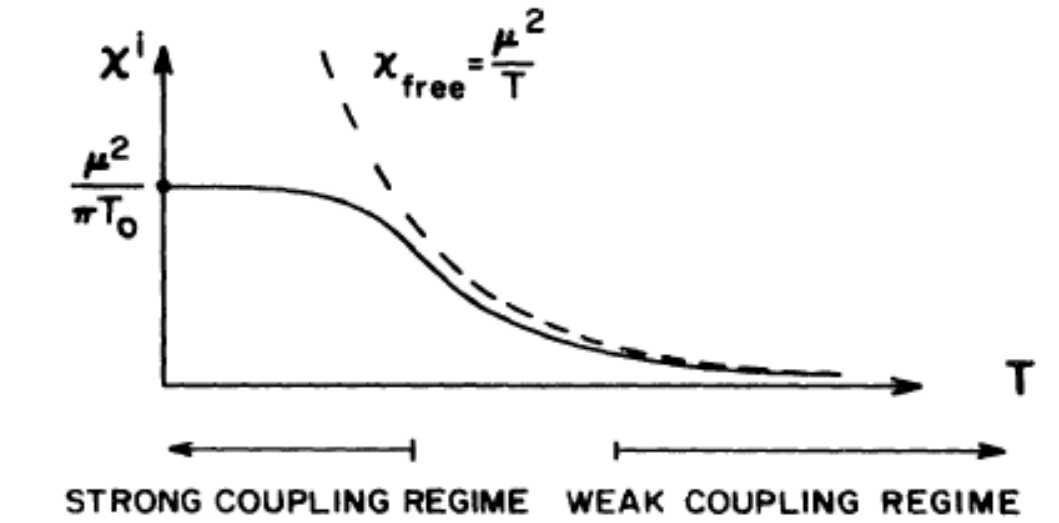}
\caption{The \textit{impurity} susceptibility $\chi^{i}$ is shown compared with 
         the free-spin susceptibility $\chi_{free}$. At high temperatures 
         $\chi^{i}$ approaches $\chi_{free}$ logarithmically on the scale set by
         $T_{K}.$ As the temperature is lowered, it goes to a finite value at 
         $T = 0$, indicative of a screened spin. Reproduced after 
         Ref. \cite{Andreietal}} \label{weak-strong}
\end{figure}

At sufficiently low impurity density we may concentrate on a single impurity 
localized, say, at $x = 0$ and study how its magnetic properties are modified 
due to its coupling with the electrons. A rough hand-waving argument of the 
physics of the system may be made based on the time scale of observation versus 
the spin-flip scattering or spin relaxation time. Denoting the spin-flip 
relaxation time by $\tau$ and the time scale of observation by $\Delta t = 
\frac{\hbar}{k_{B} T}$, where $k_{B}$ is the Boltzmann constant and $T$ is the 
temperature, then when $\Delta t\ll\tau$, the spin orientation almost remains 
constant during the time $\tau$. This condition defines the weak coupling regime 
(note that Schreiffer\cite{Schrieffer} use the interaction time, $\tau_{U}$, 
instead of $\Delta t$ of Kondo\cite{kondobook}). On the other hand when 
$\Delta t \gg \tau$, then spin-flip becomes frequent and the up and down 
orientation of spin appear equally, i.e., there is a strong fluctuation of 
impurity potential due to frequent spin flip during the time $\Delta t$. This 
condition defines the strong coupling regime, it is referred to as the screening 
or \textit{quenching} of the impurity spin since one observes vanishing impurity 
spin, leading to finite susceptibility at $T = 0$. There is a temperature 
region where $\Delta t \simeq \tau$, the corresponding temperature is often 
referred to as the Kondo temperature, denoted by $T_{K}$, after Jun Kondo who 
first study the dilute magnetic alloy problem in 1964. For $T < T_{K}$, we have 
the strong coupling regime while for $T > T_{K}$, we have the weak coupling 
regime, as shown in Fig. \ref{weak-strong}. The magnetic susceptibility changes 
from Curie behavior for $T > T_{K}$ to Pauli spin susceptibility for $T 
< T_{K}$. Thus, the impurity is said to have a moment if the susceptibility due 
to the impurity shows a $\frac{1}{k_{B} T}$ dependence (such as $\chi_{spin}$ 
in Eq. (\ref{eq51alloy}) down to $T = 0$.

\subsection{States of Magnetic Impurity in Nonmagnetic Metal Host}

Here we consider only the Hamiltonian of impurity in nonmagnetic metals and see 
how the \textit{localized} spin can be generated by the interaction with 
conduction electrons. Following Kondo,\cite{kondobook} we write the `bare'
Hamiltonian as
\begin{equation}
 H = \sum_{\vec{k}\sigma} \varepsilon_{\vec{k}} a_{\vec{k} \sigma}^{\dagger} 
      a_{\vec{k} \sigma} + V_{o} \sum_{\vec{k} \sigma} \big( a_{\vec{k} 
      \sigma}^{\dagger} a_{0 \sigma} + a_{0 \sigma}^{\dagger} a_{\vec{k} 
      \sigma} \big) + \varepsilon_{0} \sum_{\sigma} a_{0 \sigma}^{\dagger} a_{0 
      \sigma} + Ua_{0 \uparrow}^{\dagger} a_{0 \uparrow} a_{0 
      \downarrow}^{\dagger} a_{0 \downarrow} \label{kondobare}
\end{equation}
where $\varepsilon_{\vec{k}}$ is the conduction energy-band function, 
$\varepsilon_{0}$ is the impurity orbital $l = 0$ energy level, $U$ is the is
the Coulomb interaction energy between electrons, and $V_{0}$ is the
self-consistent potential. Note that $V_{0}$ is sufficient enough for fixing
the values of the impurity levels if $U$ is not large, i.e., the problem then
is simply a one-body problem. However, if $U$ is large the problem becomes a
complex many-body problem.

\subsection{Generation of localized moment}

Consider the $U$ term in Eq. (\ref{kondobare}). If the spin $\uparrow$ electron 
fills the localized orbital, i.e., $a_{0 \uparrow}^{\dagger} a_{0 \uparrow} = 
1$ then the last two terms reduces to
\begin{equation*}
 \varepsilon_{0} \sum_{\sigma} a_{0 \sigma}^{\dagger} a_{0 \sigma} + Ua_{0 
      \uparrow}^{\dagger} a_{ 0 \uparrow} a_{0 \downarrow}^{\dagger} a_{0 
      \downarrow} \Longrightarrow (\varepsilon_{0} + U) a_{0 
      \downarrow}^{\dagger} a_{0 \downarrow}
\end{equation*}
This shows that the electron down spin state is raised up in energy by the
Coulomb interaction $U$. If it happens that $(\varepsilon_{0} + U) >  
\varepsilon_{F}$ then $\Big\langle a_{0 \downarrow}^{\dagger} a_{0 
\downarrow} \Big\rangle \Longrightarrow 0$ and \textit{localized} 
$\uparrow$-spin is generated. The same argument holds if $\uparrow$ and 
$\downarrow$ are interchanged. On the other hand localized spin does not emerge 
for the following two cases: (a) $\{\varepsilon_{0}, (\varepsilon_{0} + 
U)\} < \varepsilon_{F}$ and (b) $\{\varepsilon_{0}, (\varepsilon_{0} + 
U)\} > \varepsilon_{F}$.

\subsubsection{Fluctuating localized moment}

\begin{figure}[H]
 \centering
 \includegraphics[width=6.5in]{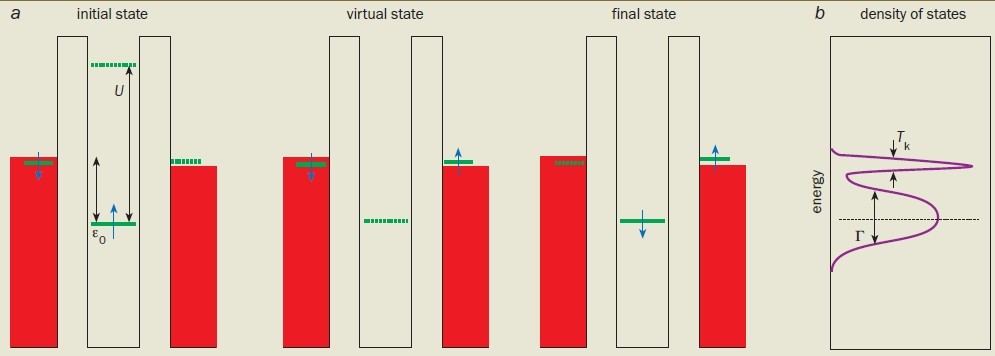}
 \caption{\small{Spin flip process: (a) Anderson model assumes just one electron 
          level, $\varepsilon_{o}$, below the Fermi energy of the metal. This 
          level is occupied by spin $\uparrow$ electron. Adding another electron 
          is prohibited by the Coulomb energy, $U$. It also cost energy, 
          $\varepsilon_{0}$, to remove the electron from the impurity (assuming 
          Fermi level is set to zero). However, quantum mechanically, the spin 
          up electron may tunnel out of the impurity site to briefly occupy a 
          classically forbidden `virtual state' outside the impurity. Then this 
          is replaced by the electron from the metal. This effectively flip the 
          spin of the impurity. (b) Many such events combine to produce the 
          appearance of of an extra resonance at the Fermi energy. This extra 
          resonance can remarkably change the conductance. [Figure reproduce 
          from Ref. \cite{KouwenhovenGlazman}]}} \label{fig3}
\end{figure}

The localized spin if present is actually fluctuating by virtue of the
spin-exchange process. A spin $\downarrow$ electron from the conduction band
may fall into the localized orbital, which has been filled by the spin
$\uparrow$ electron from the conduction band. Then and exchange process occurs
in which the spin $\uparrow$ electron goes back to the conduction band thereby
resulting in an overall spin exchange and hence a localized $\downarrow$ spin
emerges, and vice versa. This is best illustrated for quantum dots as shown in
Fig. \ref{fig3}. Note in magetic alloys, a complementary process corresponding
to that depicted for quantum dots in Fig. \ref{fig3}, can also occur in which
the spin $\downarrow$ electron from the conduction band leap to the level
$\varepsilon_{0} + U$ of the localized orbital, then fall down the level
$\varepsilon_{0}$ kicking the spin $\downarrow$ electron back to the
conduction band, again resulting in overall spin exchange.\cite{kondobook} The
most likely event between the two processes of course depends on the magnitude
of the Coulomb energy $U$ as well as on $\varepsilon_{0}$, as can be seen by
simply sliding down the $U$ and $\varepsilon_{0}$ (keeping their distance
fixed) in Fig. \ref{fig3} .

Whether indeed the localized spin is effectively present or not depends on the 
time scale of interaction, also referred to by Kondo\cite{kondobook} as the time 
scale of observation. This time scale is determined by the width of the 
localized energy level, designated as $\Delta_{0}$. If the fluctuation occurs 
much more slowly than the observation time, then the localized spin is resolved 
or is present. On the contrary, if the observation time is sufficiently greater 
than the spin fluctuation, say at low temperature, then the localized spin is no 
longer resolve and appears to have vanished. This often referred to as screened 
or compensated localized moment.

\subsection{The $s$-$d$ interaction}

In the Kondo problem, it is the two-electron interaction $U$ that makes the
fluctuating localized moment and brings the problem to a higher degree of
difficulty. When the two-electron interaction $U$ is large and is dominant in
the problem, Kondo was able to obtain an effective Hamiltonian with the
so-called $s$-$d$ interaction terms, $H_{sd}$, where $H_{sd}$ is given by%
\begin{equation*}
 H_{sd} = - J \sum_{\vec{k} \vec{k}^{\prime}} \bigg[ S_{z} \bigg( a_{k^{\prime} 
      \uparrow}^{\dagger} a_{k \uparrow} - a_{k^{\prime} \downarrow}^{\dagger} 
      a_{k \downarrow} \bigg) + S_{+} a_{k^{\prime} \downarrow}^{\dagger} a_{k 
      \uparrow} + S_{-} a_{k^{\prime} \uparrow}^{\dagger} a_{k \downarrow} 
      \bigg] + V \sum_{\vec{k} \vec{k}^{\prime} \sigma} a_{k^{\prime} 
      \sigma}^{\dagger} a_{k \sigma}
\end{equation*}
where
\begin{align*}
 J & = V_{0}^{2} \bigg( \frac{1}{\varepsilon_{0}} - \frac{1}{\varepsilon_{0} 
      + U} \bigg) < 0 \\
 V & = -\frac{V_{0}^{2}}{2} \bigg( \frac{1}{\varepsilon_{0}} + 
      \frac{1}{\varepsilon_{0} + U} \bigg)
\end{align*}
The term proportional to $J$ is called the $s$-$d$ interaction.

\subsection{Bethe \textit{Ansatz} Treatment of Exact Solution: Chiral
Gross-Neveu Model}

The Kondo problem has been interesting both in its own right where powerful
mathematical techniques and other ideas have been tested for its solution. It
has been approached by various theories, namely, perturbation theory, various
resummation techniques, S-matrix formalism, dispersion relation, 
renormalization-group techniques, etc. For reviews see Kondo\cite{kondo},
Gruner and Zawadowski,\cite{gruner} Wilson\cite{wilson}, and Nozieres. 
\cite{nozieres} A different line approach of the problem is via an exact 
diagonalization of the Hamiltonian, one using the Bethe \textit{ansatz}. 
\cite{bethe} Indeed, Bethe \textit{ansatz} has become a powerful technique in
solving the excitation spectrum of complex many-body systems.

We are here interested in the results of the exact diagonalization of the Kondo 
Hamiltonian given by Andrei. As it turns out, the Kondo model belongs to a class 
of exactly soluble models, also noted by Yang and Yang\cite{YangYang}. This was 
further shown by using the analogy to the soluble chiral Gross-Neveu model 
\cite{GrossNeveu} of elementary particle theory\cite{andrei,wiegman}. The chiral 
Gross-Neveu (or backscattering) model describes particles interacting via spin 
exchange and differs from the Kondo model only in that some of the particles are 
left-moving electrons rather than stationary impurities.

This similarity allows Andrei\cite{andrei} to take over the formalism developed 
in the diagonalization of the Gross-Neveu model \cite{andrei}; partly developed 
also by \cite{belavin} and apply it with only minor modification to the Kondo 
Hamiltonian. In general, chiral Gross-Neveu model, the Kondo model, and also the 
Heisenberg model has been shown to be very similar from the 
Bethe-\textit{ansatz} point of view, all being spin exchange models differing 
only in the kinetic properties of their constituents.

What Andrei did is to transform the Kondo Hamiltonian to the form of the chiral 
Gross-Neveu (or backscattering) model, so that one can immediately see the 
connection with the backscattering model, which describes left- and right-moving 
electrons interacting via a spin exchange. Andrei's transformed Kondo 
Hamiltonian, $\mathcal{H}_{Kondo}$, is given by by\cite{Andreietal}
\begin{align}
 \mathcal{H}_{Kondo} = & -i \sum_{\beta = 0,1} \int dx \psi_{a 
      \beta}^{\dagger}(x) \beta \partial_{x} \psi_{a \beta}(x) + J \int dx 
      \psi_{a0}^{\dagger}(x) \sigma_{ab} \psi_{b0}(x) 
      \psi_{a^{\prime}1}^{\dagger}(x) \sigma_{a^{\prime} b^{\prime}} 
      \psi_{b^{\prime}1}(x) \nonumber\\
 & + J^{\prime} \int dx \psi_{a0}^{\dagger}(x) \psi_{a0}(x) 
      \psi_{b1}^{\dagger}(x) \psi_{b1}(x)  \label{andreiH}%
\end{align}
where the $J^{\prime}$ term is the potential scattering term, whose effect is
merely to renormalize the coupling constant $J$. We have
\begin{equation*}
 \psi_{a}(x) = \left(
 \begin{array} [c]{c}
  \phi_{a}(x) \\
  \chi_{a}(x)
 \end{array}
 \right), \qquad \textrm{where components are labeled by Greek indices, e.g, } 
      \alpha
\end{equation*}
$\alpha = 1$ for electron wavefunction, $\phi_{a}(x)$, and $\alpha = 2$ for the 
impurity wavefunction, $\chi_{a}(x)$. Note that in the kinetic energy for the 
impurity, $\beta = 0$, and that the impurity has no contribution in the kinetic 
energy. The fields $\psi_{a \alpha}(x)$ are assumed to have canonical 
anti-commutation relations,
\begin{align*}
 \big\{ \psi_{a \alpha}(x), \psi_{b \beta}(y) \big\} & = 0 \\
 \big\{ \psi_{a \alpha}(x), \psi_{b \beta}^{\dagger}(y) \big\} & 
      = \delta_{ab} \delta_{\alpha \beta} \delta(x - y)
\end{align*}
In this form, the connection becomes apparent with the backscattering model,
which describes left- and right-moving electrons interacting via a spin
exchange. It is of the same form as $\mathcal{H}_{Kondo}$, with the only
difference that $\beta = \pm 1$ indicating left and right movers rather than
$\beta = 0$ or $1$, with $\beta = 1$ indicating a right-moving electron and
$\beta = 0$ indicating a stationary particle, an impurity.

Since the backscattering model (\textit{aka} the chiral Gross-Neveu) was
solved by a Bethe-\textit{ansatz} method,\cite{Andreietal, belavin} it is
clear that the Kondo model is also exactly soluble model. We will not go into
the details of Andrei's Bethe \textit{ansatz} method of exact solution of the
Kondo problem since this will take us very far from the scope of this review.

\subsection{Impurity Magnetic Susceptibility}

Here we will give the result of the magnetic susceptibility given by Andrei,
et al\cite{Andreietal} using the Bethe \textit{ansatz} method.

The impurity susceptibility attains its free value $\chi^{i} = 
\frac{\mu^{2}}{k_{B} T}$ (Curie law) up to corrections that vanish 
logarithmically at high temperatures
\begin{equation}
 \chi^{i} \Longrightarrow_{T \gg T_{0}} \frac{\mu^{2}}{k_{B} T} \Bigg\{ 1 - 
      \bigg( \ln \frac{T}{T_{K}} \bigg) - \frac{1}{2} \bigg( \ln \ln 
      \frac{T}{T_{K}} \bigg) \bigg( \ln^{2} \frac{T}{T_{K}} \bigg)^{-1} + 
      \bigg( \ln \frac{T}{T_{K}} \bigg)^{-3} \Bigg\}  \label{hiT}
\end{equation}
where a new scale $T_{K}$ has been defined by the requirement that the $\Big( 
\ln \frac{T}{T_{K}} \Big)^{-2}$ term be absent. This is equivalent to a 
normalization condition on the high temperature scale, $T_{K}$, which is
conventionally referred to as the Kondo temperature.

Consider the Curie law $\chi^{i} = \frac{\mu^{2}}{k_{B} T}$, which is the 
leading term in Eq. (\ref{hiT}). Its divergence at $T = 0$ indicates a net 
impurity spin. However, due to the strong interaction with the electrons the 
impurity spin will be quenched (screened) leading to a finite susceptibility at 
zero temperature. At $T = 0$, $\chi^{i}(T = 0)$ may be written as,
\begin{equation*}
 \chi^{i}(T = 0) = \frac{\mu^{2}}{\pi k_{B} T_{0}}
\end{equation*}
where $T_{0}$ is the scale that characterizes the low temperature regime. The
ratio
\begin{equation*}
 W = \frac{T_{k}} {T_{0}}
\end{equation*}
is a universal number. It characterizes the crossover from the weak coupling,
which is perturbatively accessible, to the strong coupling regime that has to
be constructed nonperturbatively. This was obtained numerically using
renormalization group technique by Wilson.\cite{wilson} Moreover, the exact
diagonalization of the Hamiltonian using Bethe \textit{ansatz} by Andrei et
al,\cite{Andreietal} is able to give an analytic expression for $W$.
\subsection{Kondo Effect and Nanotechnology}

The Kondo effect has found strong revival in nanoscience and nanotechnology.
Various groups around the world have exploited chip technology to fabricate
small semiconductor devices for investigating fundamental problems in physics.
One such device is the quantum dot.\cite{KouMarcus} Figure \ref{fig3}
illustrates the spin-flip process in quantum dots. Quantum dots are often
called artificial atoms since their electronic properties resemble those of
real atoms. A good report on the various research initiatives around the world
is given by Kouwenhoven and Glazman.\cite{KouwenhovenGlazman}


\begin{thebibliography}{99}
\bibitem {footnote2}Chirality pertains to the spinor wavefunctions, whereas,
          helicity pertains to the particle. For Dirac equations, chirality is 
          Lorentz invariant whereas for massive particles helicity is not 
          Lorentz invariant. Only for massless particle does chirality and 
          helicity properties coincide.
\end{thebibliography}
\end{document}